\documentclass[%
    reprint,
    showpacs,preprintnumbers,
    showkeys,
    amsmath,
    amssymb,
    aps,
    prl,
    floatfix,
    longbibliography
]{revtex4-1}

\usepackage[english]{babel}
\usepackage[utf8]{inputenc}
\usepackage{graphicx}
\usepackage{bm}
\usepackage{hyperref}
\usepackage{xcolor}
\usepackage{epsfig}
\usepackage{amsmath}
\usepackage{amsfonts}
\usepackage{amssymb}
\usepackage{amsthm}
\usepackage{tikz}
\usepackage{lipsum}
\usepackage{dsfont}
\usepackage[all]{nowidow}
\usepackage{enumerate}
\usepackage{empheq}
\usepackage{pdfpages}
\usepackage{pgffor}

\makeatletter
\AtBeginDocument{\let\LS@rot\@undefined}
\makeatother

\definecolor{refColor}{HTML}{EA00F2}
\definecolor{figColor}{HTML}{008DF2}
\definecolor{urlColor}{HTML}{00AEF2}

\hypersetup{%
    unicode=false,                 
    pdftoolbar=true,               
    pdfmenubar=true,               
    pdffitwindow=false,            
    pdfstartview={FitH},           
    pdfnewwindow=true,             
    colorlinks=true,               
    linkcolor=figColor,            
    citecolor=refColor,            
    filecolor=magenta,             
    urlcolor=urlColor              
}

\newcommand{\Bra}[1]{\left<#1\right|}
\newcommand{\Ket}[1]{\left|#1\right>}

\renewcommand{\vec}[1]{\mathbf{#1}}

\renewcommand{\vec}[1]{\boldsymbol{#1}}

\newcommand{\hc}{\mathrm{h.c.}}
\renewcommand{\ol}{\overline}
\newcommand{\rs}{\rm \scriptscriptstyle}
\newcommand{\const}{\textrm{const}}

\newcommand{\rrangle}{\rangle\rangle}
\newcommand{\llangle}{\langle\langle}

\newcommand{\ds}{\mathrm{d}s}

\newcommand{\F}{\mathcal{F}}

\newcommand{\K}{\mathcal{K}}
\renewcommand{\P}{\mathcal{P}}
\newcommand{\I}{\mathcal{I}}

\newcommand{\CaptionMark}[1]{\textit{#1}}
\newcommand{\CaptionLabel}[1]{\textbf{(#1)}}

\newcommand{\HSSH}{H_{\rs SSH}}

\newcommand{\HbSSH}{H_{\rs bSSH}}
\newcommand{\HbMC}{H_{\rs bMC}}

\newcommand{\hHSSH}{\hat{H}_{\rs SSH}}

\newcommand{\hHbSSH}{\hat{H}_{\rs bSSH}}
\newcommand{\hHbMC}{\hat{H}_{\rs bMC}}
\newcommand{\hHbB}{\hat{H}_{\rs bB}}
\newcommand{\hHbP}{\hat{H}_{\rs bP}}

\newcommand{\emode}{\tilde b}
\newcommand{\Cmode}{\emode_{\rs C}}
\newcommand{\Tmode}{\emode_{\rs T}}
\newcommand{\oi}{{\overline{i}}}

\newcommand{\bw}{\bar w}
\newcommand{\bo}{\bar \omega}
\newcommand{\bt}{\bar t}

\newcommand{\btmax}{\bt_{\rs max}}
\newcommand{\Dbw}{\Delta\bw}

\newcommand{\bwmax}{\bw_{\rs max}}
\newcommand{\bwmin}{\bw_{\rs min}}
\newcommand{\Dbwmin}{\Dbw_{\rs min}}

\newcommand{\Dobarrier}{\Delta\omega_{\rs barrier}}
\newcommand{\Dobarriermin}{\Dobarrier^{\rs min}}
\newcommand{\oedge}{\omega_{\rs edge}}
\newcommand{\obarrier}{\omega_{\rs barrier}}
\newcommand{\obarriermin}{\obarrier^{\rs min}}
\newcommand{\obarriermax}{\obarrier^{\rs max}}

\newcommand{\dom}{\delta\bar\omega}

\newcommand{\omm}{\bar\omega_-}
\newcommand{\omp}{\bar\omega_+}
\newcommand{\domax}{\dom_{\rs max}}

\newcommand{\transfer}{\mathcal{O}}
\newcommand{\AVtransfer}{\llangle\mathcal{O}\rrangle}
\newcommand{\edgeweight}{\mathcal{E}}
\newcommand{\AVedgeweight}{\llangle\mathcal{E}\rrangle}
\newcommand{\phase}{\varphi}

\newcommand{\DEedge}{\Delta E_{\rs edge}}
\newcommand{\DEbulk}{\Delta E_{\rs bulk}}
\newcommand{\DEbulkmin}{\DEbulk^{\rs min}}

\newcommand{\CL}{C_L}
\newcommand{\eL}{\varepsilon_L}

\newcommand{\Tup}{\Ket{1}_{\rs T}}
\newcommand{\Tdown}{\Ket{0}_{\rs T}}
\renewcommand{\Cup}{\Ket{1}_{\rs C}}
\newcommand{\Cdown}{\Ket{0}_{\rs C}}
\newcommand{\Caux}{\Ket{a}_{\rs C}}
\newcommand{\Taux}{\Ket{a}_{\rs T}}

\newcommand{\UCP}{\mathcal{U}_{\rs {CP}}}
\newcommand{\PiT}{\Pi_{\rs T}}
\newcommand{\PiC}{\Pi_{\rs C}}
\newcommand{\TrCT}{T_{\rs C\leftrightarrow \rs T}}
\newcommand{\UE}{\mathcal{U}_{\rs SWAP}}
\newcommand{\Cmodeup}{\Ket{1}_{\Cmode}}
\newcommand{\Tmodeup}{\Ket{1}_{\Tmode}}

\begin{document}


\title{Topological networks for quantum communication between distant qubits}

\author{Nicolai Lang}
\email{nicolai@itp3.uni-stuttgart.de}

\author{Hans Peter Büchler}

\affiliation{Institute for Theoretical Physics III and Center for Integrated Quantum Science and Technology,
University of Stuttgart, 70550 Stuttgart, Germany}

\date{\today}


\begin{abstract}
    Efficient communication between qubits relies on robust networks 
    which allow for fast and coherent transfer of quantum information.
    It seems natural to harvest the remarkable properties of systems characterized by 
    topological invariants to perform this task.  
    Here we show that a linear network of coupled bosonic degrees of freedom, characterized by topological bands,
    can be employed for the efficient exchange of quantum information over large distances.  
    Important features of our setup are that it is robust against quenched disorder,
    all relevant operations can be performed by \textit{global} variations of parameters,
    and the time required for communication between distant qubits approaches linear scaling with their distance.
    We demonstrate that our concept can be extended to an ensemble of qubits embedded in a two-dimensional 
    network to allow for communication between all of them.  
\end{abstract}


\maketitle

\section{Introduction}

Systems characterized by topological invariants are well known to exhibit unique properties
with potential applications in quantum information processing and engineering~\cite{Nayak2008}.
Ever since the first experimental observation of the integer quantum hall effect~\cite{Ando1975,Klitzing1980,Laughlin1981},
many other condensed matter systems have been identified and experimentally characterized, such as
fractional quantum Hall fluids~\cite{Tsui1982a,Stormer1983b,Willett1987} and topological insulators 
and superconductors~\cite{Konig2007,Hsieh2008,Xu2014,Schnyder2008,Kitaev2009a,Schnyder2009a,Ryu2010,Hasan2010,Qi2011}.
The latter belong to a particularly well understood family of topological systems described by non-interacting fermions,
where topological invariants can be defined on classes of random matrices~\cite{Wigner1951,Wigner1958,Dyson1962,Altland1997}. 
This concept can be straightforwardly generalized to bosonic setups as well as classical systems~\cite{Hafezi2013,Susstrunk2015,Nash2015},
where the topological features still give rise to intriguing properties such as localized and chiral edge modes. 
Here we are interested in such systems: We demonstrate that their topological properties can be 
harvested for robust and efficient transfer of quantum information over large distances.

Several different platforms for the realization of topological systems of artificial matter 
with bosonic degrees of freedom are currently explored: The construction of topological 
band structures and the observation of edge states has been achieved with photonic circuits 
in the optical~\cite{Hafezi2013,Rechtsman2013,Ling2015} and the radio-frequency~\cite{Ningyuan2015} regime, 
as well as with classical coupled harmonic oscillators~\cite{Susstrunk2015,Nash2015,Susstrunk2017}, 
and with cold atomic gases~\cite{Atala2013,Jotzu2014,Aidelsburger2014,Mancini2015,Stuhl2015,Duca2015,Lohse2015}. 
These experimental advances have been prepared and are supported by many theoretical proposals, e.g.
\cite{Haldane2008,Koch2010,Hafezi2011,Berg2011,Yannopapas2012,Kane2013,Lu2014,Kariyado2015,Wang2015a,Wang2015b,Peter2015}.
Several of the above platforms are suitable to carry a single quantized excitation with low losses and dissipation
along protected edge channels which opens the opportunity to harvest topological phenomena for 
guiding and transmitting quantum information reliably.
First approaches in this direction have been proposed~\cite{Yao2013,Dlaska2017} and primarily focus 
on the transmission of excitations along protected edge modes on the boundary of a two-dimensional, 
topologically non-trivial medium.

Here we show that a linear network of coupled bosonic degrees of freedom, characterized by topological bands,
can be employed for the highly efficient exchange of quantum information over large distances. We demonstrate 
the superiority of this setup over its topologically trivial counterparts and exemplify its application 
with the implementation of a robust quantum phase gate. 
Our proposal is based on a (quasi) one-dimensional setup, characterized by a $\mathbb{Z}$ topological index~\cite{Zak1989},
and derived from paradigmatic systems such as Kitaev's Majorana chain~\cite{Kitaev2001} and the Su-Schrieffer-Heeger (SSH) model~\cite{Su1979}.
It features symmetry protected, localized edge modes, the extend and overlap of which can be tuned via coupling parameters to
facilitate controllable communication between them.
Important features of our setup are that relevant operations can be performed by \textit{global} variations of parameters, 
its robustness with respect to the pulse shapes used for the transfer protocol, 
and that the time for the transfer scales favourably with the separation of the qubits. 
This high gain in performance is bought by more complex preparation schemes as the coupling parameters 
have to respect symmetries protecting the topological invariants~\cite{Hafezi2013,Ningyuan2015,Susstrunk2015,Susstrunk2017}.
Finally, we demonstrate that our concept can be extended to an ensemble of qubits embedded in a two-dimensional 
network of local bosonic degrees of freedom to allow for communication between all of them.  

\begin{figure*}[t]
    \centering
    \includegraphics[width=1.0\linewidth]{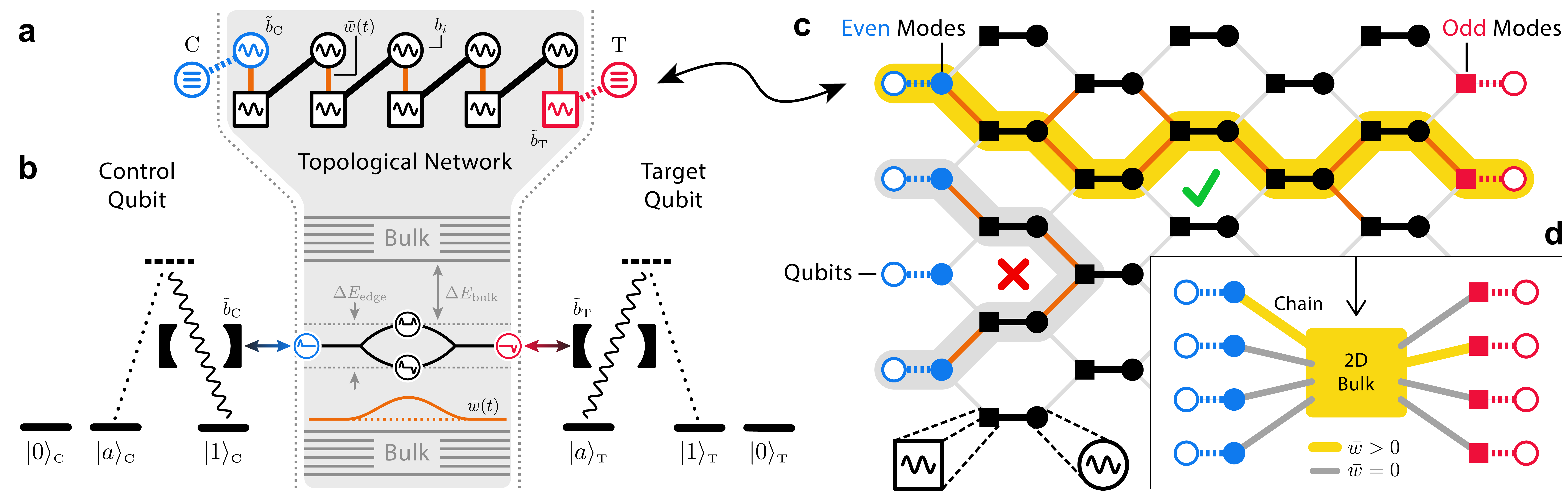}
    \caption{%
        \CaptionMark{Topological quantum networks.}
        \CaptionLabel{a}
        One-dimensional chain of bosonic modes $b_i$ with globally tunable onsite couplings $\bw(t)$ (orange); 
        derived from the SSH chain.
        It features a topological band structure with localized edge modes (in the topological phase) 
        which can be coupled to local qubits $\rm T$ and $\rm C$.
        \CaptionLabel{b}
        The local qubits are realized as three-level systems with logical states $\Ket{0}_p$, $\Ket{1}_p$ and the auxiliary state
        $\Ket{a}_p$. The state $\Caux$ ($\Tup$) can decay into $\Cup$ ($\Taux$) via an off-resonant transition and thereby 
        emits an excitation into the localized edge mode $\Cmode$ ($\Tmode$). 
        Tunnelling excitations between the two edges is facilitated by tuning the chain close to the
        phase transition via $\bw(t)$ and adiabatically decoupling bulk from edge modes.
        \CaptionLabel{c}
        Possible 2D generalization of the network on a dimerized honeycomb lattice. 
        Scattering-free transport is guaranteed by topological protection which requires a sublattice symmetry. 
        The latter is realized by directly coupling only ``even'' (filled circles) with ``odd'' modes (filled squares).
        State transfer between qubits (empty circles) of different (the same) type is possible (impossible),
        illustrated by the bold yellow (grey) path. 
        Stray couplings in the bulk (shown for the upper path) are not detrimental to the transfer fidelity.
        \CaptionLabel{d}
        Instead of locating the edge modes (filled coloured circles and squares) with their qubits at the boundary, emanating SSH chains can be used to
        separate the qubits from each other and the 2D bulk. There is no need to trace out a specific path as in
        \CaptionLabel{c}, but a weak addressability of the individual chains is sufficient whereas the couplings of the bulk can be tuned
        globally. Details are given at the end of the manuscript.
    }
\label{fig:setup-general}
\end{figure*}

We consider macroscopically separated qubits that are coupled by a linear 
quantum network, see Fig.~\ref{fig:setup-general}~(a) for an example. The quantum network itself is constructed from bosonic degrees of 
freedom with only local couplings between them, and generically described by the Hamiltonian
\begin{equation}
    \hat H_{n}=\sum_{i,j}\,b_i^\dag\,H_{ij}\,b_j\,.
    \label{CavityNetwork}
\end{equation}
Here, $b^{\dag}_{i}$ ($b_{i}$) are bosonic creation (annihilation) operators accounting for the mode at site $i$ with
$H_{i j}$ the coupling amplitudes.  
The network is designed such (see below) that at the end $p$ of each branch, a localized bosonic edge mode $\emode_p$
emerges with a controllable coupling between this mode and a local qubit.
The conceptually simplest setup to envisage is an optical network coupled to a single
atom with the level structure shown in Fig.~\ref{fig:setup-general}~(b). 
There the coupling Hamiltonian for the target qubit $\rm T$ takes the form (within the rotating wave approximation)
\begin{equation}
    \hat{H}_{\rs T}(t) = g_{\rs T}(t)\,\left[\,\Tmode^\dag\:|a\rangle\langle 1|_{\rs T}+\Tmode\:|1\rangle\langle a|_{\rs T}\,\right]\,.
    \label{edgecoupling}
\end{equation}
The coupling $g_{\rs T}(t)$ is controlled by external laser fields and allows for the application of $\pi$-pulses 
between the qubit state $\Tup$ and the edge mode $\Tmode$, i.e., the emission of a photon into the edge mode $\Tmode$  
from state $\Tup$ is accompanied by a transition into the auxiliary state $\Taux$; 
in the following, we denote such a $\pi$-pulse at edge $p$ by the unitary operation $\Pi_{p}$.  
Note that the Hamiltonian $\hat{H}_{\rs C}(t)$ for the control qubit $\rm C$ is similar, with the role of $\Cup$ and $\Caux$ exchanged. 

Several fundamental quantum information processing tasks between the qubits reduce to the transfer of edge excitations
within the linear network; 
we denote the corresponding unitary operation that describes the transfer of excitations between edges $p$ and $q$
as $T_{p \leftrightarrow q}$. 
As an example, the protocol for a controlled phase (CP) gate between
a control qubit at position ${\rm C}$ and a target qubit at position ${\rm T}$ reads
\begin{equation}
    \UCP=    
    \PiT
    \circ \TrCT
    \circ \PiC^2
    \circ \TrCT
    \circ \PiT\,.
    \label{CPgate}
\end{equation}
Another example, the transport of a control qubit to a target position, is simply
described by the protocol
\begin{equation}
    \UE=
    \mathcal{A}_{\rs C} \circ
    \PiT
    \circ \PiC
    \circ \TrCT
    \circ \PiC 
    \circ \PiT
    \circ \mathcal{A}_{\rs C}\,.
    \label{exchangegate}
\end{equation}
Here, $\mathcal{A}_{p}$ denotes the exchange of the two states  $|1\rangle_{p}$ and $|a\rangle_{p}$.
Note that this operation even performs the full exchange of the two qubits due to the linearity of the network;
a detailed discussion of these operations can be found at the end of the manuscript.

\begin{figure*}[t]
    \centering
    \includegraphics[width=1.0\linewidth]{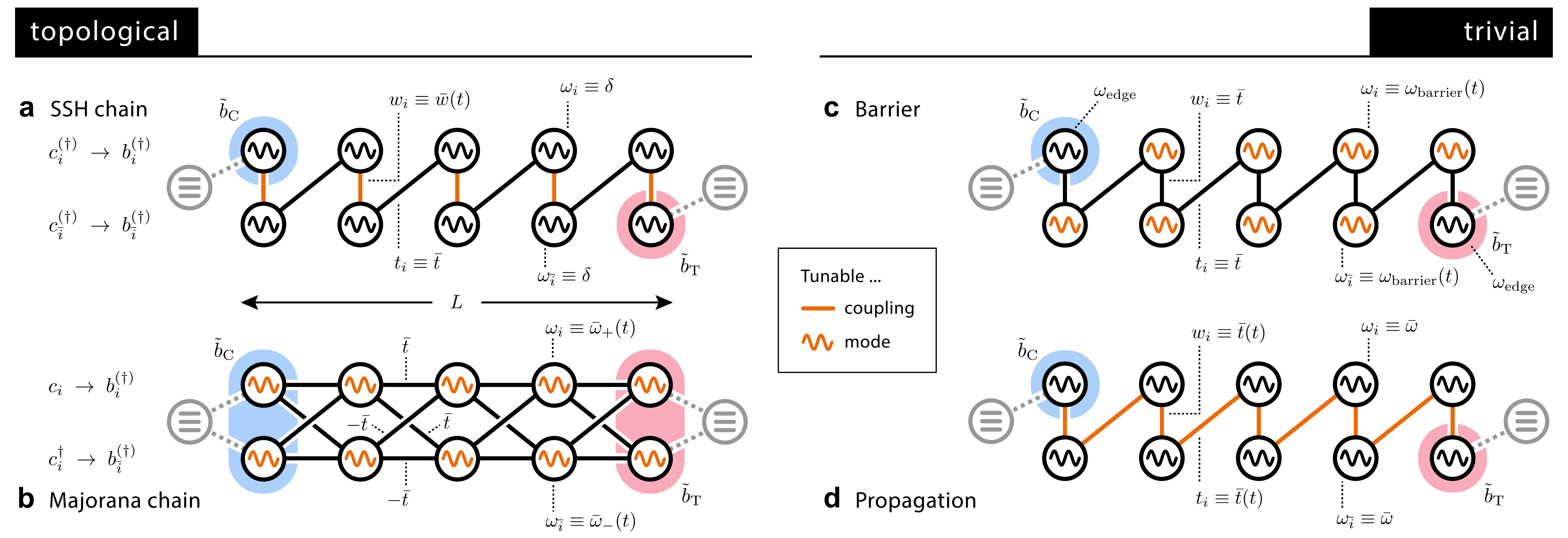}
    \caption{%
        \CaptionMark{Setups.} 
        Possible setups for state transfer via global control parameters.
        The left panel depicts networks with topological bands.
        The models in the right panel are used for comparison and feature only trivial bands.
        Each model consists of locally coupled bosonic modes with qubits coupled to the edge modes
        $\Cmode$ and $\Tmode$.
        \CaptionLabel{a} 
        The SSH chain inspired setup described by $\hHbSSH(t)$; 
        globally tunable are only the onsite couplings $\bw(t)$.
        \CaptionLabel{b} 
        Network inspired by the Majorana chain (unitarily equivalent to the SSH setup) and described by $\hHbMC(t)$;
        the homogeneous eigenfrequency differences $\dom(t)\equiv\omm(t)-\omp(t)$ are tunable.
        For details we refer the reader to Supplementary Information Section I.
        \CaptionLabel{c} 
        Model with two artificial edge modes separated by a simple tunnelling barrier of tunable
        eigenfrequencies $\obarrier(t)$; described by the Hamiltonian $\hHbB(t)$.
        \CaptionLabel{d} 
        The simplest model, based on free propagation of excitations and described by $\hHbP(t)$;
        all couplings $\bt(t)$ are tuned simultaneously.
    }
\label{fig:setup-details}
\end{figure*}

Motivated by these observations, 
we are in the following interested in the efficient transfer ($\TrCT$) of edge excitations within the linear network. 
The basic idea is most conveniently illustrated for two qubits coupled by a one-dimensional network
as illustrated in Fig.~\ref{fig:setup-general}~(a):
The structure of the network gives rise to topological bands with a gapped dispersion relation 
and entails the existence of degenerate and localized edge modes within the bulk gap.
As the existence of edge modes is topologically protected, it is robust against disorder. 
In a \textit{finite} system, the degeneracy of the topological edge states is only lifted exponentially in the edge separation. 
However, globally tuning the quantum network closer to the topological phase transition into the trivial phase 
increases overlap and finite size splitting of the edge states, and eventually allows for a $\pi$-pulse $\TrCT$ between the two edges.
This simple idea is the core of our protocol; compared to topologically trivial systems, 
it features several superior properties regarding the speed of state transfer and its robustness against disorder, 
as well as the absence of individual addressing of each part of the network. 

Finally, we would like to stress that our scheme is generic 
and one can envisage various experimental platforms for its implementation. 
In addition to the discussed optical network~\cite{Hafezi2013}, 
alternative setups are coupled optical cavities and
circuit QED systems~\cite{Houck2012,Carusotto2013} 
as well as trapped polar molecules or Rydberg atoms with a coupling mediated by dipolar exchange interactions~\cite{Peter2015}, 
while the local qubits can be artificial atoms~\cite{Lukin2001},  NV centers in diamond~\cite{Doherty2013}, 
or trapped ions~\cite{Duan2010}. 

\begin{figure*}[t]
    \centering        
    \includegraphics[width=1.0\linewidth]{figure_3}
    \caption{%
        \CaptionMark{State transfer --- Qualitative results.}
        For the four setups depicted in Fig.~\ref{fig:setup-details} and the corresponding protocols described in the text, we
        show the full time evolution of a single excitation that is initially localized in the left mode $\Cmode$ and
        transferred to the right mode $\Tmode$.
        The protocol parameters $\bwmax$ \CaptionLabel{a}, $\domax$ \CaptionLabel{b}, 
        $\obarriermin$ \CaptionLabel{c}, $\btmax$ \CaptionLabel{d}
        and the time scale $\tau$ are tuned to optimize transfer $\transfer$ and edge weight $\edgeweight$. 
        The upper row shows the amplitude of the 
        single-particle wave function under the time evolution prescribed by $\hat H_\bullet(t)$.
        The lower row parallels the time evolution by the single-particle spectrum of $\hat H_\bullet(t)$, i.e., the spectrum of
        $H_\bullet(t)$. In this work, we focus on the topological setup derived from the SSH chain \CaptionLabel{a}
        and compare it with the trivial setup of a simple tunnelling barrier \CaptionLabel{c}.
    }
\label{fig:transfer-qualitative}
\end{figure*}

\section{Results}

\subsection{Topological network}

We start with a description of the requirements on the quantum network $\hat H_n$ to exhibit topologically protected edge modes in  
a one-dimensional chain with two edges, as illustrated in Fig.~\ref{fig:setup-general}~(a); the generalization to 2D networks is 
discussed at the end of the manuscript. 

The most prominent paradigmatic model in one-dimension is the Majorana chain which exemplifies the concept of symmetry protected 
topological phases, originally formulated for spinless fermions with a mean-field $p$-wave pairing term~\cite{Kitaev2001}.
This model is closely related to the Su-Schrieffer-Heeger (SSH) model~\cite{Su1979} in the single-particle picture.  
It turns out that the necessary steps to translate these models into our
bosonic quantum network language are more conveniently performed for the SSH model; 
the discussion of the Majorana chain and its relation to the SSH chain 
is presented in the Supplementary Information Sections I and II. 

The SSH model on a chain with $L$ sites and open boundaries is described by the Hamiltonian
\begin{equation}
    \hHSSH = \sum_{i=1}^{L}\,w_i\: {c_{i}}^{\dag}  c_{\ol i}+\sum_{i=1}^{L-1}\,t_i\:{c_{\ol i}}^{\dag} c^{}_{i+1}+\hc
    \label{eq:SSH}
\end{equation}
with the two fermion operators $c_i$ and $c_\oi$ on each site.  
Note that the indices $i$ label the ``upper'' fermionic modes whereas bar-ed indices $\oi$ denote the ``lower'' ones, see
Fig.~\ref{fig:setup-details}~(a);  we will use upper-case indices $I$ if we refer to both indifferently.
Here, $w_i$ and $t_i$ are the hopping amplitudes. For a uniform system with $w_{i}\equiv \bw$ and
$t_{i}=\bt$, one obtains a gapless point for $\bw=\bt$ separating a topological  phase for $\bw < \bt$ from the trivial 
phase for  $\bw > \bt$. The former features topologically protected edge modes which are fermionic in nature. 
The second-quantized Hamiltonian can be encoded by a matrix $\HSSH$ via
$\hat H_{\rs SSH} = \vec\Phi^\dag \HSSH \vec \Phi$
with the pseudo spinor $\vec\Phi={\left(c_1,c_{\ol 1},\dots,c_L,c_{\ol L}\right)}^T$.
For real hopping amplitudes $w_{i}$ and  $t_{i}$, the Hamiltonian exhibits time reversal symmetry
$T=\K$ where $\K$ denotes complex conjugation. Furthermore, time reversal
$T$ together with the sublattice symmetry $S=U_C^{\rs SSH}$,
represented by the unitary
\begin{equation}
    U_C^{\rs SSH}=\mathds{1}_{L\times L}\otimes
    \begin{pmatrix}
        1 & 0 \\
        0 & -1 \\ 
    \end{pmatrix}\,,
    \label{eq:sshsym}
\end{equation}
yields the particle-hole (PH) symmetry $C=\K U_C^{\rs SSH}$ with $C^2=+\mathds{1}$. 
Hence, the SSH chain is in symmetry class BDI of the Altland-Zirnbauer classification 
\cite{Wigner1951,Wigner1958,Dyson1962,Altland1997}. 
In one dimension, this allows for the definition of a $\mathbb{Z}$ topological invariant
\cite{Schnyder2008,Kitaev2009a,Schnyder2009a,Ryu2010} 
which is responsible for the emergence of the disorder-resilient edge modes bound to the open ends of the chain  in the topological phase. 

The implementation of an analogue system with bosonic degrees of freedom is straightforward:
We replace the fermionic operators by bosonic ones, i.e., $ c_I^{(\dag)}\rightarrow b_I^{(\dag)}$.
The bosonic Hamiltonian takes the form 
\begin{equation}
    \hHbSSH \equiv \vec\xi^\dag\left[\HSSH+\delta\:\mathds{1}\right]\vec\xi\equiv\vec\xi^\dag \HbSSH\vec\xi
\end{equation}
with $\vec\xi={\left(b_1,b_{\ol 1},\ldots , b_L,b_{\ol L}\right)}^T$ and 
$\HbSSH = \HSSH + \delta \mathds{1}$. 
The constant positive energy shift $\delta>0$ is required to enforce positivity on the matrix $H_{\rs bSSH}$
and accounts for the energy $\omega_{I}$ of each bosonic mode $b_{I}$.  The bosonic Hamiltonian $\hHbSSH$ 
features the same single-particle band structure as the original fermionic chain, and exhibits the same topological properties 
and topological quantum numbers.  Therefore, it gives rise to the same edge modes.
Note that these are statements about \textit{single}-particle physics where statistics is not relevant.
To satisfy the PH symmetry, one must respect the sublattice symmetry (\ref{eq:sshsym}) that protects the topological
invariant. This is equivalent to the constraints
\begin{equation}
    \omega_i=\delta=\omega_\oi
    \label{eq:SSHsymmetries}
\end{equation}
for all sites $i$ and $\oi$, i.e., all bosonic modes must have the same energy.
Note that there are no constraints on the couplings $w_i$ and $t_i$.

To complete the picture, we point out that the bosonic realization of the Majorana chain $\HbMC$ 
is unitarily equivalent to that of the SSH chain $\HbSSH$; this is shown in the Supplementary Information Section II.
However, despite their unitary equivalence, 
the bosonic networks impose different symmetry constraints on the coupling Hamiltonians. 
Therefore, depending on the experimental constraints on protecting the symmetry, 
it may be advantageous to implement one or the other of the unitary equivalent models. 

\begin{figure*}[t]   
    \centering
    \includegraphics[width=1.0\linewidth]{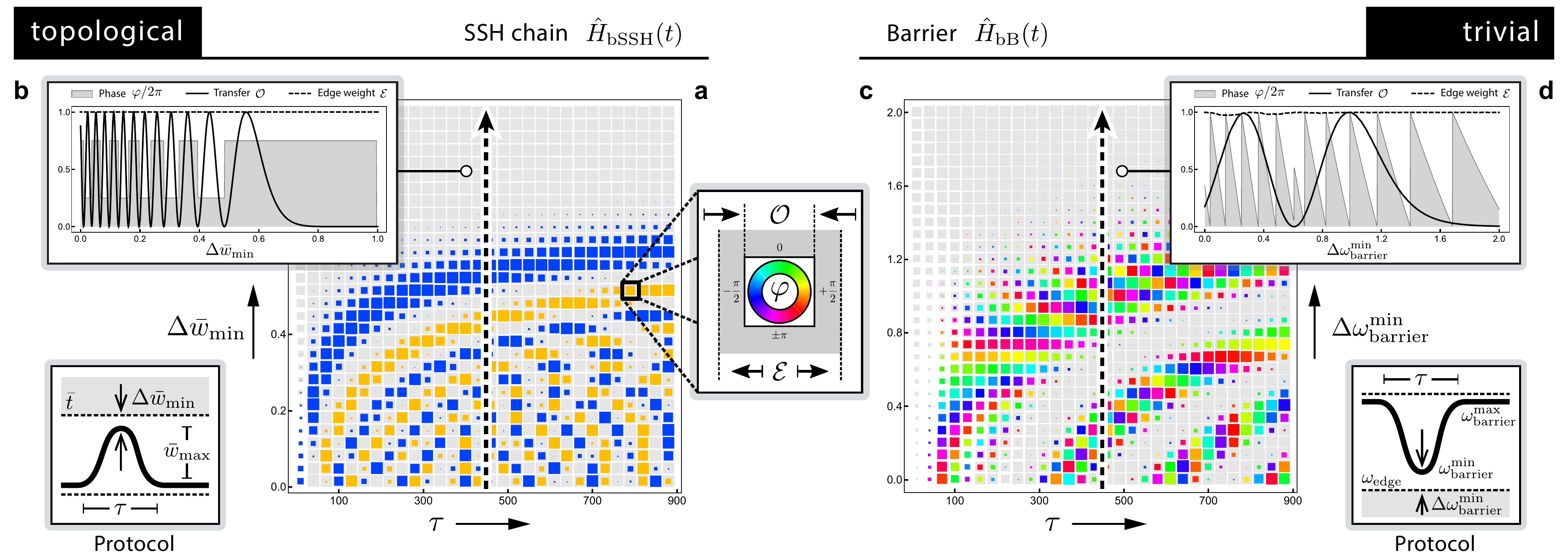} 
    \caption{%
        \CaptionMark{State transfer --- Quantitative results.}
        In \CaptionLabel{a} and \CaptionLabel{c} we plot the figures of merit for transfers driven by 
        $\hHbSSH (t)$ and $\hHbB (t)$ in dependence of the protocol timescale $\tau$ and the distance from
        criticality, namely $\Dbwmin=\bt-\bwmax$ and $\Dobarriermin=\obarriermin-\oedge$.
        The diameter of the grey background squares encodes the edge weight $\edgeweight$; a thinning of the grey background
        tiling therefore indicates a loss of adiabaticity. In the shown parameter regimes, however, 
        the edge weight is almost everywhere close to unity as there is barely any loss to bulk excitations 
        (except for regions of fast protocols close to criticality).
        The diameter (colour) of the coloured squares encodes the transfer $\transfer$ (phase $\phase$) after the protocol
        reached its final state ($\phase$ is measured in the rotating frame of the localized edge modes).
        \CaptionLabel{a} shows results for a topological SSH setup of size $L=5$.         
        \CaptionLabel{c} shows the corresponding data for a trivial tunnelling barrier setup of size $L=5$.         
        In \CaptionLabel{b} and \CaptionLabel{d} we plot $\transfer$, $\phase$, and $\edgeweight$ along the dashed slices in
        \CaptionLabel{a} and \CaptionLabel{c}, respectively.
        Note that the phase is fixed for the topological setup: $\phase=\pm\pi/2$.
    }
\label{fig:transfer-quantitative}
\end{figure*}

\subsection{Protocol for state transfer}

Next, we discuss the protocol for state transfer. One of the key features of the protocol is that we require only
global (translational invariant), time-dependent tuning of the hopping amplitudes $w_{i}$; in particular, single-site addressability 
and control is not required. The goal of the protocol is to coherently transfer a single quantized excitation from one of the localized 
edge modes to the other by means of an adiabatic variation of the couplings $\bw$ in $\HbSSH$.

The crucial point we exploit for state transfer is that in \textit{finite} systems and in the topological phase 
(for $0<\bw<\bt$), there is a finite overlap between the edge modes due to their
exponential extension into the bulk. While deep in the topological phase this overlap is exponentially 
suppressed, it can be be strongly enhanced by tuning $\bw$
closer to $\bt$ from below, allowing for tunnelling between the macroscopically separated edge modes.
In order to prevent scattering into bulk modes, edge- and bulk physics have to be adiabatically decoupled. 
This can be achieved by tuning $\bw$ smoothly (and slowly, see below) towards the topological transition and return to the 
``sweet spot'' $\bw=0$ afterwards to relocalize (and thereby decouple) the edge modes.
To this end, we introduce a time dependent hopping rate $\bw (t) = \bwmax\,\F (t)$ 
giving rise to the time-dependent network Hamiltonian $\hHbSSH (t)$ with perfectly localized edge modes at $t=0$ and $t= \tau$.  
For simplicity, we choose for the adiabatic process the smooth pulse shape
\begin{equation}
    \F (t)=\sin^2\left(\pi\,t/\tau\right) \quad \text{for} \quad 0\leq t\leq\tau\,.
    \label{eq:pulse}
\end{equation}
Here, $\tau$ denotes the characteristic time scale for the pulse. 
The exact pulse shape does have influence on the performance of the protocol, and setup-specific optimizations may yield
quantitatively better results, see below.

We analyse the transfer efficiency and its dependence on the parameters $\tau$ and $\bwmax$ by 
numerically evaluating the full unitary time evolution. 
We start with an excitation in the left edge mode $\Cmode$,
\begin{equation}
    \Ket{\Psi_0}=\Ket{1}_{1}\otimes\Ket{0,\dots,0}_{\rs bulk}\otimes\Ket{0}_{\ol L}\equiv \Ket{1,\vec 0,0}\,,
    \label{}
\end{equation}
and are interested in the transfer to the right edge mode $\Tmode$, i.e., the state  $\Ket{0,\vec 0,1}$.
The transfer is characterized by the overlap
\begin{equation}
    \Bra{0,\vec 0,1}U_\tau(\bwmax)\Ket{1,\vec 0,0}\equiv \sqrt{\transfer}\,e^{i\phase}\,.
    \label{}
\end{equation}
Here,  $\transfer\geq 0$ denotes the \textit{transfer fidelity},
while $\phase$ is the relative phase accumulated during the adiabatic process.
$U_\tau(\bwmax)$ is the unitary time evolution operator at time $t=\tau$ 
which depends parametrically on $\bwmax$.
To quantify the degree of adiabaticity, we introduce another characteristic parameter 
which describes the total edge mode population,
\begin{equation}
    \edgeweight=\transfer +
    |\Bra{1,\vec 0,0}U_\tau(\bwmax)\Ket{1,\vec 0,0}|^2\,.
    \label{}
\end{equation}
Deviations of $\edgeweight$ from unity indicate undesired losses into bulk modes.  

The qualitative results for the transfer are shown in Fig.~\ref{fig:transfer-qualitative}~(a) for an optimized set of parameters
$\tau$ and $\bwmax$, i.e., slow transfer with $\tau \gg \bt^{-1}$ ($\hbar=1$). As expected, we find perfect transfer and decoupling  
of edge and bulk modes. The overall performance is quantified by  $\transfer$, $\phase$ and $\edgeweight$, and  depends 
on how close the protocol parameter $\bwmax$ is to the critical value $\bt$, the size of the system $L$, 
and the global time scale $\tau$; see Fig.~\ref{fig:transfer-quantitative}~(a) and~(b) for a chain of length $L=5$.
The edge weight $\edgeweight$ (grey background tiles) equals unity almost everywhere, 
except for very fast protocols and tiny bulk-edge gaps. 
We observe quite generally that for a smooth pulse shape like $\F$, the adiabatic bulk-edge 
decoupling is rather generically established in the topological setup.  
The size of coloured squares denotes the transfer $\transfer$, 
while the colour accounts for the value of the phase $\phase$ accumulated during the transfer
(measured in the rotating frame of the localized edge modes). 
We find several disjoint branches with $\transfer \approx 1$ corresponding to an increasing number of 
round trips of the excitation; see Fig.~\ref{fig:transfer-quantitative}~(b). 
The outermost branch allows for the fastest and most robust transfer, 
and is therefore the desired parameter regime to perform quantum operations. 
However, the most striking property of this setup is a fixed phase $\phase$ accumulated
during a transfer, i.e., $\phase = \pm \pi/2$. The sign depends on the number of round trips and on the parity of the chain length 
$L$, see Supplementary Information Sections IV and V for an explanation. 
This remarkable feature is a peculiarity of the PH symmetric topological setup and in general violated for other setups
(see comparison below). 
A motivation for the relation of PH symmetry and fixed phase is presented in Supplementary Information Section IV.

As a concluding remark, note that there may be residual couplings $\bwmin\ll\bwmax$ that cannot be switched off for $t<0$ and $t>\tau$.
Weak residual couplings $\bwmin$ (compared to $\bt$) can be tolerable on the relevant timescales 
as they are exponentially suppressed with the qubit distance $L$ in the topological setup, 
while controlled coupling is always possible for $\bwmax\to\bt$.

\subsection{Scaling and adiabaticity}

\begin{figure}[t] 
    \centering
    \includegraphics[width=1.0\linewidth]{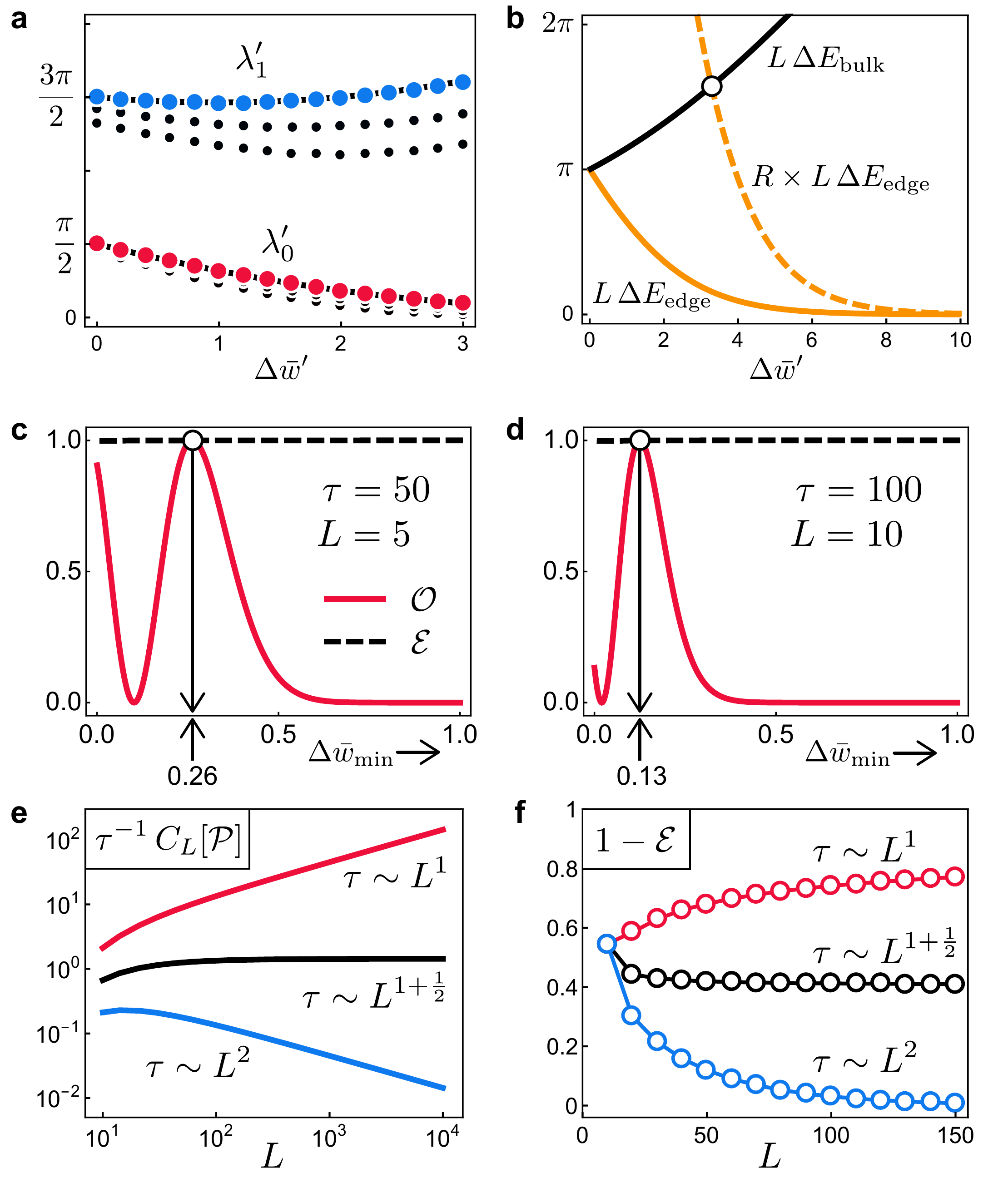}
    \caption{%
        \CaptionMark{Scaling and adiabaticity.}
        \CaptionLabel{a}
        Rescaled lowest eigenvalues $\lambda_i'$ ($i=0,1$) of $\HbSSH$ as function of the rescaled coupling
        $\Dbw'=L(\bt-\bw)$. Solid lines denote solutions of equation~(\ref{eq:rescaled_spec_limit}) exact for
        $L\to\infty$ whereas circles denote finite size results for $L=5,10$ (black) and $L=200$ (red/blue).
        \CaptionLabel{b}
        Energy scales $L\,\DEedge=2\lambda_0'$ (solid yellow) and
        $L\,\DEbulk=|\lambda_1'-\lambda_0'|$ (solid black) for $L\to\infty$ calculated from the results in \CaptionLabel{a}.
        Fixing the ratio $R=\DEbulk/\DEedge$ (here $R=10$) determines $\Dbw'$ via
        the intersection marked with a circle (here $\Dbw'\approx 3.3$).
        \CaptionLabel{c}
        Simulations of transfer fidelity $\transfer$ (solid red) and edge weight $\edgeweight$ (dashed black) 
        for system size $L=5$, protocol timescale $\tau=50$ and pulse $\bw (t)=(\bt-\Dbwmin)\cdot\F (t)$
        as function of $\Dbwmin$. Optimal transfer for fixed $L$ and $\tau$ is found numerically for $\Dbwmin\approx 0.26$ with
        bulk loss $1-\edgeweight\approx 1\cdot 10^{-4}$.
        \CaptionLabel{d}
        The same as in \CaptionLabel{c} for doubled size $L=10$ and timescale $\tau=100$.
        Now, optimal transfer is achieved for $\Dbwmin\approx 0.26/2=0.13$ with
        bulk loss $1-\edgeweight\approx 2\cdot 10^{-5}$.
        \CaptionLabel{e}
        Rigorous upper bounds $\tau^{-1}\CL[\P]$ for $\P(s)=\F(s)=\sin^2(\pi s)$ and $\tau=\tau_0\cdot L^{1+\alpha}$
        with $\alpha=0,\frac{1}{2},1$ and $\tau_0=100$, $\Dbwmin'=3.3$. A scaling $\tau\sim L^{1+\frac{1}{2}}$
        yields constant bulk losses for $L\to\infty$.
        \CaptionLabel{f}
        Simulations of the bulk losses $0\leq 1-\edgeweight\leq 1$ for the parameters in \CaptionLabel{e} without tuning for
        optimal transfer. We find that the scaling follows the corresponding upper bounds. Note that the loss was chosen large
        ($\sim 50\%$ for $L=10$) for illustrative purposes and can be controlled via $\tau_0$ 
        (here $\tau_0=1,0.3,0.1$ for $\alpha=0,\frac{1}{2},1$).
    }
\label{fig:scaling}
\end{figure}

An important aspect for quantum information processing over large distances is the
scalability of the protocol with separation $L$ between the qubits. 
We identify the two relevant time scales of the transfer protocol: The inverse 
edge mode splitting $\DEedge^{-1}$  which determines the time for a state 
transfer between the two edge states, and second,  the inverse of the bulk-edge 
separation $ \DEbulk^{-1}$ which gives a lower bound on the protocol time scale 
due to the required adiabatic bulk-edge decoupling.  

We start by considering the scaling of these energies when the topological phase transition 
$\bw=\bt$ is approached from the topological phase $\bw < \bt$. In the limit  $L\to\infty$, 
the eigenvalues of $\HbSSH$ derive from the transcendental equation (see Supplementary Information Section V for the derivation)
\begin{equation}
    \frac{\Dbw'-\sqrt{\Dbw'^2-\lambda'^2}}{\Dbw'+\sqrt{\Dbw'^2-\lambda'^2}} = e^{-2\sqrt{\Dbw'^2-\lambda'^2}}
    \label{eq:rescaled_spec_limit}
\end{equation}
with $\Dbw'/L =\bt-\bw$ measuring the distance to the topological phase transition. 
The lowest two solutions $\lambda_0'$ and $\lambda_1'$ of equation~(\ref{eq:rescaled_spec_limit})
determine the  relevant energies  $\DEedge=2\lambda_0'/L$ 
and $\DEbulk=|\lambda_1'-\lambda_0'|/L$ , see Fig.~\ref{fig:scaling}~(a) and~(b). 
Notably, both energies scale as $\sim 1/L$ with the result that their ratio  $R=\DEbulk/\DEedge$
saturates for large $L$.  For a fixed ratio $R$ of the two energy scales, it is therefore required to
approach the critical point as $ \bt-\bw \sim 1/L$. 
E.g., for $R=10$ one finds $ \bt-\bw \approx 3.3/L$, see Fig.~\ref{fig:scaling}~(b).

This result demonstrates that if one requires an adiabatic protocol $\bw(t)$
with a fixed minimum ratio $R^{\rs min}=\DEbulk^{\rs min}/\DEedge^{\rs max}$ at the minimal distance $\Dbwmin=\bt-\bwmax$ 
from the critical coupling, then the time $\tau$ for the protocol scales as $\tau \sim L$;
see Fig.~\ref{fig:scaling}~(c) and (d) for simulations. 
The latter corresponds to the optimal scaling achievable 
since the Lieb-Robinson bound predicts a finite propagation speed for information~\cite{Lieb1972}.

However, we still need to adiabatically decouple bulk from edge modes since losses to the bulk cannot be 
refocused in edge modes via a global tuning of parameters. A common (and conservative) estimate for 
adiabaticity then reads $\tau \gtrsim {\left(\DEbulkmin\right)}^{-2}$, which leads to the non-optimal scaling 
condition $\tau \sim L^2$. We demonstrate in the following, that a much better scaling is achievable
by a rigorous estimation of the adiabaticity condition. 
To this end, we parametrize the time with  $s=t/\tau$, $s\in[0,1]$ and make the ansatz 
$\bw(t)=\bwmax\cdot\P(s)$, where the generic pulse $\P:[0,1]\to[0,1]$ and 
its first derivative vanish for $s=0,1$, and $\P(1/2)=1$.

Then, the non-adiabatic losses to the bulk can be rigorously upper-bounded \cite{Jansen2007} by 
\begin{equation}
    1-\edgeweight\leq \left(\frac{\CL[\P]}{\tau}\right)^2,
\end{equation}
where
\begin{equation}
    \CL[\P]=
    \int_0^1\ds\,\frac{C_1\,|\I''|}{\left(\eL+\I\right)^2}
    +\int_0^1\ds\,\frac{C_2\,|\I'|^2}{\left(\eL+\I\right)^3}
    \label{eq:adiabatic_bound}
\end{equation}
with $\I\equiv 1-\P$, $C_{1,2}$ numerical constants, and $\eL=\Dbwmin'/(L-\Dbwmin')$;
see the Supplementary Information Section VI for details.
Note that $\I(1/2)=0$ and $\eL\sim 1/L$ so that $\CL[\P]$ diverges for $L\to\infty$ in general.
In order to bound the bulk losses, the scaling of $\tau$ has to match the scaling of $\CL[\P]$.

For $\P(s)=\F(s)=\sin^2(\pi s)$ we find $\CL[\F]\sim L^{1+\frac{1}{2}}$ so that a scaling of $\tau\sim L^{1+\frac{1}{2}}$ is
necessary for bulk-edge decoupling in the limit of long chains, see Fig.~\ref{fig:scaling}~(e). 
This is better than the quadratic scaling expected from the minimal gap $\DEbulkmin$.
Unfortunately, the optimal scaling $\tau \sim L$ allowed by the Lieb-Robinson bound cannot be reached by the
unoptimized pulse $\F$. However, in Supplementary Information Section VI we prove that there is a sequence of polynomial pulses $\P_n$ 
such that $\CL[\P_n]\sim L^{1+\frac{1}{n}}$ for $n\geq 2$ even integers, i.e., the scaling can be drastically improved by pulse
optimization so that linear scaling can be approached to an arbitrary degree. 
Numerical simulations of the bulk losses for various scalings of $\tau$ reveal that they indeed follow the
prescribed scaling of the rigorous upper bounds, 
see Fig.~\ref{fig:scaling}~(f).
As a final remark, we stress that the coefficients in $\CL[\P_n]$ become larger with $n$, 
i.e., there is a pay-off between scaling and offset.
Thus one may even benefit from pulses with poor scaling if only chains of fixed length are considered.

\subsection{Benchmarking against topologically trivial setups}

To unveil the characteristic features of the topological setup, we contrast it with 
two similar but topologically trivial networks, see Fig.~\ref{fig:setup-details}~(c) and~(d). 
The simplest approach to envisage is based on initially decoupled modes at 
fixed frequency $\omega_{I}\equiv\bo$, a homogeneous tuning of all couplings 
$w_i=t_i\equiv \bt(t)= \btmax\,\F (t)$, 
and employs the free bulk propagation of the initially localized edge modes. 
As shown in Fig.~\ref{fig:transfer-qualitative}~(d), this approach fails to relocalize the excitation
at the opposite edge due to the propagation via bulk modes; such a protocol would require
either fine tuning of the pulse shape via optimal control and/or local addressability of all couplings
within the network. It is therefore not competitive against the topological setup.

A more sophisticated approach mimics the presence of localized edge modes by a large 
tunnel barrier: The two modes at the edge have fixed frequency
$\omega_1=\omega_{\ol L}=\oedge=\const$, and are separated from each other 
by a ``potential barrier'' of modes with tunable frequencies $\obarrier (t)$ and fixed couplings $\bt= \bw$. 
In analogy to the topological setup, this network exhibits exponentially localized edge modes. 
Transfer is again achieved by lowering the excitation gap to the bulk modes to allow for tunnelling between the edges. 
The protocol of this scheme reads
\begin{equation}
    \obarrier (t)=\obarriermax+(\obarriermin-\obarriermax)\cdot\F (t)
    \label{}
\end{equation}
where $\obarriermax \gg \oedge$ will be kept fixed and $\obarriermin > \oedge$ 
is a tunable protocol parameter. 
The bosonic network Hamiltonian of this scheme is denoted by $\hHbB (t)$.

\begin{figure}[t]   
    \centering
    \includegraphics[width=1.0\linewidth]{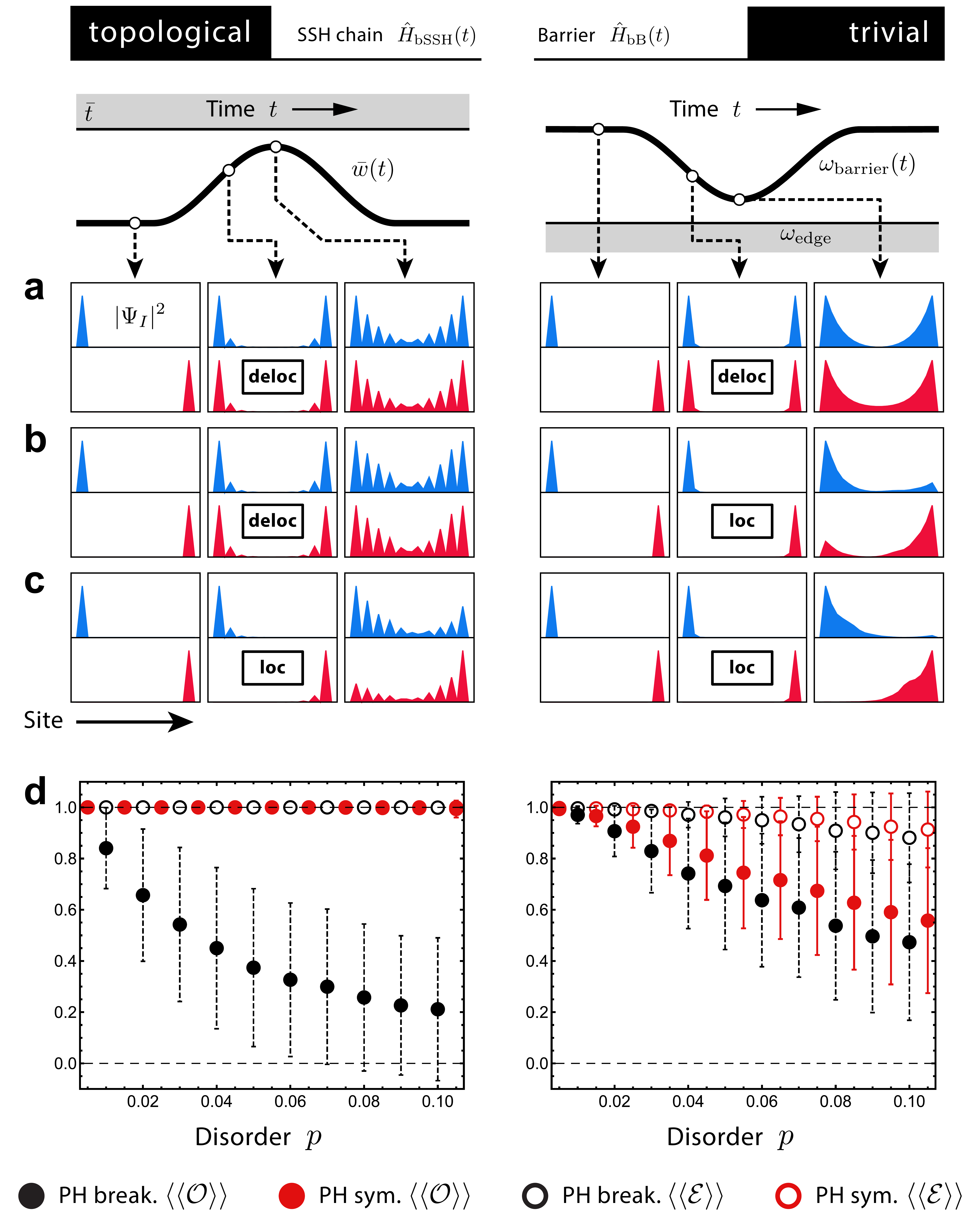} 
    \caption{%
        \CaptionMark{Effects of disorder.}
        In \CaptionLabel{a-c} we show the spatial amplitudes of the two edge modes (blue \& red)
        at three different times during the protocols of the SSH setup
        (left) and the barrier setup (right) for \CaptionLabel{a} no disorder, \CaptionLabel{b} PH symmetric disorder, 
        and \CaptionLabel{c} PH breaking disorder.
        Note the (de-)~localization of the edge states for PH symmetric disorder in \CaptionLabel{b}.
        In \CaptionLabel{d} we show the transfer $\AVtransfer$ (bullets) and edge weight
        $\AVedgeweight$ (circles) for PH breaking (black) and symmetric (red) disorder.
        The averages are computed from $N=1000$ samples for a chain of length $L=5$
        with a retuning of $\tau$ for every single disorder realization to optimize transfer.
        The error bars denote one standard deviation of the sample.
    }
\label{fig:disorder}
\end{figure}

As shown in Fig.~\ref{fig:transfer-qualitative}~(c), the tunnelling approach still allows for near perfect transfer for optimal parameters
and long times. However, the qualitative comparison in Fig.~\ref{fig:transfer-quantitative}~(c) shows that the trivial tunnelling
approach requires longer time scales of the protocol and is more sensitive to bulk losses. 
Even then the adiabatic decoupling is much harder to achieve with $\hHbB (t)$ than with $\hHbSSH (t)$,
as the plots of $\edgeweight$ along the dashed cuts in Fig.~\ref{fig:transfer-quantitative}~(b) and~(d) reveal.

However, the most striking difference is the phase accumulated during the protocol: for the trivial
setup, it is highly sensitive to both parameters. This is expected for a generic adiabatic protocol and is 
in stark contrast to the topological setup. 
The reason for this qualitative difference is rooted in the PH symmetry 
of the SSH setup which gives rise to the symmetric band structure depicted in 
Fig.~\ref{fig:transfer-qualitative}~(a), as opposed to the asymmetric band structure of the 
barrier setup in Fig.~\ref{fig:transfer-qualitative}~(c). 
As a consequence, we find that even for the ideal, topologically trivial setup, adiabatic protocols are unsuitable
as the sensitivity of the phase increases for longer wires: a transfer preserving quantum coherence requires 
fine-tuning of the shape of the transfer pulse. 
This effect becomes even more drastic in the presence of disorder.

\subsection{Effects of disorder and symmetry protection}

The unique features of the topological setup become even more apparent in the
presence of disorder and/or imperfections in the preparation. Here we focus on
quenched disorder on the time scales for a transfer. The disorder is described
as Gaussian noise with dimensionless standard deviation $p$ acting on
the parameters in the Hamiltonian, i.e., for the onsite hopping we have
$\llangle w_{i}\rrangle = \bw$ and
$\llangle w_{i}^2-\bw^2\rrangle = p^2 \bw^2$ with
$\llangle \bullet \rrangle$ the disorder average. In the following, two classes 
of disorder will be of interest: 
\textit{PH symmetric} disorder affects only mode couplings, but assumes perfect mode frequencies; 
recall equation~(\ref{eq:SSHsymmetries}).
In contrast, \textit{PH breaking} disorder affects both mode couplings and frequencies. 

For the topological trivial setup, both types of disorder give rise to Anderson localization of the ``artificial''
edge states, see Fig.~\ref{fig:disorder}~(a)--(c). The transfer protocol, however, relies on the delocalization of the edge modes
which is prohibited by Anderson localization. As a consequence, disorder leads to 
a significant reduction of transfer fidelity and increased bulk losses, Fig.~\ref{fig:disorder}~(d). 
Furthermore, the phase $\phase$ accumulated during the transfer strongly fluctuates for each disorder realization;
more details on this aspect are given in Supplementary Information Section IV.

In contrast, for the topological SSH setup, the \textit{PH symmetric} disorder respects the protecting symmetry.
Then, Anderson localization of the edge modes is forbidden by a topological obstruction~\cite{Schnyder2008,Schnyder2009a,Ryu2010} 
and the required overlap between the two edge modes can be established, see Fig.~\ref{fig:disorder}~(b). 
As a consequence, the transfer can still be performed perfectly with a fixed phase $\phase = \pm \pi/2$. 
However, this requires that for each disorder realization one is allowed to adapt the transfer time $\tau$ of the protocol. 
In an experimental setup this corresponds, for example, to imperfections in sample preparation 
which can be overcome by calibrating the setup and the transfer protocol beforehand.
In turn, the \textit{PH breaking} disorder leads also in the topological setup to localization of the edge modes
and a reduction of transfer fidelity, see Fig.~\ref{fig:disorder}~(d).

\subsection{Application: Controlled-phase gate}

\begin{figure}[t]
    \centering  
    \includegraphics[width=1.0\linewidth]{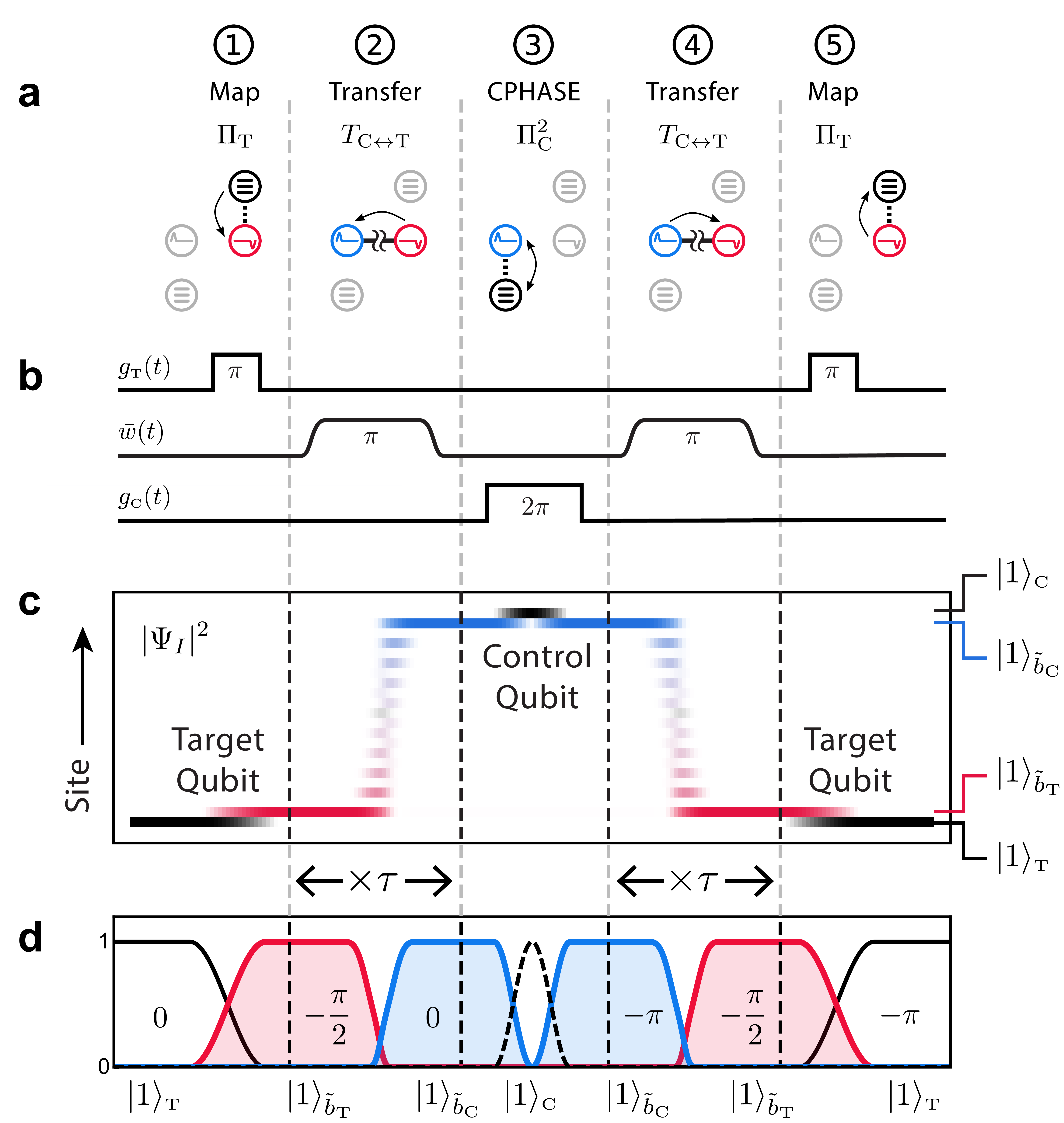}
    \caption{%
        \CaptionMark{CPHASE gate --- Pulse sequence \& Results.}
        \CaptionLabel{a} 
        Schematic illustration of the five steps needed to perform the CP gate $\UCP$~(\ref{CPgate}) on two remote qubits, 
        see setup in Fig.~\ref{fig:setup-general}~(a) and~(b).
        \CaptionLabel{b} 
        The complete pulse sequence consisting of
        two $\pi$ pulses $\PiT$ to map the target qubit to and from edge mode $\Tmode$, a $2\pi$ pulse $\PiC^2$ to perform the actual CP
        gate, and two edge mode tunnelling pulses $\TrCT$.
        \CaptionLabel{c} 
        Numerical single-particle evolution for the two-qubit basis state $\Cup\Tup$ for a chain of length $L=10$.
        The density plot encodes the squared single-particle amplitude where the 
        upper and lower edges correspond to the logical states $\Cup$ and $\Tup$ with their
        adjacent boundary modes $\Cmodeup=\Ket{1,\vec 0,0}=\Cmode^\dag\Ket{0}$ 
        and $\Tmodeup=\Ket{0,\vec 0,1}=\Tmode^\dag\Ket{0}$.
        \CaptionLabel{d}
        Square of the absolute value of the overlaps with $\Tup$ (solid black), $\Tmodeup$ (solid red), 
        $\Cmodeup$ (solid blue), and $\Cup$ (dashed black).
        The relative phases w.r.t.\ the target qubit are shown as insets.
        Note that the scales for the Rabi pulses and the topological state transfer differ by a factor of $\tau$.
    }
\label{fig:cphase}
\end{figure}

As an application, we demonstrate how the proposed state transfer, protected by PH symmetry, can be employed for a
controlled-phase (CP) gate between two remote qubits that are coupled to the local edge modes of the topological SSH network.
The complete protocol for a CP gate between the target qubit $\rm T$ and the control qubit $\rm C$ is based on 
a well-known scheme that makes use of auxiliary levels $\Ket{a}_{\rs T/\rs C}$~\cite{Haack2010}. 
We focus on the setup shown in Fig.~\ref{fig:setup-general}~(b) for the qubits and the auxiliary states with 
the coupling Hamiltonian between the qubits and the edge modes of the SSH chain given by equation~(\ref{edgecoupling}). 
The full protocol for the CP gate follows the procedure in equation~(\ref{CPgate})
and is described in the following, see Fig.~\ref{fig:cphase}.

First, the full sequence $\UCP$ leaves the states $\Cdown\Tdown$ and $\Cup\Tdown$ invariant 
because there are no excitations in the network.
On the other hand, for $\Cdown\Tup$ and $\Cup\Tup$
the first $\pi$-pulse $\PiT$ on the target qubit maps the state $\Tup$ to a bosonic excitation  
in the right edge mode $\Tmode$ with the phase $-\pi/2$.
It is important that this operation is performed slowly compared to the energy gap to bulk excitations in the SSH chain:
Then, energy conservation allows one to only address the coupling to the edge states and suppress admixture of bulk excitations.  
The subsequent transfer of the excitation to the left edge $\TrCT$ implies an additional phase $\pi/2+L\pi$.
The full Rabi cycle $\PiC^2$ provides a phase $\pi$ if and only if the control qubit is in state $\Cup$. 
The subsequent transfer back $\TrCT$ and the $\pi$-pulse $\PiT$ provide additional phases $\pi/2+L\pi$ and $-\pi/2$. 
Therefore, the full protocol implements the mapping $\Cup\Tup\,\rightarrow\,- \Cup\Tup$ while all orthogonal states remain invariant. 
Given quantum coherence during the protocol, this realizes a controlled phase gate with phase $\pi$.  
A full numerical time-evolution for the state $\Cup\Tup$ is shown in Fig.~\ref{fig:cphase}~(c) and~(d), confirming the above argumentation.

Finally, we point out that the linearity of the network implies that the transfer takes
place for each excitation of edge modes independently. I.e., if $\Ket{n_{\rs C},n_{\rs T}}$ denotes the state with $n_{\rs C}$ 
excitations in the left edge mode $\Cmode$ and $n_{\rs T}$ excitations in the right edge mode $\Tmode$, 
the transfer operation $\TrCT$ implements the mapping 
$\Ket{n_{\rs C},n_{\rs T}}\,\rightarrow\,{(\pm i)}^{n_{\rs C}+n_{\rs T}}\,\Ket{n_{\rs T},n_{\rs C}}$ (for even/odd $L$).
This observation immediately implies that the unitary operation $\UE$ in equation~(\ref{exchangegate}) swaps the qubits.

\subsection{Extension to 2D networks of coupled qubits}

An important aspect of the SSH chain is its symmetry class BDI (due to the real hopping amplitudes $w_{i}$ and $t_{i}$)
so that the setup is characterized by a $\mathbb{Z}$ topological invariant.
This allows us to extend the analysis to two-dimensional setups by placing several SSH chains parallel to each other
and adding couplings between them, see Fig.~\ref{fig:setup-general}~(c) for a possible realization. 
As long as these couplings are real and respect the sublattice symmetry, the setup is still topologically protected
and each chain endpoint carries an edge mode.  
Possible relaxations of the symmetry constraints are presented in Supplementary Information Section III.

Due to the sublattice symmetry, there are two types of modes, ``even'' and ``odd'' ones, 
depending on how they transform under $S=U_C^{\rs SSH}$ [recall equation~(\ref{eq:sshsym})].
Edge mode pairs of \textit{different} types can communicate efficiently 
by tuning the couplings along a connecting path that resembles the one-dimensional SSH setup; 
this is illustrated in Fig.~\ref{fig:setup-general}~(c).
Because of the bulk gap, this procedure is very robust and couplings that deviate from the desired path (``stray couplings'')
have no detrimental effect on the state transfer, as long as they respect the symmetries and do not couple to other edge modes.
This can be guaranteed by a modification of the setup
such that qubits and edge modes are relocated at the end of one-dimensional chains that emanate from the 2D network, see
Fig.~\ref{fig:setup-general}~(d).
Remarkably, the setup allows for an enhancement of 
the edge mode overlap by tuning the couplings of the 2D bulk \textit{globally} 
instead of tracing out a particular path that connects the qubits.
Therefore the minimal experimental requirement is the individual addressability 
of each branch that connects an edge mode to the 2D bulk.
Since the edge modes can now be separated, this constraint is very weak, 
and in general already satisfied by the requirement of local gate operations on the qubit. 

Note that coupling edge modes of the \textit{same} type is obstructed by the sublattice symmetry,
as indicated by the lower path in Fig.~\ref{fig:setup-general}~(c) and proven in Supplementary Information Section III.
However, the implementation of an exchange of qubits via $\UE$ facilitates the application of
the controlled phase gate $\UCP$ between any pair in the network: 
If the two qubits couple to edge modes of \textit{different} types, one can directly perform
the CP gate between them. 
Conversely, if the qubits couple to edge modes of the \textit{same} type, 
one first performs an exchange $\UE$ with an arbitrary qubit of the opposite type,
applies the CP gate, and maps the qubit back by another exchange.

\section{Discussion}

We have demonstrated that a topological network consisting of linearly coupled bosonic degrees of freedom, 
capable of carrying single quantized excitations, allows for efficient quantum communication between distant qubits. 
Weak addressability of each branch within the network is sufficient
and no local addressability of individual sites is required. 
Remarkably, the time scale for the operations scales almost linearly with the distance between the qubits.
Furthermore, topological protection guarantees robustness against quenched disorder in the setup by 
evading Anderson localization of the edge modes. 
In summary, we have shown that the unique properties of (quasi) one-dimensional topological systems can be harvested for efficient 
quantum communication between qubits. 
These benefits come with the price of higher complexity in realization and preparation
as the coupling parameters have to respect the symmetries protecting the topological invariants.





\section{Acknowledgements}
This research has received funding from the European Research Council (ERC) 
under the European Union’s Horizon 2020 research and innovation programme (grant agreement No 681208).
This research was supported in part by the National Science Foundation under Grant No. NSF PHY-1125915.









\begin{thebibliography}{58}%
\makeatletter
\providecommand \@ifxundefined [1]{%
 \@ifx{#1\undefined}
}%
\providecommand \@ifnum [1]{%
 \ifnum #1\expandafter \@firstoftwo
 \else \expandafter \@secondoftwo
 \fi
}%
\providecommand \@ifx [1]{%
 \ifx #1\expandafter \@firstoftwo
 \else \expandafter \@secondoftwo
 \fi
}%
\providecommand \natexlab [1]{#1}%
\providecommand \enquote  [1]{``#1''}%
\providecommand \bibnamefont  [1]{#1}%
\providecommand \bibfnamefont [1]{#1}%
\providecommand \citenamefont [1]{#1}%
\providecommand \href@noop [0]{\@secondoftwo}%
\providecommand \href [0]{\begingroup \@sanitize@url \@href}%
\providecommand \@href[1]{\@@startlink{#1}\@@href}%
\providecommand \@@href[1]{\endgroup#1\@@endlink}%
\providecommand \@sanitize@url [0]{\catcode `\\12\catcode `\$12\catcode
  `\&12\catcode `\#12\catcode `\^12\catcode `\_12\catcode `\%12\relax}%
\providecommand \@@startlink[1]{}%
\providecommand \@@endlink[0]{}%
\providecommand \url  [0]{\begingroup\@sanitize@url \@url }%
\providecommand \@url [1]{\endgroup\@href {#1}{\urlprefix }}%
\providecommand \urlprefix  [0]{URL }%
\providecommand \Eprint [0]{\href }%
\providecommand \doibase [0]{http://dx.doi.org/}%
\providecommand \selectlanguage [0]{\@gobble}%
\providecommand \bibinfo  [0]{\@secondoftwo}%
\providecommand \bibfield  [0]{\@secondoftwo}%
\providecommand \translation [1]{[#1]}%
\providecommand \BibitemOpen [0]{}%
\providecommand \bibitemStop [0]{}%
\providecommand \bibitemNoStop [0]{.\EOS\space}%
\providecommand \EOS [0]{\spacefactor3000\relax}%
\providecommand \BibitemShut  [1]{\csname bibitem#1\endcsname}%
\let\auto@bib@innerbib\@empty
\bibitem [{\citenamefont {Nayak}\ \emph {et~al.}(2008)\citenamefont {Nayak},
  \citenamefont {Simon}, \citenamefont {Stern}, \citenamefont {Freedman},\ and\
  \citenamefont {Sarma}}]{Nayak2008}%
  \BibitemOpen
  \bibfield  {author} {\bibinfo {author} {\bibfnamefont {Chetan}\ \bibnamefont
  {Nayak}}, \bibinfo {author} {\bibfnamefont {Steven~H.}\ \bibnamefont
  {Simon}}, \bibinfo {author} {\bibfnamefont {Ady}\ \bibnamefont {Stern}},
  \bibinfo {author} {\bibfnamefont {Michael}\ \bibnamefont {Freedman}}, \ and\
  \bibinfo {author} {\bibfnamefont {Sankar~Das}\ \bibnamefont {Sarma}},\
  }\bibfield  {title} {\enquote {\bibinfo {title} {Non-abelian anyons and
  topological quantum computation},}\ }\href {\doibase
  10.1103/revmodphys.80.1083} {\bibfield  {journal} {\bibinfo  {journal}
  {Reviews of Modern Physics}\ }\textbf {\bibinfo {volume} {80}},\ \bibinfo
  {pages} {1083--1159} (\bibinfo {year} {2008})}\BibitemShut {NoStop}%
\bibitem [{\citenamefont {Ando}\ \emph {et~al.}(1975)\citenamefont {Ando},
  \citenamefont {Matsumoto},\ and\ \citenamefont {Uemura}}]{Ando1975}%
  \BibitemOpen
  \bibfield  {author} {\bibinfo {author} {\bibfnamefont {Tsuneya}\ \bibnamefont
  {Ando}}, \bibinfo {author} {\bibfnamefont {Yukio}\ \bibnamefont {Matsumoto}},
  \ and\ \bibinfo {author} {\bibfnamefont {Yasutada}\ \bibnamefont {Uemura}},\
  }\bibfield  {title} {\enquote {\bibinfo {title} {Theory of hall effect in a
  two-dimensional electron system},}\ }\href {\doibase 10.1143/jpsj.39.279}
  {\bibfield  {journal} {\bibinfo  {journal} {Journal of the Physical Society
  of Japan}\ }\textbf {\bibinfo {volume} {39}},\ \bibinfo {pages} {279--288}
  (\bibinfo {year} {1975})}\BibitemShut {NoStop}%
\bibitem [{\citenamefont {v.~Klitzing}\ \emph {et~al.}(1980)\citenamefont
  {v.~Klitzing}, \citenamefont {Dorda},\ and\ \citenamefont
  {Pepper}}]{Klitzing1980}%
  \BibitemOpen
  \bibfield  {author} {\bibinfo {author} {\bibfnamefont {K.}~\bibnamefont
  {v.~Klitzing}}, \bibinfo {author} {\bibfnamefont {G.}~\bibnamefont {Dorda}},
  \ and\ \bibinfo {author} {\bibfnamefont {M.}~\bibnamefont {Pepper}},\
  }\bibfield  {title} {\enquote {\bibinfo {title} {New method for high-accuracy
  determination of the fine-structure constant based on quantized hall
  resistance},}\ }\href {\doibase 10.1103/physrevlett.45.494} {\bibfield
  {journal} {\bibinfo  {journal} {Physical Review Letters}\ }\textbf {\bibinfo
  {volume} {45}},\ \bibinfo {pages} {494--497} (\bibinfo {year}
  {1980})}\BibitemShut {NoStop}%
\bibitem [{\citenamefont {Laughlin}(1981)}]{Laughlin1981}%
  \BibitemOpen
  \bibfield  {author} {\bibinfo {author} {\bibfnamefont {R.~B.}\ \bibnamefont
  {Laughlin}},\ }\bibfield  {title} {\enquote {\bibinfo {title} {Quantized hall
  conductivity in two dimensions},}\ }\href {\doibase 10.1103/physrevb.23.5632}
  {\bibfield  {journal} {\bibinfo  {journal} {Physical Review B}\ }\textbf
  {\bibinfo {volume} {23}},\ \bibinfo {pages} {5632--5633} (\bibinfo {year}
  {1981})}\BibitemShut {NoStop}%
\bibitem [{\citenamefont {Tsui}\ \emph {et~al.}(1982)\citenamefont {Tsui},
  \citenamefont {Stormer},\ and\ \citenamefont {Gossard}}]{Tsui1982a}%
  \BibitemOpen
  \bibfield  {author} {\bibinfo {author} {\bibfnamefont {D.~C.}\ \bibnamefont
  {Tsui}}, \bibinfo {author} {\bibfnamefont {H.~L.}\ \bibnamefont {Stormer}}, \
  and\ \bibinfo {author} {\bibfnamefont {A.~C.}\ \bibnamefont {Gossard}},\
  }\bibfield  {title} {\enquote {\bibinfo {title} {Two-dimensional
  magnetotransport in the extreme quantum limit},}\ }\href {\doibase
  10.1103/physrevlett.48.1559} {\bibfield  {journal} {\bibinfo  {journal}
  {Physical Review Letters}\ }\textbf {\bibinfo {volume} {48}},\ \bibinfo
  {pages} {1559--1562} (\bibinfo {year} {1982})}\BibitemShut {NoStop}%
\bibitem [{\citenamefont {Stormer}\ \emph {et~al.}(1983)\citenamefont
  {Stormer}, \citenamefont {Chang}, \citenamefont {Tsui}, \citenamefont
  {Hwang}, \citenamefont {Gossard},\ and\ \citenamefont
  {Wiegmann}}]{Stormer1983b}%
  \BibitemOpen
  \bibfield  {author} {\bibinfo {author} {\bibfnamefont {H.~L.}\ \bibnamefont
  {Stormer}}, \bibinfo {author} {\bibfnamefont {A.}~\bibnamefont {Chang}},
  \bibinfo {author} {\bibfnamefont {D.~C.}\ \bibnamefont {Tsui}}, \bibinfo
  {author} {\bibfnamefont {J.~C.~M.}\ \bibnamefont {Hwang}}, \bibinfo {author}
  {\bibfnamefont {A.~C.}\ \bibnamefont {Gossard}}, \ and\ \bibinfo {author}
  {\bibfnamefont {W.}~\bibnamefont {Wiegmann}},\ }\bibfield  {title} {\enquote
  {\bibinfo {title} {Fractional quantization of the hall effect},}\ }\href
  {\doibase 10.1103/physrevlett.50.1953} {\bibfield  {journal} {\bibinfo
  {journal} {Physical Review Letters}\ }\textbf {\bibinfo {volume} {50}},\
  \bibinfo {pages} {1953--1956} (\bibinfo {year} {1983})}\BibitemShut {NoStop}%
\bibitem [{\citenamefont {Willett}\ \emph {et~al.}(1987)\citenamefont
  {Willett}, \citenamefont {Eisenstein}, \citenamefont {Störmer},
  \citenamefont {Tsui}, \citenamefont {Gossard},\ and\ \citenamefont
  {English}}]{Willett1987}%
  \BibitemOpen
  \bibfield  {author} {\bibinfo {author} {\bibfnamefont {R.}~\bibnamefont
  {Willett}}, \bibinfo {author} {\bibfnamefont {J.~P.}\ \bibnamefont
  {Eisenstein}}, \bibinfo {author} {\bibfnamefont {H.~L.}\ \bibnamefont
  {Störmer}}, \bibinfo {author} {\bibfnamefont {D.~C.}\ \bibnamefont {Tsui}},
  \bibinfo {author} {\bibfnamefont {A.~C.}\ \bibnamefont {Gossard}}, \ and\
  \bibinfo {author} {\bibfnamefont {J.~H.}\ \bibnamefont {English}},\
  }\bibfield  {title} {\enquote {\bibinfo {title} {Observation of an
  even-denominator quantum number in the fractional quantum hall effect},}\
  }\href {\doibase 10.1103/physrevlett.59.1776} {\bibfield  {journal} {\bibinfo
   {journal} {Physical Review Letters}\ }\textbf {\bibinfo {volume} {59}},\
  \bibinfo {pages} {1776--1779} (\bibinfo {year} {1987})}\BibitemShut {NoStop}%
\bibitem [{\citenamefont {Konig}\ \emph {et~al.}(2007)\citenamefont {Konig},
  \citenamefont {Wiedmann}, \citenamefont {Brune}, \citenamefont {Roth},
  \citenamefont {Buhmann}, \citenamefont {Molenkamp}, \citenamefont {Qi},\ and\
  \citenamefont {Zhang}}]{Konig2007}%
  \BibitemOpen
  \bibfield  {author} {\bibinfo {author} {\bibfnamefont {M.}~\bibnamefont
  {Konig}}, \bibinfo {author} {\bibfnamefont {S.}~\bibnamefont {Wiedmann}},
  \bibinfo {author} {\bibfnamefont {C.}~\bibnamefont {Brune}}, \bibinfo
  {author} {\bibfnamefont {A.}~\bibnamefont {Roth}}, \bibinfo {author}
  {\bibfnamefont {H.}~\bibnamefont {Buhmann}}, \bibinfo {author} {\bibfnamefont
  {L.~W.}\ \bibnamefont {Molenkamp}}, \bibinfo {author} {\bibfnamefont {X.-L.}\
  \bibnamefont {Qi}}, \ and\ \bibinfo {author} {\bibfnamefont {S.-C.}\
  \bibnamefont {Zhang}},\ }\bibfield  {title} {\enquote {\bibinfo {title}
  {Quantum spin hall insulator state in {HgTe} quantum wells},}\ }\href
  {\doibase 10.1126/science.1148047} {\bibfield  {journal} {\bibinfo  {journal}
  {Science}\ }\textbf {\bibinfo {volume} {318}},\ \bibinfo {pages} {766--770}
  (\bibinfo {year} {2007})}\BibitemShut {NoStop}%
\bibitem [{\citenamefont {Hsieh}\ \emph {et~al.}(2008)\citenamefont {Hsieh},
  \citenamefont {Qian}, \citenamefont {Wray}, \citenamefont {Xia},
  \citenamefont {Hor}, \citenamefont {Cava},\ and\ \citenamefont
  {Hasan}}]{Hsieh2008}%
  \BibitemOpen
  \bibfield  {author} {\bibinfo {author} {\bibfnamefont {D.}~\bibnamefont
  {Hsieh}}, \bibinfo {author} {\bibfnamefont {D.}~\bibnamefont {Qian}},
  \bibinfo {author} {\bibfnamefont {L.}~\bibnamefont {Wray}}, \bibinfo {author}
  {\bibfnamefont {Y.}~\bibnamefont {Xia}}, \bibinfo {author} {\bibfnamefont
  {Y.~S.}\ \bibnamefont {Hor}}, \bibinfo {author} {\bibfnamefont {R.~J.}\
  \bibnamefont {Cava}}, \ and\ \bibinfo {author} {\bibfnamefont {M.~Z.}\
  \bibnamefont {Hasan}},\ }\bibfield  {title} {\enquote {\bibinfo {title} {A
  topological dirac insulator in a quantum spin hall phase},}\ }\href {\doibase
  10.1038/nature06843} {\bibfield  {journal} {\bibinfo  {journal} {Nature}\
  }\textbf {\bibinfo {volume} {452}},\ \bibinfo {pages} {970--974} (\bibinfo
  {year} {2008})}\BibitemShut {NoStop}%
\bibitem [{\citenamefont {Xu}\ \emph {et~al.}(2014)\citenamefont {Xu},
  \citenamefont {Miotkowski}, \citenamefont {Liu}, \citenamefont {Tian},
  \citenamefont {Nam}, \citenamefont {Alidoust}, \citenamefont {Hu},
  \citenamefont {Shih}, \citenamefont {Hasan},\ and\ \citenamefont
  {Chen}}]{Xu2014}%
  \BibitemOpen
  \bibfield  {author} {\bibinfo {author} {\bibfnamefont {Yang}\ \bibnamefont
  {Xu}}, \bibinfo {author} {\bibfnamefont {Ireneusz}\ \bibnamefont
  {Miotkowski}}, \bibinfo {author} {\bibfnamefont {Chang}\ \bibnamefont {Liu}},
  \bibinfo {author} {\bibfnamefont {Jifa}\ \bibnamefont {Tian}}, \bibinfo
  {author} {\bibfnamefont {Hyoungdo}\ \bibnamefont {Nam}}, \bibinfo {author}
  {\bibfnamefont {Nasser}\ \bibnamefont {Alidoust}}, \bibinfo {author}
  {\bibfnamefont {Jiuning}\ \bibnamefont {Hu}}, \bibinfo {author}
  {\bibfnamefont {Chih-Kang}\ \bibnamefont {Shih}}, \bibinfo {author}
  {\bibfnamefont {M.~Zahid}\ \bibnamefont {Hasan}}, \ and\ \bibinfo {author}
  {\bibfnamefont {Yong~P.}\ \bibnamefont {Chen}},\ }\bibfield  {title}
  {\enquote {\bibinfo {title} {Observation of topological surface state quantum
  hall effect in an intrinsic three-dimensional topological insulator},}\
  }\href {\doibase 10.1038/nphys3140} {\bibfield  {journal} {\bibinfo
  {journal} {Nature Physics}\ }\textbf {\bibinfo {volume} {10}},\ \bibinfo
  {pages} {956--963} (\bibinfo {year} {2014})}\BibitemShut {NoStop}%
\bibitem [{\citenamefont {Schnyder}\ \emph {et~al.}(2008)\citenamefont
  {Schnyder}, \citenamefont {Ryu}, \citenamefont {Furusaki},\ and\
  \citenamefont {Ludwig}}]{Schnyder2008}%
  \BibitemOpen
  \bibfield  {author} {\bibinfo {author} {\bibfnamefont {Andreas~P.}\
  \bibnamefont {Schnyder}}, \bibinfo {author} {\bibfnamefont {Shinsei}\
  \bibnamefont {Ryu}}, \bibinfo {author} {\bibfnamefont {Akira}\ \bibnamefont
  {Furusaki}}, \ and\ \bibinfo {author} {\bibfnamefont {Andreas W.~W.}\
  \bibnamefont {Ludwig}},\ }\bibfield  {title} {\enquote {\bibinfo {title}
  {Classification of topological insulators and superconductors in three
  spatial dimensions},}\ }\href {\doibase 10.1103/physrevb.78.195125}
  {\bibfield  {journal} {\bibinfo  {journal} {Physical Review B}\ }\textbf
  {\bibinfo {volume} {78}},\ \bibinfo {pages} {195125} (\bibinfo {year}
  {2008})}\BibitemShut {NoStop}%
\bibitem [{\citenamefont {Kitaev}(2009)}]{Kitaev2009a}%
  \BibitemOpen
  \bibfield  {author} {\bibinfo {author} {\bibfnamefont {Alexei}\ \bibnamefont
  {Kitaev}},\ }\bibfield  {title} {\enquote {\bibinfo {title} {Periodic table
  for topological insulators and superconductors},}\ }in\ \href {\doibase
  10.1063/1.3149495} {\emph {\bibinfo {booktitle} {{AIP} Conference
  Proceedings}}},\ Vol.\ \bibinfo {volume} {1134}\ (\bibinfo  {publisher}
  {{AIP}},\ \bibinfo {year} {2009})\ pp.\ \bibinfo {pages} {22--30}\BibitemShut
  {NoStop}%
\bibitem [{\citenamefont {Schnyder}\ \emph {et~al.}(2009)\citenamefont
  {Schnyder}, \citenamefont {Ryu}, \citenamefont {Furusaki},\ and\
  \citenamefont {Ludwig}}]{Schnyder2009a}%
  \BibitemOpen
  \bibfield  {author} {\bibinfo {author} {\bibfnamefont {Andreas~P.}\
  \bibnamefont {Schnyder}}, \bibinfo {author} {\bibfnamefont {Shinsei}\
  \bibnamefont {Ryu}}, \bibinfo {author} {\bibfnamefont {Akira}\ \bibnamefont
  {Furusaki}}, \ and\ \bibinfo {author} {\bibfnamefont {Andreas W.~W.}\
  \bibnamefont {Ludwig}},\ }\bibfield  {title} {\enquote {\bibinfo {title}
  {Classification of topological insulators and superconductors},}\ }in\ \href
  {\doibase 10.1063/1.3149481} {\emph {\bibinfo {booktitle} {{AIP} Conference
  Proceedings}}},\ Vol.\ \bibinfo {volume} {1134}\ (\bibinfo  {publisher}
  {{AIP}},\ \bibinfo {year} {2009})\ pp.\ \bibinfo {pages} {10--21}\BibitemShut
  {NoStop}%
\bibitem [{\citenamefont {Ryu}\ \emph {et~al.}(2010)\citenamefont {Ryu},
  \citenamefont {Schnyder}, \citenamefont {Furusaki},\ and\ \citenamefont
  {Ludwig}}]{Ryu2010}%
  \BibitemOpen
  \bibfield  {author} {\bibinfo {author} {\bibfnamefont {Shinsei}\ \bibnamefont
  {Ryu}}, \bibinfo {author} {\bibfnamefont {Andreas~P}\ \bibnamefont
  {Schnyder}}, \bibinfo {author} {\bibfnamefont {Akira}\ \bibnamefont
  {Furusaki}}, \ and\ \bibinfo {author} {\bibfnamefont {Andreas W~W}\
  \bibnamefont {Ludwig}},\ }\bibfield  {title} {\enquote {\bibinfo {title}
  {Topological insulators and superconductors: tenfold way and dimensional
  hierarchy},}\ }\href {\doibase 10.1088/1367-2630/12/6/065010} {\bibfield
  {journal} {\bibinfo  {journal} {New Journal of Physics}\ }\textbf {\bibinfo
  {volume} {12}},\ \bibinfo {pages} {065010} (\bibinfo {year}
  {2010})}\BibitemShut {NoStop}%
\bibitem [{\citenamefont {Hasan}\ and\ \citenamefont {Kane}(2010)}]{Hasan2010}%
  \BibitemOpen
  \bibfield  {author} {\bibinfo {author} {\bibfnamefont {M.~Z.}\ \bibnamefont
  {Hasan}}\ and\ \bibinfo {author} {\bibfnamefont {C.~L.}\ \bibnamefont
  {Kane}},\ }\bibfield  {title} {\enquote {\bibinfo {title} {Colloquium:
  Topological insulators},}\ }\href {\doibase 10.1103/revmodphys.82.3045}
  {\bibfield  {journal} {\bibinfo  {journal} {Reviews of Modern Physics}\
  }\textbf {\bibinfo {volume} {82}},\ \bibinfo {pages} {3045--3067} (\bibinfo
  {year} {2010})}\BibitemShut {NoStop}%
\bibitem [{\citenamefont {Qi}\ and\ \citenamefont {Zhang}(2011)}]{Qi2011}%
  \BibitemOpen
  \bibfield  {author} {\bibinfo {author} {\bibfnamefont {Xiao-Liang}\
  \bibnamefont {Qi}}\ and\ \bibinfo {author} {\bibfnamefont {Shou-Cheng}\
  \bibnamefont {Zhang}},\ }\bibfield  {title} {\enquote {\bibinfo {title}
  {Topological insulators and superconductors},}\ }\href {\doibase
  10.1103/revmodphys.83.1057} {\bibfield  {journal} {\bibinfo  {journal}
  {Reviews of Modern Physics}\ }\textbf {\bibinfo {volume} {83}},\ \bibinfo
  {pages} {1057--1110} (\bibinfo {year} {2011})}\BibitemShut {NoStop}%
\bibitem [{\citenamefont {Wigner}(1951)}]{Wigner1951}%
  \BibitemOpen
  \bibfield  {author} {\bibinfo {author} {\bibfnamefont {Eugene~P.}\
  \bibnamefont {Wigner}},\ }\bibfield  {title} {\enquote {\bibinfo {title} {On
  the statistical distribution of the widths and spacings of nuclear resonance
  levels},}\ }\href {\doibase 10.1017/s0305004100027237} {\bibfield  {journal}
  {\bibinfo  {journal} {Mathematical Proceedings of the Cambridge Philosophical
  Society}\ }\textbf {\bibinfo {volume} {47}},\ \bibinfo {pages} {790}
  (\bibinfo {year} {1951})}\BibitemShut {NoStop}%
\bibitem [{\citenamefont {Wigner}(1958)}]{Wigner1958}%
  \BibitemOpen
  \bibfield  {author} {\bibinfo {author} {\bibfnamefont {Eugene~P.}\
  \bibnamefont {Wigner}},\ }\bibfield  {title} {\enquote {\bibinfo {title} {On
  the distribution of the roots of certain symmetric matrices},}\ }\href
  {\doibase 10.2307/1970008} {\bibfield  {journal} {\bibinfo  {journal} {The
  Annals of Mathematics}\ }\textbf {\bibinfo {volume} {67}},\ \bibinfo {pages}
  {325} (\bibinfo {year} {1958})}\BibitemShut {NoStop}%
\bibitem [{\citenamefont {Dyson}(1962)}]{Dyson1962}%
  \BibitemOpen
  \bibfield  {author} {\bibinfo {author} {\bibfnamefont {Freeman~J.}\
  \bibnamefont {Dyson}},\ }\bibfield  {title} {\enquote {\bibinfo {title} {The
  threefold way. algebraic structure of symmetry groups and ensembles in
  quantum mechanics},}\ }\href {\doibase 10.1063/1.1703863} {\bibfield
  {journal} {\bibinfo  {journal} {Journal of Mathematical Physics}\ }\textbf
  {\bibinfo {volume} {3}},\ \bibinfo {pages} {1199--1215} (\bibinfo {year}
  {1962})}\BibitemShut {NoStop}%
\bibitem [{\citenamefont {Altland}\ and\ \citenamefont
  {Zirnbauer}(1997)}]{Altland1997}%
  \BibitemOpen
  \bibfield  {author} {\bibinfo {author} {\bibfnamefont {Alexander}\
  \bibnamefont {Altland}}\ and\ \bibinfo {author} {\bibfnamefont {Martin~R.}\
  \bibnamefont {Zirnbauer}},\ }\bibfield  {title} {\enquote {\bibinfo {title}
  {Nonstandard symmetry classes in mesoscopic normal-superconducting hybrid
  structures},}\ }\href {\doibase 10.1103/physrevb.55.1142} {\bibfield
  {journal} {\bibinfo  {journal} {Physical Review B}\ }\textbf {\bibinfo
  {volume} {55}},\ \bibinfo {pages} {1142--1161} (\bibinfo {year}
  {1997})}\BibitemShut {NoStop}%
\bibitem [{\citenamefont {Hafezi}\ \emph {et~al.}(2013)\citenamefont {Hafezi},
  \citenamefont {Mittal}, \citenamefont {Fan}, \citenamefont {Migdall},\ and\
  \citenamefont {Taylor}}]{Hafezi2013}%
  \BibitemOpen
  \bibfield  {author} {\bibinfo {author} {\bibfnamefont {M.}~\bibnamefont
  {Hafezi}}, \bibinfo {author} {\bibfnamefont {S.}~\bibnamefont {Mittal}},
  \bibinfo {author} {\bibfnamefont {J.}~\bibnamefont {Fan}}, \bibinfo {author}
  {\bibfnamefont {A.}~\bibnamefont {Migdall}}, \ and\ \bibinfo {author}
  {\bibfnamefont {J.~M.}\ \bibnamefont {Taylor}},\ }\bibfield  {title}
  {\enquote {\bibinfo {title} {Imaging topological edge states in silicon
  photonics},}\ }\href {\doibase 10.1038/nphoton.2013.274} {\bibfield
  {journal} {\bibinfo  {journal} {Nature Photonics}\ }\textbf {\bibinfo
  {volume} {7}},\ \bibinfo {pages} {1001--1005} (\bibinfo {year}
  {2013})}\BibitemShut {NoStop}%
\bibitem [{\citenamefont {Susstrunk}\ and\ \citenamefont
  {Huber}(2015)}]{Susstrunk2015}%
  \BibitemOpen
  \bibfield  {author} {\bibinfo {author} {\bibfnamefont {R.}~\bibnamefont
  {Susstrunk}}\ and\ \bibinfo {author} {\bibfnamefont {S.~D.}\ \bibnamefont
  {Huber}},\ }\bibfield  {title} {\enquote {\bibinfo {title} {Observation of
  phononic helical edge states in a mechanical topological insulator},}\ }\href
  {\doibase 10.1126/science.aab0239} {\bibfield  {journal} {\bibinfo  {journal}
  {Science}\ }\textbf {\bibinfo {volume} {349}},\ \bibinfo {pages} {47--50}
  (\bibinfo {year} {2015})}\BibitemShut {NoStop}%
\bibitem [{\citenamefont {Nash}\ \emph {et~al.}(2015)\citenamefont {Nash},
  \citenamefont {Kleckner}, \citenamefont {Read}, \citenamefont {Vitelli},
  \citenamefont {Turner},\ and\ \citenamefont {Irvine}}]{Nash2015}%
  \BibitemOpen
  \bibfield  {author} {\bibinfo {author} {\bibfnamefont {Lisa~M.}\ \bibnamefont
  {Nash}}, \bibinfo {author} {\bibfnamefont {Dustin}\ \bibnamefont {Kleckner}},
  \bibinfo {author} {\bibfnamefont {Alismari}\ \bibnamefont {Read}}, \bibinfo
  {author} {\bibfnamefont {Vincenzo}\ \bibnamefont {Vitelli}}, \bibinfo
  {author} {\bibfnamefont {Ari~M.}\ \bibnamefont {Turner}}, \ and\ \bibinfo
  {author} {\bibfnamefont {William T.~M.}\ \bibnamefont {Irvine}},\ }\bibfield
  {title} {\enquote {\bibinfo {title} {Topological mechanics of gyroscopic
  metamaterials},}\ }\href {\doibase 10.1073/pnas.1507413112} {\bibfield
  {journal} {\bibinfo  {journal} {Proceedings of the National Academy of
  Sciences}\ }\textbf {\bibinfo {volume} {112}},\ \bibinfo {pages}
  {14495--14500} (\bibinfo {year} {2015})}\BibitemShut {NoStop}%
\bibitem [{\citenamefont {Rechtsman}\ \emph {et~al.}(2013)\citenamefont
  {Rechtsman}, \citenamefont {Zeuner}, \citenamefont {Plotnik}, \citenamefont
  {Lumer}, \citenamefont {Podolsky}, \citenamefont {Dreisow}, \citenamefont
  {Nolte}, \citenamefont {Segev},\ and\ \citenamefont
  {Szameit}}]{Rechtsman2013}%
  \BibitemOpen
  \bibfield  {author} {\bibinfo {author} {\bibfnamefont {Mikael~C.}\
  \bibnamefont {Rechtsman}}, \bibinfo {author} {\bibfnamefont {Julia~M.}\
  \bibnamefont {Zeuner}}, \bibinfo {author} {\bibfnamefont {Yonatan}\
  \bibnamefont {Plotnik}}, \bibinfo {author} {\bibfnamefont {Yaakov}\
  \bibnamefont {Lumer}}, \bibinfo {author} {\bibfnamefont {Daniel}\
  \bibnamefont {Podolsky}}, \bibinfo {author} {\bibfnamefont {Felix}\
  \bibnamefont {Dreisow}}, \bibinfo {author} {\bibfnamefont {Stefan}\
  \bibnamefont {Nolte}}, \bibinfo {author} {\bibfnamefont {Mordechai}\
  \bibnamefont {Segev}}, \ and\ \bibinfo {author} {\bibfnamefont {Alexander}\
  \bibnamefont {Szameit}},\ }\bibfield  {title} {\enquote {\bibinfo {title}
  {Photonic floquet topological insulators},}\ }\href {\doibase
  10.1038/nature12066} {\bibfield  {journal} {\bibinfo  {journal} {Nature}\
  }\textbf {\bibinfo {volume} {496}},\ \bibinfo {pages} {196--200} (\bibinfo
  {year} {2013})}\BibitemShut {NoStop}%
\bibitem [{\citenamefont {Ling}\ \emph {et~al.}(2015)\citenamefont {Ling},
  \citenamefont {Xiao}, \citenamefont {Chan}, \citenamefont {Yu},\ and\
  \citenamefont {Fung}}]{Ling2015}%
  \BibitemOpen
  \bibfield  {author} {\bibinfo {author} {\bibfnamefont {C.~W.}\ \bibnamefont
  {Ling}}, \bibinfo {author} {\bibfnamefont {Meng}\ \bibnamefont {Xiao}},
  \bibinfo {author} {\bibfnamefont {C.~T.}\ \bibnamefont {Chan}}, \bibinfo
  {author} {\bibfnamefont {S.~F.}\ \bibnamefont {Yu}}, \ and\ \bibinfo {author}
  {\bibfnamefont {K.~H.}\ \bibnamefont {Fung}},\ }\bibfield  {title} {\enquote
  {\bibinfo {title} {Topological edge plasmon modes between diatomic chains of
  plasmonic nanoparticles},}\ }\href {\doibase 10.1364/oe.23.002021} {\bibfield
   {journal} {\bibinfo  {journal} {Optics Express}\ }\textbf {\bibinfo {volume}
  {23}},\ \bibinfo {pages} {2021} (\bibinfo {year} {2015})}\BibitemShut
  {NoStop}%
\bibitem [{\citenamefont {Ningyuan}\ \emph {et~al.}(2015)\citenamefont
  {Ningyuan}, \citenamefont {Owens}, \citenamefont {Sommer}, \citenamefont
  {Schuster},\ and\ \citenamefont {Simon}}]{Ningyuan2015}%
  \BibitemOpen
  \bibfield  {author} {\bibinfo {author} {\bibfnamefont {Jia}\ \bibnamefont
  {Ningyuan}}, \bibinfo {author} {\bibfnamefont {Clai}\ \bibnamefont {Owens}},
  \bibinfo {author} {\bibfnamefont {Ariel}\ \bibnamefont {Sommer}}, \bibinfo
  {author} {\bibfnamefont {David}\ \bibnamefont {Schuster}}, \ and\ \bibinfo
  {author} {\bibfnamefont {Jonathan}\ \bibnamefont {Simon}},\ }\bibfield
  {title} {\enquote {\bibinfo {title} {Time- and site-resolved dynamics in a
  topological circuit},}\ }\href {\doibase 10.1103/physrevx.5.021031}
  {\bibfield  {journal} {\bibinfo  {journal} {Physical Review X}\ }\textbf
  {\bibinfo {volume} {5}},\ \bibinfo {pages} {021031} (\bibinfo {year}
  {2015})}\BibitemShut {NoStop}%
\bibitem [{\citenamefont {Süsstrunk}\ \emph {et~al.}(2017)\citenamefont
  {Süsstrunk}, \citenamefont {Zimmermann},\ and\ \citenamefont
  {Huber}}]{Susstrunk2017}%
  \BibitemOpen
  \bibfield  {author} {\bibinfo {author} {\bibfnamefont {Roman}\ \bibnamefont
  {Süsstrunk}}, \bibinfo {author} {\bibfnamefont {Philipp}\ \bibnamefont
  {Zimmermann}}, \ and\ \bibinfo {author} {\bibfnamefont {Sebastian~D}\
  \bibnamefont {Huber}},\ }\bibfield  {title} {\enquote {\bibinfo {title}
  {Switchable topological phonon channels},}\ }\href {\doibase
  10.1088/1367-2630/aa591c} {\bibfield  {journal} {\bibinfo  {journal} {New
  Journal of Physics}\ }\textbf {\bibinfo {volume} {19}},\ \bibinfo {pages}
  {015013} (\bibinfo {year} {2017})}\BibitemShut {NoStop}%
\bibitem [{\citenamefont {Atala}\ \emph {et~al.}(2013)\citenamefont {Atala},
  \citenamefont {Aidelsburger}, \citenamefont {Barreiro}, \citenamefont
  {Abanin}, \citenamefont {Kitagawa}, \citenamefont {Demler},\ and\
  \citenamefont {Bloch}}]{Atala2013}%
  \BibitemOpen
  \bibfield  {author} {\bibinfo {author} {\bibfnamefont {Marcos}\ \bibnamefont
  {Atala}}, \bibinfo {author} {\bibfnamefont {Monika}\ \bibnamefont
  {Aidelsburger}}, \bibinfo {author} {\bibfnamefont {Julio~T.}\ \bibnamefont
  {Barreiro}}, \bibinfo {author} {\bibfnamefont {Dmitry}\ \bibnamefont
  {Abanin}}, \bibinfo {author} {\bibfnamefont {Takuya}\ \bibnamefont
  {Kitagawa}}, \bibinfo {author} {\bibfnamefont {Eugene}\ \bibnamefont
  {Demler}}, \ and\ \bibinfo {author} {\bibfnamefont {Immanuel}\ \bibnamefont
  {Bloch}},\ }\bibfield  {title} {\enquote {\bibinfo {title} {Direct
  measurement of the zak phase in topological bloch bands},}\ }\href {\doibase
  10.1038/nphys2790} {\bibfield  {journal} {\bibinfo  {journal} {Nature
  Physics}\ }\textbf {\bibinfo {volume} {9}},\ \bibinfo {pages} {795--800}
  (\bibinfo {year} {2013})}\BibitemShut {NoStop}%
\bibitem [{\citenamefont {Jotzu}\ \emph {et~al.}(2014)\citenamefont {Jotzu},
  \citenamefont {Messer}, \citenamefont {Desbuquois}, \citenamefont {Lebrat},
  \citenamefont {Uehlinger}, \citenamefont {Greif},\ and\ \citenamefont
  {Esslinger}}]{Jotzu2014}%
  \BibitemOpen
  \bibfield  {author} {\bibinfo {author} {\bibfnamefont {Gregor}\ \bibnamefont
  {Jotzu}}, \bibinfo {author} {\bibfnamefont {Michael}\ \bibnamefont {Messer}},
  \bibinfo {author} {\bibfnamefont {R{\'{e}}mi}\ \bibnamefont {Desbuquois}},
  \bibinfo {author} {\bibfnamefont {Martin}\ \bibnamefont {Lebrat}}, \bibinfo
  {author} {\bibfnamefont {Thomas}\ \bibnamefont {Uehlinger}}, \bibinfo
  {author} {\bibfnamefont {Daniel}\ \bibnamefont {Greif}}, \ and\ \bibinfo
  {author} {\bibfnamefont {Tilman}\ \bibnamefont {Esslinger}},\ }\bibfield
  {title} {\enquote {\bibinfo {title} {Experimental realization of the
  topological haldane model with ultracold fermions},}\ }\href {\doibase
  10.1038/nature13915} {\bibfield  {journal} {\bibinfo  {journal} {Nature}\
  }\textbf {\bibinfo {volume} {515}},\ \bibinfo {pages} {237--240} (\bibinfo
  {year} {2014})}\BibitemShut {NoStop}%
\bibitem [{\citenamefont {Aidelsburger}\ \emph {et~al.}(2014)\citenamefont
  {Aidelsburger}, \citenamefont {Lohse}, \citenamefont {Schweizer},
  \citenamefont {Atala}, \citenamefont {Barreiro}, \citenamefont
  {Nascimb{\`{e}}ne}, \citenamefont {Cooper}, \citenamefont {Bloch},\ and\
  \citenamefont {Goldman}}]{Aidelsburger2014}%
  \BibitemOpen
  \bibfield  {author} {\bibinfo {author} {\bibfnamefont {M.}~\bibnamefont
  {Aidelsburger}}, \bibinfo {author} {\bibfnamefont {M.}~\bibnamefont {Lohse}},
  \bibinfo {author} {\bibfnamefont {C.}~\bibnamefont {Schweizer}}, \bibinfo
  {author} {\bibfnamefont {M.}~\bibnamefont {Atala}}, \bibinfo {author}
  {\bibfnamefont {J.~T.}\ \bibnamefont {Barreiro}}, \bibinfo {author}
  {\bibfnamefont {S.}~\bibnamefont {Nascimb{\`{e}}ne}}, \bibinfo {author}
  {\bibfnamefont {N.~R.}\ \bibnamefont {Cooper}}, \bibinfo {author}
  {\bibfnamefont {I.}~\bibnamefont {Bloch}}, \ and\ \bibinfo {author}
  {\bibfnamefont {N.}~\bibnamefont {Goldman}},\ }\bibfield  {title} {\enquote
  {\bibinfo {title} {Measuring the chern number of hofstadter bands with
  ultracold bosonic atoms},}\ }\href {\doibase 10.1038/nphys3171} {\bibfield
  {journal} {\bibinfo  {journal} {Nature Physics}\ }\textbf {\bibinfo {volume}
  {11}},\ \bibinfo {pages} {162--166} (\bibinfo {year} {2014})}\BibitemShut
  {NoStop}%
\bibitem [{\citenamefont {Mancini}\ \emph {et~al.}(2015)\citenamefont
  {Mancini}, \citenamefont {Pagano}, \citenamefont {Cappellini}, \citenamefont
  {Livi}, \citenamefont {Rider}, \citenamefont {Catani}, \citenamefont {Sias},
  \citenamefont {Zoller}, \citenamefont {Inguscio}, \citenamefont {Dalmonte},\
  and\ \citenamefont {Fallani}}]{Mancini2015}%
  \BibitemOpen
  \bibfield  {author} {\bibinfo {author} {\bibfnamefont {M.}~\bibnamefont
  {Mancini}}, \bibinfo {author} {\bibfnamefont {G.}~\bibnamefont {Pagano}},
  \bibinfo {author} {\bibfnamefont {G.}~\bibnamefont {Cappellini}}, \bibinfo
  {author} {\bibfnamefont {L.}~\bibnamefont {Livi}}, \bibinfo {author}
  {\bibfnamefont {M.}~\bibnamefont {Rider}}, \bibinfo {author} {\bibfnamefont
  {J.}~\bibnamefont {Catani}}, \bibinfo {author} {\bibfnamefont
  {C.}~\bibnamefont {Sias}}, \bibinfo {author} {\bibfnamefont {P.}~\bibnamefont
  {Zoller}}, \bibinfo {author} {\bibfnamefont {M.}~\bibnamefont {Inguscio}},
  \bibinfo {author} {\bibfnamefont {M.}~\bibnamefont {Dalmonte}}, \ and\
  \bibinfo {author} {\bibfnamefont {L.}~\bibnamefont {Fallani}},\ }\bibfield
  {title} {\enquote {\bibinfo {title} {Observation of chiral edge states with
  neutral fermions in synthetic hall ribbons},}\ }\href {\doibase
  10.1126/science.aaa8736} {\bibfield  {journal} {\bibinfo  {journal}
  {Science}\ }\textbf {\bibinfo {volume} {349}},\ \bibinfo {pages} {1510--1513}
  (\bibinfo {year} {2015})}\BibitemShut {NoStop}%
\bibitem [{\citenamefont {Stuhl}\ \emph {et~al.}(2015)\citenamefont {Stuhl},
  \citenamefont {Lu}, \citenamefont {Aycock}, \citenamefont {Genkina},\ and\
  \citenamefont {Spielman}}]{Stuhl2015}%
  \BibitemOpen
  \bibfield  {author} {\bibinfo {author} {\bibfnamefont {B.~K.}\ \bibnamefont
  {Stuhl}}, \bibinfo {author} {\bibfnamefont {H.-I.}\ \bibnamefont {Lu}},
  \bibinfo {author} {\bibfnamefont {L.~M.}\ \bibnamefont {Aycock}}, \bibinfo
  {author} {\bibfnamefont {D.}~\bibnamefont {Genkina}}, \ and\ \bibinfo
  {author} {\bibfnamefont {I.~B.}\ \bibnamefont {Spielman}},\ }\bibfield
  {title} {\enquote {\bibinfo {title} {Visualizing edge states with an atomic
  bose gas in the quantum hall regime},}\ }\href {\doibase
  10.1126/science.aaa8515} {\bibfield  {journal} {\bibinfo  {journal}
  {Science}\ }\textbf {\bibinfo {volume} {349}},\ \bibinfo {pages} {1514--1518}
  (\bibinfo {year} {2015})}\BibitemShut {NoStop}%
\bibitem [{\citenamefont {Duca}\ \emph {et~al.}(2015)\citenamefont {Duca},
  \citenamefont {Li}, \citenamefont {Reitter}, \citenamefont {Bloch},
  \citenamefont {Schleier-Smith},\ and\ \citenamefont {Schneider}}]{Duca2015}%
  \BibitemOpen
  \bibfield  {author} {\bibinfo {author} {\bibfnamefont {L.}~\bibnamefont
  {Duca}}, \bibinfo {author} {\bibfnamefont {T.}~\bibnamefont {Li}}, \bibinfo
  {author} {\bibfnamefont {M.}~\bibnamefont {Reitter}}, \bibinfo {author}
  {\bibfnamefont {I.}~\bibnamefont {Bloch}}, \bibinfo {author} {\bibfnamefont
  {M.}~\bibnamefont {Schleier-Smith}}, \ and\ \bibinfo {author} {\bibfnamefont
  {U.}~\bibnamefont {Schneider}},\ }\bibfield  {title} {\enquote {\bibinfo
  {title} {An aharonov-bohm interferometer for determining bloch band
  topology},}\ }\href {\doibase 10.1126/science.1259052} {\bibfield  {journal}
  {\bibinfo  {journal} {Science}\ }\textbf {\bibinfo {volume} {347}},\ \bibinfo
  {pages} {288--292} (\bibinfo {year} {2015})}\BibitemShut {NoStop}%
\bibitem [{\citenamefont {Lohse}\ \emph {et~al.}(2015)\citenamefont {Lohse},
  \citenamefont {Schweizer}, \citenamefont {Zilberberg}, \citenamefont
  {Aidelsburger},\ and\ \citenamefont {Bloch}}]{Lohse2015}%
  \BibitemOpen
  \bibfield  {author} {\bibinfo {author} {\bibfnamefont {M.}~\bibnamefont
  {Lohse}}, \bibinfo {author} {\bibfnamefont {C.}~\bibnamefont {Schweizer}},
  \bibinfo {author} {\bibfnamefont {O.}~\bibnamefont {Zilberberg}}, \bibinfo
  {author} {\bibfnamefont {M.}~\bibnamefont {Aidelsburger}}, \ and\ \bibinfo
  {author} {\bibfnamefont {I.}~\bibnamefont {Bloch}},\ }\bibfield  {title}
  {\enquote {\bibinfo {title} {A thouless quantum pump with ultracold bosonic
  atoms in an optical superlattice},}\ }\href {\doibase 10.1038/nphys3584}
  {\bibfield  {journal} {\bibinfo  {journal} {Nature Physics}\ }\textbf
  {\bibinfo {volume} {12}},\ \bibinfo {pages} {350--354} (\bibinfo {year}
  {2015})}\BibitemShut {NoStop}%
\bibitem [{\citenamefont {Haldane}\ and\ \citenamefont
  {Raghu}(2008)}]{Haldane2008}%
  \BibitemOpen
  \bibfield  {author} {\bibinfo {author} {\bibfnamefont {F.~D.~M.}\
  \bibnamefont {Haldane}}\ and\ \bibinfo {author} {\bibfnamefont
  {S.}~\bibnamefont {Raghu}},\ }\bibfield  {title} {\enquote {\bibinfo {title}
  {Possible realization of directional optical waveguides in photonic crystals
  with broken time-reversal symmetry},}\ }\href {\doibase
  10.1103/physrevlett.100.013904} {\bibfield  {journal} {\bibinfo  {journal}
  {Physical Review Letters}\ }\textbf {\bibinfo {volume} {100}},\ \bibinfo
  {pages} {013904} (\bibinfo {year} {2008})}\BibitemShut {NoStop}%
\bibitem [{\citenamefont {Koch}\ \emph {et~al.}(2010)\citenamefont {Koch},
  \citenamefont {Houck}, \citenamefont {Hur},\ and\ \citenamefont
  {Girvin}}]{Koch2010}%
  \BibitemOpen
  \bibfield  {author} {\bibinfo {author} {\bibfnamefont {Jens}\ \bibnamefont
  {Koch}}, \bibinfo {author} {\bibfnamefont {Andrew~A.}\ \bibnamefont {Houck}},
  \bibinfo {author} {\bibfnamefont {Karyn~Le}\ \bibnamefont {Hur}}, \ and\
  \bibinfo {author} {\bibfnamefont {S.~M.}\ \bibnamefont {Girvin}},\ }\bibfield
   {title} {\enquote {\bibinfo {title} {Time-reversal-symmetry breaking in
  circuit-{QED}-based photon lattices},}\ }\href {\doibase
  10.1103/physreva.82.043811} {\bibfield  {journal} {\bibinfo  {journal}
  {Physical Review A}\ }\textbf {\bibinfo {volume} {82}},\ \bibinfo {pages}
  {043811} (\bibinfo {year} {2010})}\BibitemShut {NoStop}%
\bibitem [{\citenamefont {Hafezi}\ \emph {et~al.}(2011)\citenamefont {Hafezi},
  \citenamefont {Demler}, \citenamefont {Lukin},\ and\ \citenamefont
  {Taylor}}]{Hafezi2011}%
  \BibitemOpen
  \bibfield  {author} {\bibinfo {author} {\bibfnamefont {Mohammad}\
  \bibnamefont {Hafezi}}, \bibinfo {author} {\bibfnamefont {Eugene~A.}\
  \bibnamefont {Demler}}, \bibinfo {author} {\bibfnamefont {Mikhail~D.}\
  \bibnamefont {Lukin}}, \ and\ \bibinfo {author} {\bibfnamefont {Jacob~M.}\
  \bibnamefont {Taylor}},\ }\bibfield  {title} {\enquote {\bibinfo {title}
  {Robust optical delay lines with topological protection},}\ }\href {\doibase
  10.1038/nphys2063} {\bibfield  {journal} {\bibinfo  {journal} {Nature
  Physics}\ }\textbf {\bibinfo {volume} {7}},\ \bibinfo {pages} {907--912}
  (\bibinfo {year} {2011})}\BibitemShut {NoStop}%
\bibitem [{\citenamefont {Berg}\ \emph {et~al.}(2011)\citenamefont {Berg},
  \citenamefont {Joel}, \citenamefont {Koolyk},\ and\ \citenamefont
  {Prodan}}]{Berg2011}%
  \BibitemOpen
  \bibfield  {author} {\bibinfo {author} {\bibfnamefont {Nina}\ \bibnamefont
  {Berg}}, \bibinfo {author} {\bibfnamefont {Kira}\ \bibnamefont {Joel}},
  \bibinfo {author} {\bibfnamefont {Miriam}\ \bibnamefont {Koolyk}}, \ and\
  \bibinfo {author} {\bibfnamefont {Emil}\ \bibnamefont {Prodan}},\ }\bibfield
  {title} {\enquote {\bibinfo {title} {Topological phonon modes in filamentary
  structures},}\ }\href {\doibase 10.1103/physreve.83.021913} {\bibfield
  {journal} {\bibinfo  {journal} {Physical Review E}\ }\textbf {\bibinfo
  {volume} {83}},\ \bibinfo {pages} {021913} (\bibinfo {year}
  {2011})}\BibitemShut {NoStop}%
\bibitem [{\citenamefont {Yannopapas}(2012)}]{Yannopapas2012}%
  \BibitemOpen
  \bibfield  {author} {\bibinfo {author} {\bibfnamefont {Vassilios}\
  \bibnamefont {Yannopapas}},\ }\bibfield  {title} {\enquote {\bibinfo {title}
  {Topological photonic bands in two-dimensional networks of metamaterial
  elements},}\ }\href {\doibase 10.1088/1367-2630/14/11/113017} {\bibfield
  {journal} {\bibinfo  {journal} {New Journal of Physics}\ }\textbf {\bibinfo
  {volume} {14}},\ \bibinfo {pages} {113017} (\bibinfo {year}
  {2012})}\BibitemShut {NoStop}%
\bibitem [{\citenamefont {Kane}\ and\ \citenamefont
  {Lubensky}(2013)}]{Kane2013}%
  \BibitemOpen
  \bibfield  {author} {\bibinfo {author} {\bibfnamefont {C.~L.}\ \bibnamefont
  {Kane}}\ and\ \bibinfo {author} {\bibfnamefont {T.~C.}\ \bibnamefont
  {Lubensky}},\ }\bibfield  {title} {\enquote {\bibinfo {title} {Topological
  boundary modes in isostatic lattices},}\ }\href {\doibase 10.1038/nphys2835}
  {\bibfield  {journal} {\bibinfo  {journal} {Nature Physics}\ }\textbf
  {\bibinfo {volume} {10}},\ \bibinfo {pages} {39--45} (\bibinfo {year}
  {2013})}\BibitemShut {NoStop}%
\bibitem [{\citenamefont {Lu}\ \emph {et~al.}(2014)\citenamefont {Lu},
  \citenamefont {Joannopoulos},\ and\ \citenamefont
  {Solja{\v{c}}i{\'{c}}}}]{Lu2014}%
  \BibitemOpen
  \bibfield  {author} {\bibinfo {author} {\bibfnamefont {Ling}\ \bibnamefont
  {Lu}}, \bibinfo {author} {\bibfnamefont {John~D.}\ \bibnamefont
  {Joannopoulos}}, \ and\ \bibinfo {author} {\bibfnamefont {Marin}\
  \bibnamefont {Solja{\v{c}}i{\'{c}}}},\ }\bibfield  {title} {\enquote
  {\bibinfo {title} {Topological photonics},}\ }\href {\doibase
  10.1038/nphoton.2014.248} {\bibfield  {journal} {\bibinfo  {journal} {Nature
  Photonics}\ }\textbf {\bibinfo {volume} {8}},\ \bibinfo {pages} {821--829}
  (\bibinfo {year} {2014})}\BibitemShut {NoStop}%
\bibitem [{\citenamefont {Kariyado}\ and\ \citenamefont
  {Hatsugai}(2015)}]{Kariyado2015}%
  \BibitemOpen
  \bibfield  {author} {\bibinfo {author} {\bibfnamefont {Toshikaze}\
  \bibnamefont {Kariyado}}\ and\ \bibinfo {author} {\bibfnamefont {Yasuhiro}\
  \bibnamefont {Hatsugai}},\ }\bibfield  {title} {\enquote {\bibinfo {title}
  {Manipulation of dirac cones in mechanical graphene},}\ }\href {\doibase
  10.1038/srep18107} {\bibfield  {journal} {\bibinfo  {journal} {Scientific
  Reports}\ }\textbf {\bibinfo {volume} {5}},\ \bibinfo {pages} {18107}
  (\bibinfo {year} {2015})}\BibitemShut {NoStop}%
\bibitem [{\citenamefont {Wang}\ \emph
  {et~al.}(2015{\natexlab{a}})\citenamefont {Wang}, \citenamefont {Luan},\ and\
  \citenamefont {Zhang}}]{Wang2015a}%
  \BibitemOpen
  \bibfield  {author} {\bibinfo {author} {\bibfnamefont {Yao-Ting}\
  \bibnamefont {Wang}}, \bibinfo {author} {\bibfnamefont {Pi-Gang}\
  \bibnamefont {Luan}}, \ and\ \bibinfo {author} {\bibfnamefont {Shuang}\
  \bibnamefont {Zhang}},\ }\bibfield  {title} {\enquote {\bibinfo {title}
  {Coriolis force induced topological order for classical mechanical
  vibrations},}\ }\href {\doibase 10.1088/1367-2630/17/7/073031} {\bibfield
  {journal} {\bibinfo  {journal} {New Journal of Physics}\ }\textbf {\bibinfo
  {volume} {17}},\ \bibinfo {pages} {073031} (\bibinfo {year}
  {2015}{\natexlab{a}})}\BibitemShut {NoStop}%
\bibitem [{\citenamefont {Wang}\ \emph
  {et~al.}(2015{\natexlab{b}})\citenamefont {Wang}, \citenamefont {Lu},\ and\
  \citenamefont {Bertoldi}}]{Wang2015b}%
  \BibitemOpen
  \bibfield  {author} {\bibinfo {author} {\bibfnamefont {Pai}\ \bibnamefont
  {Wang}}, \bibinfo {author} {\bibfnamefont {Ling}\ \bibnamefont {Lu}}, \ and\
  \bibinfo {author} {\bibfnamefont {Katia}\ \bibnamefont {Bertoldi}},\
  }\bibfield  {title} {\enquote {\bibinfo {title} {Topological phononic
  crystals with one-way elastic edge waves},}\ }\href {\doibase
  10.1103/physrevlett.115.104302} {\bibfield  {journal} {\bibinfo  {journal}
  {Physical Review Letters}\ }\textbf {\bibinfo {volume} {115}},\ \bibinfo
  {pages} {104302} (\bibinfo {year} {2015}{\natexlab{b}})}\BibitemShut
  {NoStop}%
\bibitem [{\citenamefont {Peter}\ \emph {et~al.}(2015)\citenamefont {Peter},
  \citenamefont {Yao}, \citenamefont {Lang}, \citenamefont {Huber},
  \citenamefont {Lukin},\ and\ \citenamefont {Büchler}}]{Peter2015}%
  \BibitemOpen
  \bibfield  {author} {\bibinfo {author} {\bibfnamefont {David}\ \bibnamefont
  {Peter}}, \bibinfo {author} {\bibfnamefont {Norman~Y.}\ \bibnamefont {Yao}},
  \bibinfo {author} {\bibfnamefont {Nicolai}\ \bibnamefont {Lang}}, \bibinfo
  {author} {\bibfnamefont {Sebastian~D.}\ \bibnamefont {Huber}}, \bibinfo
  {author} {\bibfnamefont {Mikhail~D.}\ \bibnamefont {Lukin}}, \ and\ \bibinfo
  {author} {\bibfnamefont {Hans~Peter}\ \bibnamefont {Büchler}},\ }\bibfield
  {title} {\enquote {\bibinfo {title} {Topological bands with a chern number
  {C=2} by dipolar exchange interactions},}\ }\href {\doibase
  10.1103/physreva.91.053617} {\bibfield  {journal} {\bibinfo  {journal}
  {Physical Review A}\ }\textbf {\bibinfo {volume} {91}},\ \bibinfo {pages}
  {053617} (\bibinfo {year} {2015})}\BibitemShut {NoStop}%
\bibitem [{\citenamefont {Yao}\ \emph {et~al.}(2013)\citenamefont {Yao},
  \citenamefont {Laumann}, \citenamefont {Gorshkov}, \citenamefont {Weimer},
  \citenamefont {Jiang}, \citenamefont {Cirac}, \citenamefont {Zoller},\ and\
  \citenamefont {Lukin}}]{Yao2013}%
  \BibitemOpen
  \bibfield  {author} {\bibinfo {author} {\bibfnamefont {N.Y.}\ \bibnamefont
  {Yao}}, \bibinfo {author} {\bibfnamefont {C.R.}\ \bibnamefont {Laumann}},
  \bibinfo {author} {\bibfnamefont {A.V.}\ \bibnamefont {Gorshkov}}, \bibinfo
  {author} {\bibfnamefont {H.}~\bibnamefont {Weimer}}, \bibinfo {author}
  {\bibfnamefont {L.}~\bibnamefont {Jiang}}, \bibinfo {author} {\bibfnamefont
  {J.I.}\ \bibnamefont {Cirac}}, \bibinfo {author} {\bibfnamefont
  {P.}~\bibnamefont {Zoller}}, \ and\ \bibinfo {author} {\bibfnamefont {M.D.}\
  \bibnamefont {Lukin}},\ }\bibfield  {title} {\enquote {\bibinfo {title}
  {Topologically protected quantum state transfer in a chiral spin liquid},}\
  }\href {\doibase 10.1038/ncomms2531} {\bibfield  {journal} {\bibinfo
  {journal} {Nature Communications}\ }\textbf {\bibinfo {volume} {4}},\
  \bibinfo {pages} {1585} (\bibinfo {year} {2013})}\BibitemShut {NoStop}%
\bibitem [{\citenamefont {Dlaska}\ \emph {et~al.}(2017)\citenamefont {Dlaska},
  \citenamefont {Vermersch},\ and\ \citenamefont {Zoller}}]{Dlaska2017}%
  \BibitemOpen
  \bibfield  {author} {\bibinfo {author} {\bibfnamefont {C}~\bibnamefont
  {Dlaska}}, \bibinfo {author} {\bibfnamefont {B}~\bibnamefont {Vermersch}}, \
  and\ \bibinfo {author} {\bibfnamefont {P}~\bibnamefont {Zoller}},\ }\bibfield
   {title} {\enquote {\bibinfo {title} {Robust quantum state transfer via
  topologically protected edge channels in dipolar arrays},}\ }\href {\doibase
  10.1088/2058-9565/2/1/015001} {\bibfield  {journal} {\bibinfo  {journal}
  {Quantum Science and Technology}\ }\textbf {\bibinfo {volume} {2}},\ \bibinfo
  {pages} {015001} (\bibinfo {year} {2017})}\BibitemShut {NoStop}%
\bibitem [{\citenamefont {Zak}(1989)}]{Zak1989}%
  \BibitemOpen
  \bibfield  {author} {\bibinfo {author} {\bibfnamefont {J.}~\bibnamefont
  {Zak}},\ }\bibfield  {title} {\enquote {\bibinfo {title} {Berry's phase for
  energy bands in solids},}\ }\href {\doibase 10.1103/physrevlett.62.2747}
  {\bibfield  {journal} {\bibinfo  {journal} {Physical Review Letters}\
  }\textbf {\bibinfo {volume} {62}},\ \bibinfo {pages} {2747--2750} (\bibinfo
  {year} {1989})}\BibitemShut {NoStop}%
\bibitem [{\citenamefont {Kitaev}(2001)}]{Kitaev2001}%
  \BibitemOpen
  \bibfield  {author} {\bibinfo {author} {\bibfnamefont {A~Yu}\ \bibnamefont
  {Kitaev}},\ }\bibfield  {title} {\enquote {\bibinfo {title} {Unpaired
  majorana fermions in quantum wires},}\ }\href {\doibase
  10.1070/1063-7869/44/10s/s29} {\bibfield  {journal} {\bibinfo  {journal}
  {Physics-Uspekhi}\ }\textbf {\bibinfo {volume} {44}},\ \bibinfo {pages}
  {131--136} (\bibinfo {year} {2001})}\BibitemShut {NoStop}%
\bibitem [{\citenamefont {Su}\ \emph {et~al.}(1979)\citenamefont {Su},
  \citenamefont {Schrieffer},\ and\ \citenamefont {Heeger}}]{Su1979}%
  \BibitemOpen
  \bibfield  {author} {\bibinfo {author} {\bibfnamefont {W.~P.}\ \bibnamefont
  {Su}}, \bibinfo {author} {\bibfnamefont {J.~R.}\ \bibnamefont {Schrieffer}},
  \ and\ \bibinfo {author} {\bibfnamefont {A.~J.}\ \bibnamefont {Heeger}},\
  }\bibfield  {title} {\enquote {\bibinfo {title} {Solitons in
  polyacetylene},}\ }\href {\doibase 10.1103/physrevlett.42.1698} {\bibfield
  {journal} {\bibinfo  {journal} {Physical Review Letters}\ }\textbf {\bibinfo
  {volume} {42}},\ \bibinfo {pages} {1698--1701} (\bibinfo {year}
  {1979})}\BibitemShut {NoStop}%
\bibitem [{\citenamefont {Houck}\ \emph {et~al.}(2012)\citenamefont {Houck},
  \citenamefont {Türeci},\ and\ \citenamefont {Koch}}]{Houck2012}%
  \BibitemOpen
  \bibfield  {author} {\bibinfo {author} {\bibfnamefont {Andrew~A.}\
  \bibnamefont {Houck}}, \bibinfo {author} {\bibfnamefont {Hakan~E.}\
  \bibnamefont {Türeci}}, \ and\ \bibinfo {author} {\bibfnamefont {Jens}\
  \bibnamefont {Koch}},\ }\bibfield  {title} {\enquote {\bibinfo {title}
  {On-chip quantum simulation with superconducting circuits},}\ }\href
  {\doibase 10.1038/nphys2251} {\bibfield  {journal} {\bibinfo  {journal}
  {Nature Physics}\ }\textbf {\bibinfo {volume} {8}},\ \bibinfo {pages}
  {292--299} (\bibinfo {year} {2012})}\BibitemShut {NoStop}%
\bibitem [{\citenamefont {Carusotto}\ and\ \citenamefont
  {Ciuti}(2013)}]{Carusotto2013}%
  \BibitemOpen
  \bibfield  {author} {\bibinfo {author} {\bibfnamefont {Iacopo}\ \bibnamefont
  {Carusotto}}\ and\ \bibinfo {author} {\bibfnamefont {Cristiano}\ \bibnamefont
  {Ciuti}},\ }\bibfield  {title} {\enquote {\bibinfo {title} {Quantum fluids of
  light},}\ }\href {\doibase 10.1103/revmodphys.85.299} {\bibfield  {journal}
  {\bibinfo  {journal} {Reviews of Modern Physics}\ }\textbf {\bibinfo {volume}
  {85}},\ \bibinfo {pages} {299--366} (\bibinfo {year} {2013})}\BibitemShut
  {NoStop}%
\bibitem [{\citenamefont {Lukin}\ \emph {et~al.}(2001)\citenamefont {Lukin},
  \citenamefont {Fleischhauer}, \citenamefont {Cote}, \citenamefont {Duan},
  \citenamefont {Jaksch}, \citenamefont {Cirac},\ and\ \citenamefont
  {Zoller}}]{Lukin2001}%
  \BibitemOpen
  \bibfield  {author} {\bibinfo {author} {\bibfnamefont {M.~D.}\ \bibnamefont
  {Lukin}}, \bibinfo {author} {\bibfnamefont {M.}~\bibnamefont {Fleischhauer}},
  \bibinfo {author} {\bibfnamefont {R.}~\bibnamefont {Cote}}, \bibinfo {author}
  {\bibfnamefont {L.~M.}\ \bibnamefont {Duan}}, \bibinfo {author}
  {\bibfnamefont {D.}~\bibnamefont {Jaksch}}, \bibinfo {author} {\bibfnamefont
  {J.~I.}\ \bibnamefont {Cirac}}, \ and\ \bibinfo {author} {\bibfnamefont
  {P.}~\bibnamefont {Zoller}},\ }\bibfield  {title} {\enquote {\bibinfo {title}
  {Dipole blockade and quantum information processing in mesoscopic atomic
  ensembles},}\ }\href {\doibase 10.1103/physrevlett.87.037901} {\bibfield
  {journal} {\bibinfo  {journal} {Physical Review Letters}\ }\textbf {\bibinfo
  {volume} {87}},\ \bibinfo {pages} {037901} (\bibinfo {year}
  {2001})}\BibitemShut {NoStop}%
\bibitem [{\citenamefont {Doherty}\ \emph {et~al.}(2013)\citenamefont
  {Doherty}, \citenamefont {Manson}, \citenamefont {Delaney}, \citenamefont
  {Jelezko}, \citenamefont {Wrachtrup},\ and\ \citenamefont
  {Hollenberg}}]{Doherty2013}%
  \BibitemOpen
  \bibfield  {author} {\bibinfo {author} {\bibfnamefont {Marcus~W.}\
  \bibnamefont {Doherty}}, \bibinfo {author} {\bibfnamefont {Neil~B.}\
  \bibnamefont {Manson}}, \bibinfo {author} {\bibfnamefont {Paul}\ \bibnamefont
  {Delaney}}, \bibinfo {author} {\bibfnamefont {Fedor}\ \bibnamefont
  {Jelezko}}, \bibinfo {author} {\bibfnamefont {Jörg}\ \bibnamefont
  {Wrachtrup}}, \ and\ \bibinfo {author} {\bibfnamefont {Lloyd~C.L.}\
  \bibnamefont {Hollenberg}},\ }\bibfield  {title} {\enquote {\bibinfo {title}
  {The nitrogen-vacancy colour centre in diamond},}\ }\href {\doibase
  10.1016/j.physrep.2013.02.001} {\bibfield  {journal} {\bibinfo  {journal}
  {Physics Reports}\ }\textbf {\bibinfo {volume} {528}},\ \bibinfo {pages}
  {1--45} (\bibinfo {year} {2013})}\BibitemShut {NoStop}%
\bibitem [{\citenamefont {Duan}\ and\ \citenamefont {Monroe}(2010)}]{Duan2010}%
  \BibitemOpen
  \bibfield  {author} {\bibinfo {author} {\bibfnamefont {L.-M.}\ \bibnamefont
  {Duan}}\ and\ \bibinfo {author} {\bibfnamefont {C.}~\bibnamefont {Monroe}},\
  }\bibfield  {title} {\enquote {\bibinfo {title} {Colloquium: Quantum networks
  with trapped ions},}\ }\href {\doibase 10.1103/revmodphys.82.1209} {\bibfield
   {journal} {\bibinfo  {journal} {Reviews of Modern Physics}\ }\textbf
  {\bibinfo {volume} {82}},\ \bibinfo {pages} {1209--1224} (\bibinfo {year}
  {2010})}\BibitemShut {NoStop}%
\bibitem [{\citenamefont {Lieb}\ and\ \citenamefont
  {Robinson}(1972)}]{Lieb1972}%
  \BibitemOpen
  \bibfield  {author} {\bibinfo {author} {\bibfnamefont {Elliott~H.}\
  \bibnamefont {Lieb}}\ and\ \bibinfo {author} {\bibfnamefont {Derek~W.}\
  \bibnamefont {Robinson}},\ }\bibfield  {title} {\enquote {\bibinfo {title}
  {The finite group velocity of quantum spin systems},}\ }\href {\doibase
  10.1007/BF01645779} {\bibfield  {journal} {\bibinfo  {journal}
  {Communications in Mathematical Physics}\ }\textbf {\bibinfo {volume} {28}},\
  \bibinfo {pages} {251--257} (\bibinfo {year} {1972})}\BibitemShut {NoStop}%
\bibitem [{\citenamefont {Jansen}\ \emph {et~al.}(2007)\citenamefont {Jansen},
  \citenamefont {Ruskai},\ and\ \citenamefont {Seiler}}]{Jansen2007}%
  \BibitemOpen
  \bibfield  {author} {\bibinfo {author} {\bibfnamefont {Sabine}\ \bibnamefont
  {Jansen}}, \bibinfo {author} {\bibfnamefont {Mary-Beth}\ \bibnamefont
  {Ruskai}}, \ and\ \bibinfo {author} {\bibfnamefont {Ruedi}\ \bibnamefont
  {Seiler}},\ }\bibfield  {title} {\enquote {\bibinfo {title} {Bounds for the
  adiabatic approximation with applications to quantum computation},}\ }\href
  {\doibase 10.1063/1.2798382} {\bibfield  {journal} {\bibinfo  {journal}
  {Journal of Mathematical Physics}\ }\textbf {\bibinfo {volume} {48}},\
  \bibinfo {pages} {102111} (\bibinfo {year} {2007})}\BibitemShut {NoStop}%
\bibitem [{\citenamefont {Haack}\ \emph {et~al.}(2010)\citenamefont {Haack},
  \citenamefont {Helmer}, \citenamefont {Mariantoni}, \citenamefont
  {Marquardt},\ and\ \citenamefont {Solano}}]{Haack2010}%
  \BibitemOpen
  \bibfield  {author} {\bibinfo {author} {\bibfnamefont {G.}~\bibnamefont
  {Haack}}, \bibinfo {author} {\bibfnamefont {F.}~\bibnamefont {Helmer}},
  \bibinfo {author} {\bibfnamefont {M.}~\bibnamefont {Mariantoni}}, \bibinfo
  {author} {\bibfnamefont {F.}~\bibnamefont {Marquardt}}, \ and\ \bibinfo
  {author} {\bibfnamefont {E.}~\bibnamefont {Solano}},\ }\bibfield  {title}
  {\enquote {\bibinfo {title} {Resonant quantum gates in circuit quantum
  electrodynamics},}\ }\href {\doibase 10.1103/physrevb.82.024514} {\bibfield
  {journal} {\bibinfo  {journal} {Physical Review B}\ }\textbf {\bibinfo
  {volume} {82}},\ \bibinfo {pages} {024514} (\bibinfo {year}
  {2010})}\BibitemShut {NoStop}%
\end{thebibliography}
%



\section*{Supplementary material}

Nicolai Lang[*] and Hans Peter Büchler
*Institute for Theoretical Physics III and Center for Integrated Quantum Science and Technology,
University of Stuttgart, 70550 Stuttgart, Germany*
(Dated: September 8, 2017)

This supplementary material contains detailed calculations for results presented in the manuscript and extends some aspects more thoroughly: In Sec. I we discuss an alternative implementation of the one-dimensional topological network derived from the Majorana chain. In Sec. II we comment on the relations between Majorana- and SSH chain in the context of bosonic networks. In Sec. III we give a more detailed account on the role of symmetries and their importance for two-dimensional networks.

In Sec. IV we present additional results for the effects of quenched disorder on the transfer. In Sec. V we derive analytic expressions for the spectrum of the SSH/Majorana chain with open boundaries, discuss the scaling of low-lying eigenstates for $L \to \infty$ close to the topological phase transition, and present a detailed derivation of the universal behaviour when approaching the transition from the topological phase. In Sec. VI we derive rigorous bounds on the non-adiabatic bulk losses and compare them with simulations.

## I.  MAJORANA CHAIN MODEL

The original Majorana chain Hamiltonian for spinless fermions on an open chain of length $L$ reads [S1]

$$\hat{H}_{\text{MC}} = \sum_{i=1}^{L-1} \left( w_i \, c_i^\dagger c_{i+1} - \Delta_i \, c_i c_{i+1} + \text{h.c.} \right) + \sum_{i=1}^{L} \mu_i \left( c_i^\dagger c_i - \frac{1}{2} \right), \tag{S1}$$

where $c_i^\dagger$ ($c_i$) denote *fermionic* creation (annihilation) operators, $w_i$ is the tunnelling amplitude, $\Delta_i$ the superconducting gap (which can always be gauged real), and $\mu_i$ denotes the chemical potential, all of which can be, in principle, site-dependent.

For homogeneous parameters $\mu_i \equiv \bar{\mu}$ and $w_i \equiv \bar{w} = \bar{\Delta} \equiv \Delta_i$ one finds two gapped phases for $|\bar{\mu}| \lessgtr 2|\bar{w}|$ connected by a gapless spectrum at $|\bar{\mu}| = 2|\bar{w}|$ indicating a phase transition. The latter is of topological nature as the symmetries of both phases coincide: For the topological phase, $|\bar{\mu}| < 2|\bar{w}|$, one finds two degenerate edge modes that give rise to a two-fold ground state degeneracy; hence the modes are identified as Majorana bound states. In the trivial phase, $|\bar{\mu}| > 2|\bar{w}|$, the ground state is unique and no localized edge modes are present.

In the $2L$-dimensional space of Nambu spinors $\boldsymbol{\Psi} = (c_1^\dagger, c_1, \ldots, c_L^\dagger, c_L)^T$ the Hamiltonian can be encoded by a Hermitian matrix $H_{\text{MC}}$ via

$$\hat{H}_{\text{MC}} = \frac{1}{2} \, \boldsymbol{\Psi}^\dagger H_{\text{MC}} \boldsymbol{\Psi} \,. \tag{S2}$$

This matrix features an *intrinsic* (antilinear) particle-hole (PH) symmetry $C \, H_{\text{MC}} \, C^{-1} = -H_{\text{MC}}$ with $C = \mathcal{K} U_C^{\text{MC}}$ and $C^2 = +\mathbb{1}$. Here, $\mathcal{K}$ denotes complex conjugation and the unitary $U_C^{\text{MC}}$ takes the form

$$U_C^{\text{MC}} = \mathbb{1}_{L \times L} \otimes \begin{pmatrix} 0 & 1 \\ 1 & 0 \end{pmatrix}, \tag{S3}$$

i.e., the PH symmetry acts as $c_i^\dagger \leftrightarrow c_i$ on Nambu space. Hence the Majorana chain is in symmetry class D of the Altland-Zirnbauer classification [S2–S5]. In one dimension, this allows for the definition of a $\mathbb{Z}_2$ topological invariant $\nu$ [S6–S9] which is responsible for the emergence of the disorder-resilient edge modes bound to the open ends of the chain whenever $\nu \neq 0 \mod 2 \Leftrightarrow |\bar{\mu}| < 2|\bar{w}|$, i.e., in the topological phase. If all tunnelling amplitudes $w_i$ and superconducting pairings $\Delta_i$ are real, the Hamiltonian features additionally the time-reversal symmetry $T = \mathcal{K}$ with $T^2 = +\mathbb{1}$. Then the model can be considered as an element of the symmetry class BDI with a $\mathbb{Z}$ topological index [S6–S9].

The implementation of an analogue system with bosonic degrees of freedom follows a straightforward procedure: First, each fermionic site in the original Majorana chain is replaced by *two* bosonic modes. The two

---

[*] nicolai@itp3.uni-stuttgart.de

bosonic modes at site $i$ are denoted by the bosonic operators $b_i^{(\dagger)}$ and $b_{\bar{i}}^{(\dagger)}$,

$$c_i \to b_i^{(\dagger)} \quad \text{and} \quad c_i^\dagger \to b_{\bar{i}}^{(\dagger)}. \tag{S4}$$

Note that the original indices $i = 1, \ldots, L$ label now the "upper" sites whereas bar-ed indices $\bar{i} = \bar{1}, \ldots, \bar{L}$ denote the "lower" sites, see Fig. 2 (b) of the main text. By this construction, the Majorana chain translates to a linear chain of length $L$ with *two* quantum harmonic oscillators per site. The bosonic Hamiltonian takes the form

$$\hat{H}_{\text{bMC}} \equiv \boldsymbol{\xi}^\dagger \left[H_{\text{MC}} + \delta \mathbb{1}\right] \boldsymbol{\xi} \equiv \boldsymbol{\xi}^\dagger H_{\text{bMC}} \boldsymbol{\xi} \tag{S5}$$

with $\boldsymbol{\xi} = (b_1, b_{\bar{1}}, \ldots b_i, b_{\bar{i}}, \ldots, b_L, b_{\bar{L}})^T$ and $H_{\text{bMC}} = H_{\text{MC}} + \delta \mathbb{1}$ the matrix appearing in equation (S5). The constant positive energy shift $\delta > 0$ is required to enforce positivity on the matrix $H_{\text{bMC}}$; its value can be chosen arbitrarily (as long as $H_{\text{bMC}} > 0$) and does not change the topological properties (e.g., the existence of edge modes).

To make the new interpretation in terms of bosonic modes explicit, we substitute the parameters in $H_{\text{bMC}}$, inherited from the Majorana chain (S1), as follows:

$$H_{\text{bMC}} = \begin{pmatrix} \delta - \mu_1 & 0 & -w_1 & -\Delta_1 \\ 0 & \delta + \mu_1 & \Delta_1 & w_1 \\ -w_1 & \Delta_1 & \ddots & \ddots \\ -\Delta_1 & w_1 & \ddots & \ddots \end{pmatrix} \longrightarrow \begin{pmatrix} \omega_1 & 0 & t_{1,2} & t_{1,\bar{2}} \\ 0 & \omega_{\bar{1}} & t_{\bar{1},2} & t_{\bar{1},\bar{2}} \\ t_{1,2} & t_{\bar{1},2} & \ddots & \ddots \\ t_{1,\bar{2}} & t_{\bar{1},\bar{2}} & \ddots & \ddots \end{pmatrix} = H_{\text{bMC}} \tag{S6}$$

The diagonal elements of $H_{\text{bMC}}$ describe the eigenfrequencies $\omega_I$ ($I = i, \bar{i}$) of the local modes $b_I$, while the off-diagonal elements capture the real hopping amplitudes $t_{I,J}$ connecting modes $b_I$ and $b_J$. E.g., $\omega_{\bar{1}}$ denotes the eigenfrequency of the lower mode $b_{\bar{1}}$ and $t_{1,\bar{2}}$ the coupling between the upper mode $b_1$ and the lower mode $b_{\bar{2}}$, see Fig. 2 (b) of the main text.

Using the identifications in equation (S6), one finds that the role of the chemical potential $2\mu_i$ is now played by the difference $\delta\omega_i = \omega_{\bar{i}} - \omega_i$ of the local mode frequencies at each site, while the horizontal hopping amplitudes $t_{i,i+1}$ and $t_{\bar{i},\overline{i+1}}$ are identified with the fermion tunnelling rate $w_i$, and the diagonal amplitudes $t_{i,\overline{i+1}}$ and $t_{\bar{i},i+1}$ stem from the superconducting order parameter $\Delta_i$.

The bosonic Hamiltonian $\hat{H}_{\text{bMC}}$ features the same single-particle band structure as the original Majorana chain, and exhibits the same topological properties and topological quantum numbers. Therefore it gives rise to the same edge modes. Note that these are statements about *single*-particle physics where statistics is not relevant. Furthermore, it is crucial to stress a conceptual difference between the fermionic Hamiltonian $\hat{H}_{\text{MC}}$ and its bosonic descendant $\hat{H}_{\text{bMC}}$: The former is described by $L$ fermionic modes and only in Nambu space an artificial mode doubling occurs, giving rise to the intrinsic particle hole symmetry $C$. The bosonic setup is truly described by $2L$ *independent* bosonic modes, as the mode doubling at each lattice site is required for the implementation. This has immediate consequences for the interpretation of the topology-protecting PH symmetry as well: The latter—inherent to any (fermionic) Bogoliubov-de Gennes Hamiltonian in Nambu space—is converted to a non-trivial *real* symmetry of the new bosonic theory. Either the identifications in equation (S6), or the symmetry relation $C H_{\text{bMC}} C^{-1} = -H_{\text{bMC}} + 2\delta \mathbb{1}$, give rise to the local constraints

$$\omega_i + \omega_{\bar{i}} = 2\delta \tag{S7a}$$
$$t_{i,\overline{i+1}} = -t_{\bar{i},i+1} \tag{S7b}$$
$$t_{i,i+1} = -t_{\bar{i},\overline{i+1}}. \tag{S7c}$$

Note that the value of $\delta$ merely determines the global energy scale whereas its site-independence is crucial for PH symmetry. All these constraints can be fulfilled by pairwise, *local* fine-tuning of the modes and their couplings to nearest neighbours.

For the purpose of state transfer, we choose homogeneous frequencies $\bar{\omega}_- \equiv \omega_i$ and $\bar{\omega}_+ \equiv \omega_{\bar{i}}$ the difference of which is (globally) tunable: $\delta\bar{\omega}(t) = \bar{\omega}_-(t) - \bar{\omega}_+(t)$. In contrast to the SSH chain setup from the main text, here the mode couplings are fixed at $t_{i,i+1} \equiv \bar{t} \equiv t_{i,\overline{i+1}}$ and $t_{\bar{i},\overline{i+1}} \equiv -\bar{t} \equiv t_{\bar{i},i+1}$. Note that the spatial homogeneity assumed for couplings and mode frequencies is not essential due to the topological band structure, as long as the local symmetries (S7) are respected.

For $\delta\bar{\omega} = 0$, one finds the flat-band "sweet spot" of perfectly localized (and degenerate) edge modes $\tilde{b}_{\text{C}} \propto b_1 + b_{\bar{1}}$ and $\tilde{b}_{\text{T}} \propto b_L - b_{\bar{L}}$. At the critical point $\delta\bar{\omega}_{\text{crit}} = 4\bar{t}$, the spectrum becomes gapless and the topological phase transition occurs. For $\delta\bar{\omega} > \delta\bar{\omega}_{\text{crit}}$ the chain becomes gapped again but features no edge modes any more.

Therefore the protocol for state transfer reads

$$\delta\bar{\omega}(t) = \delta\bar{\omega}_{\text{max}} \cdot \mathcal{F}(t) \tag{S8}$$

with the protocol parameter $\delta\bar{\omega}_{\text{max}} < \delta\bar{\omega}_{\text{crit}}$. Together with the previously given fixed values of mode couplings,

this defines the time-dependent network Hamiltonian $\hat{H}_{\text{bMC}}(t)$ with perfectly localized edge modes at $t = 0$ and $t = \tau$.

Due to the unitary equivalence of $\hat{H}_{\text{bMC}}$ and $\hat{H}_{\text{bSSH}}$ (see below), their characteristics regarding state transfer are the same. However, their practical requirements differ as $\hat{H}_{\text{bMC}}$ ($\hat{H}_{\text{bSSH}}$) relies on tunable (fixed) mode frequencies and fixed (tunable) mode couplings. Their local symmetry constraints [cf. (S7)] are also different and may (or may not) be suited for specific implementations.

## II. RELATION OF MAJORANA- AND SSH MODEL

The single-particle theories for the Majorana chain $\hat{H}_{\text{MC}}$ (described above) and the SSH setup $\hat{H}_{\text{SSH}}$ (described in the main text),

$$H_{\text{MC}} = \begin{pmatrix} -\mu_1 & 0 & -w_1 & -\Delta_1 \\ 0 & +\mu_1 & \Delta_1 & w_1 \\ -w_1 & \Delta_1 & \ddots & \ddots \\ -\Delta_1 & w_1 & \ddots & \ddots \end{pmatrix} \quad \text{and} \quad H_{\text{SSH}} = \begin{pmatrix} 0 & w_1 & 0 & 0 \\ w_1 & 0 & t_1 & 0 \\ 0 & t_1 & \ddots & \ddots \\ 0 & 0 & \ddots & \ddots \end{pmatrix}, \tag{S9}$$

can be related by the unitary transformation $M_L = \mathbb{1}_{L\times L} \otimes M_1$, where

$$M_1 = \frac{1}{\sqrt{2}} \begin{pmatrix} 1 & 1 \\ -1 & 1 \end{pmatrix}, \tag{S10}$$

via

$$M_L H_{\text{MC}} M_L^\dagger = \begin{pmatrix} 0 & \mu_1 & 0 & w_1 - \Delta_1 \\ \mu_1 & 0 & w_1 + \Delta_1 & 0 \\ 0 & w_1 + \Delta_1 & \ddots & \ddots \\ w_1 - \Delta_1 & 0 & \ddots & \ddots \end{pmatrix}. \tag{S11}$$

For the special case $\Delta_i = w_i$, one has $H_{\text{SSH}} = M_L H_{\text{MC}} M_L^\dagger$ with the identifications $\mu_i \leftrightarrow w_i$ and $2\Delta_i \leftrightarrow t_i$. In this case, the *bosonic* many-body theories $\hat{H}_{\text{bSSH}}$ and $\hat{H}_{\text{bMC}}$ are unitarily equivalent—which is why we focus on the conceptually simpler $\hat{H}_{\text{bSSH}}$ in this work. In contrast, this unitary equivalence holds only for the single-particle spectrum for the *fermionic* versions $\hat{H}_{\text{SSH}}$ and $\hat{H}_{\text{MC}}$ but not for the many-body theories since the former acts on $2L$ and the latter on $L$ fermionic modes.

Finally, we should mention that in the context of the fermionic Majorana chain, another unitary transformation is commonly used, namely

$$M_1' = \frac{1}{\sqrt{2}} \begin{pmatrix} 1 & 1 \\ -i & i \end{pmatrix} = \begin{pmatrix} 1 & 0 \\ 0 & i \end{pmatrix} \cdot M_1 \tag{S12}$$

which transforms the fermion algebra $\{c_i, c_j^\dagger\} = \delta_{ij}$ via $\gamma_{2i-1} = \frac{c_i + c_i^\dagger}{\sqrt{2}}$ and $\gamma_{2i} = i\frac{c_i - c_i^\dagger}{\sqrt{2}}$ into the eponymous Majorana algebra $\{\gamma_i, \gamma_j\} = \delta_{ij}$ with self-adjoint (Majorana) fermions $\gamma_i^\dagger = \gamma_i$. Then ($\Delta_i = w_i$)

$$M_L' H_{\text{MC}} M_L'^\dagger = \begin{pmatrix} 0 & -i\mu_1 & 0 & 0 \\ i\mu_1 & 0 & i2\Delta_1 & 0 \\ 0 & -i2\Delta_1 & \ddots & \ddots \\ 0 & 0 & \ddots & \ddots \end{pmatrix} \tag{S13}$$

encodes the Majorana chain in terms of "Majorana dimers" $\gamma_i \gamma_{i+1}$ which parallels the SSH chain.

## III. SYMMETRY PROTECTION AND NETWORKS

In the main text, we argued that the envisioned networks, such as the one shown in Fig. 1 (c), require two types of edge modes (dubbed "even" and "odd") and that direct transfer is only possible between pairs of different

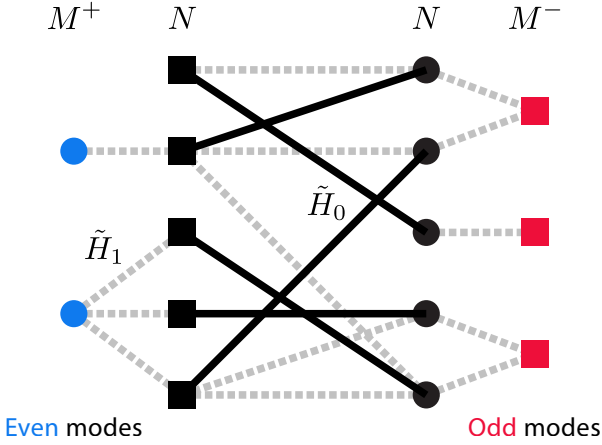

Supplementary Figure S1. *Generic network.* The Hamiltonian $\tilde{H}(t)$ defines a coupling graph with modes as vertices and (complex) weighted edges. The sublattice symmetry corresponds to the bipartiteness of this graph, grouping its vertices into "even" (circles) and "odd" (squares) with only edges between the classes. The couplings of $\tilde{H}_0$ ($\tilde{H}_1$) are indicated as solid black (dashed grey) edges. The $M = M^+ + M^-$ zero-energy edge states do not belong to the dimerized bulk of $\tilde{H}_0$ and are shown as coloured vertices.

types. Here we show the reason for this constraint for the most generic class of networks that allows for topologically protected, localized (zero-dimensional) edge modes.

Therefore we consider systems/protocols of the form

$$\tilde{H}(t) = \bar{t}\,\tilde{H}_0 + \bar{w}(t)\,\tilde{H}_1 \qquad (S14)$$

with $\bar{w}(t) = \bar{w}_{\max} \cdot \mathcal{P}(t)$ where the generic pulse $\mathcal{P}(t)$ is only required to obey $\mathcal{P}(0) = 0 = \mathcal{P}(\tau)$. The parameter $\bar{w}_{\max}$ is unconstrained for the following discussion. Here, $\tilde{H}_0$ encodes an arbitrary, completely dimerized setup of $2N$ (bulk) modes where the tunnelling amplitudes are required to be of order 1 to account for a bulk gap of order $\bar{t}$; apart from this, they may be disordered and/or complex. Now add $M$ *additional* modes which remain uncoupled by $\tilde{H}_0$ (these are the localized zero-energy edge modes). The Hamiltonian $\tilde{H}_1$ may now couple all $2N + M$ modes, again with possibly disordered/complex amplitudes of order 1.

The only generic symmetry we demand is the sublattice symmetry $U_S \tilde{H} U_S^\dagger = -\tilde{H}$ with representation

$$U_S = \mathbb{1}_{N \times N} \otimes \begin{pmatrix} 1 & 0 \\ 0 & -1 \end{pmatrix} \oplus \mathrm{diag}\,(s_1, \ldots, s_M)\,, \qquad (S15)$$

because time-reversal $T = \mathcal{K}$ is broken due to the (possibly) complex couplings. The transformations $s_i = \pm 1$ of the $M$ edge modes are determined by their coupling to the bulk via $\tilde{H}_1$. This places $\tilde{H}(t)$ in symmetry class AIII of the Altland-Zirnbauer classification [S2–S5] and allows for a $\mathbb{Z}$ topological index [S6–S9] in one dimension.

The representation (S15) classifies all $2N + M$ modes into two classes: The modes multiplied by $+1$ ($-1$) when acted upon by $U_S$ will be called "even" ("odd"). The sublattice symmetry suggests the illustrative interpretation of $\tilde{H}$ as adjacency matrix of a graph with modes as vertices and (complex) weighted edges, see Supplementary Fig. S1. Then, $U_S \tilde{H} U_S^\dagger = -\tilde{H}$ is equivalent to the statement that this coupling graph has to be *bipartite*, i.e., only edges between the two classes of "even" and "odd" vertices are allowed.

If we sort the mode basis/vertices into these classes, the single-particle Hamiltonian/adjacency matrix has the generic form

$$\tilde{H} = \begin{pmatrix} 0 & A \\ A^\dagger & 0 \end{pmatrix} \qquad (S16)$$

with complex $(n^+ \times n^-)$-matrix $A$. Then it follows that $\tilde{H}$ has at least $|n^+ - n^-|$ zero eigenvalues. Indeed, squaring $\tilde{H}$ yields

$$\tilde{H}^2 = \begin{pmatrix} AA^\dagger & 0 \\ 0 & A^\dagger A \end{pmatrix}. \qquad (S17)$$

Let w.l.o.g. $n^- \geq n^+$. Then, the $(n^+ \times n^+)$-matrix $AA^\dagger$ has rank $\mathrm{rank}\,(AA^\dagger) \leq n^+$ (which is a trivial bound since $n^- \geq n^+$). However, the $(n^- \times n^-)$-matrix $A^\dagger A$ yields the non-trivial bound $\mathrm{rank}\,(A^\dagger A) \leq n^+ \leq n^-$ due to its composite structure. Therefore $\mathrm{rank}\,(\tilde{H}^2) \leq 2n^+$ and for the corank we find $\mathrm{corank}\,(\tilde{H}^2) \geq n^+ + n^- - 2n^+ = |n^- - n^+|$ and we are done.

Coming back to our setup of $2N$ dimerized bulk modes coupled with $M$ edge modes via $\tilde{H}_1$, we realize that $n^\pm = N + M^\pm$ with $M = M^+ + M^-$, where the sublattice symmetric couplings $\tilde{H}_1$ determine the class of each edge mode (i.e., $s_i = \pm 1$ and thereby $M^\pm$). Then we just showed that $\tilde{H}(t)$ has (at least)

$$|n^+ - n^-| = |M^+ - M^-| \qquad (S18)$$

exact zero energy modes at any time.

We discuss three special cases:

- If $\tilde{H}_1$ connects two edge modes of *opposite* type, we have $|M^+ - M^-| = |1 - 1| = 0$ protected zero-modes. I.e., those modes will generically gap out during $0 < t < \tau$ and facilitate state transfer between them. The considered SSH chain setup is a minimal example of this case.

- If $\tilde{H}_1$ connects two edge modes of the *same* type, we have $|M^+ - M^-| = |2 - 0| = 2$ protected zero-modes. Assuming adiabatically decoupled bulk modes, initial edge excitations are bound to this two-dimensional zero-energy subspace for all times. The only possible unitary acting on this subspace stems from the non-abelian Berry connection $\mathcal{A}_{kl}^{\bar{w}} = i\,\langle \Psi_k(\bar{w})|\,\partial_{\bar{w}}\,|\Psi_l(\bar{w})\rangle$ with $\Psi_l(\bar{w})$ the two zero-energy states ($l = 1, 2$) for coupling $\bar{w}$.

The geometric unitary is then given by (the exponential of) the integral of the Berry connection along the path traced by $\bar{w}(t)$ in parameter space. In our case, the latter is simply connected and one-dimensional so that all loop integrals vanish identically, i.e., there is no holonomic transformation of edge modes possible. We conclude that, despite their delocalization for $0 < t < \tau$, all excitations end up in their initial edge mode and there is no transfer.

- If $\tilde{H}_1$ connects three edge modes, two "even" and one "odd", we have $|M^+ - M^-| = |2 - 1| = 1$ protected zero-mode. For $0 < t < \tau$, the three-dimensional edge manifold gaps out and allows for transfer between the two "even" modes via the "odd" mode. Note that this allows for a "transistor-like" setup where two modes talk to each other only if a third "gate"-mode is present.

It is important to note that the presence/absence of time-reversal symmetry has no effect on the existence of protected edge modes as both symmetry classes BDI and AIII allow for $\mathbb{Z}$ topological invariants in one dimension. The locked transfer phase $\varphi = \pm\pi/2$, however, requires time-reversal symmetry because disordered coupling phases obviously randomize $\varphi$ in the absence of additional symmetries.

As a final remark, note that one could alternatively require $\tilde{H}$ to be PH symmetric, i.e., $U_S \tilde{H}^* U_S^\dagger = -\tilde{H}$, instead of imposing the sublattice symmetry $U_S \tilde{H} U_S^\dagger = -\tilde{H}$. Then, its generic form were

$$\tilde{H}_{\text{PH}} = \begin{pmatrix} iA_+ & B \\ B^T & iA_- \end{pmatrix} \quad (\text{S19})$$

with the antisymmetric real matrices $A_\pm$ and the arbitrary real matrix $B$, cf. equation (S16). I.e., couplings between modes of different (the same) parities must be real (imaginary). There are two reasons why this relaxation is not practical for our purposes:

1. With broken time-reversal symmetry, $\tilde{H}_{\text{PH}}$ belongs to the symmetry class D which only allows for $\mathbb{Z}_2$ topological invariants in one dimension. That is, there are no longer arbitrary numbers of protected, localized edge modes—an essential feature for generalizations such as the 2D network presented in the main text.

2. The coupling structure (S16) is much more natural than the PH symmetric version (S19): In the first case, one has just to ensure that no modes of the same parity couple while arbitrary couplings between modes of different parities are allowed. In contrast, the PH symmetry allows for couplings between arbitrary modes which, however, must satisfy certain (rather unnatural) reality conditions.

Note that the three symmetry classes BDI, D, and AIII exhaust all possibilities to construct topologically protected edge modes in one-dimensional systems with representations of PH and time-reversal that square to unity, i.e., $C^2 = +\mathbb{1}$ and $T^2 = +\mathbb{1}$.

## IV. INFLUENCE OF DISORDER

### A. Additional numerics

In Fig. 6 of the main text we show the spatial amplitudes of the edge modes for the topological and the trivial setup for (a) clean setups, (b) PH symmetric disorder, and (c) PH breaking disorder. Depending on $\bar{w}(t)$, the delocalization of eigenmodes varies between perfectly localized and delocalized. The edge modes reveal a striking difference for PH symmetric disorder: While *any* kind of disorder localizes the "artificial" edge modes of the trivial setup, there is no localization in the topological setup for PH symmetric disorder. This is a distinctive feature of topologically protected edge states in general: Anderson localization is forbidden for the latter as long as the protective symmetries are kept intact (here the PH symmetry). Due to this feature, the topological setups (SSH and Majorana) are outclassing the trivial barrier setup as tunnelling between the edges directly relies on their overlap close to criticality.

To substantiate this claim, we sampled both setups for PH symmetric and breaking disorder and computed their average figures of merit $\langle\langle\mathcal{O}\rangle\rangle$ and $\langle\langle\mathcal{E}\rangle\rangle$. The results for fixed protocols are shown in Supplementary Fig. S2 (a) as a function of the disorder strength $p$. *Fixed* means that for a given system size $L$ and timescale $\tau$, the protocol parameter is tuned such that—in the clean system—the transfer is maximized. This procedure captures effects of slow, uncorrelated drift in the constituent's parameters on the transfer performance if *no* fine tuning of the protocol is performed on a regular basis. The results reveal almost perfect bulk-edge decoupling for the topological setup, irrespective of disorder type and strength. In contrast, the trivial setup shows scattering into the bulk for $p > 0.06$, with only quantitative differences between PH symmetric- and breaking disorder. The differences in transfer $\langle\langle\mathcal{O}\rangle\rangle$ are much more pronounced for the topological than for the trivial setup: For modest PH symmetric disorder, $p < 0.04$, the former still transfers almost the complete population into the right-hand edge mode; in the trivial setup, the transfer drops considerably even for weak disorder. Moreover, there is no qualitative difference between PH breaking and symmetric disorder for the trivial setup.

However, the fundamental difference between the topological and the trivial setup becomes apparent if we allow for the protocol to adapt to each disorder realizations by tuning the timescale $\tau$ in order to maximize the transfer fidelity. The results are shown in Supplementary Fig. S2 (b) [reproduced from the main text for con-

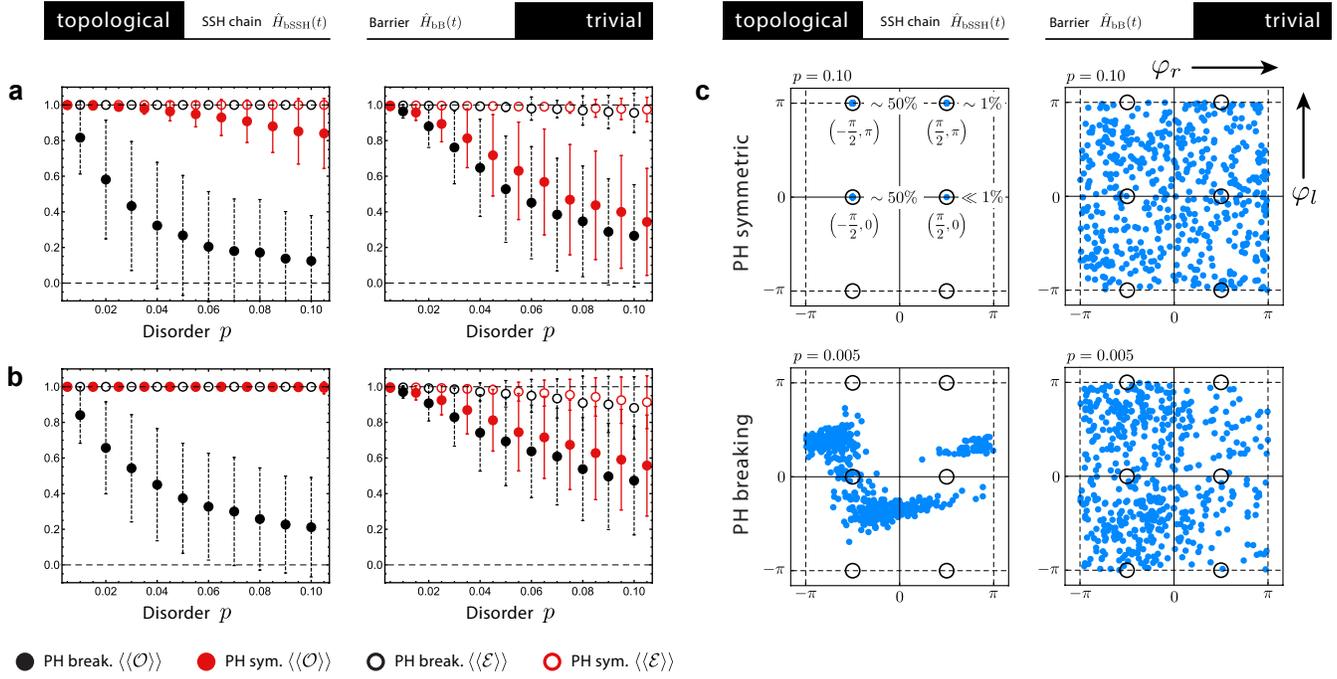

Supplementary Figure S2. *Effects of disorder — Details.* **(a)** Comparison of the transfer $\langle\langle\mathcal{O}\rangle\rangle$ (bullets) and edge weight $\langle\langle\mathcal{E}\rangle\rangle$ (circles) for PH breaking (black) and symmetric (red) disorder. Both, topological and trivial protocol parameters are optimized for clean systems of length $L = 5$ with $\tau = 400$; the latter being fixed for all disorder realizations. The averages are computed from $N = 5000$ samples; the error bars denote one standard deviation of the sample. The left column shows results for the topological SSH chain setup, the right column for the trivial tunnelling approach. **(b)** The same data over $N = 1000$ samples with a retuning of $\tau$ for every single disorder realization to optimize transfer. **(c)** Samples for $N = 500$ disorder realizations of the right-hand edge mode phase $\varphi_r = \varphi$ and its left-hand counterpart $\varphi_l$ for PH symmetric (upper row) and breaking (lower row) disorder for the topological setup (left column) and the trivial setup (right column). The phases are measured in the rotating frame of the left-hand edge mode at $t = 0$. Note that the disorder rate was chosen large ($p = 0.10$) for PH symmetric and small ($p = 0.005$) for PH breaking disorder for illustrative purposes. Details are given in the text.

venience] and reveal a fundamental difference between topological and trivial setup: The already modest losses of transfer for PH symmetric disorder can be cancelled completely by adapting the pulse length $\tau$. This is not possible for PH breaking disorder where the edge mode localization suppresses tunnelling exponentially. Again, the trivial setup does not reveal qualitative differences between PH symmetric and breaking disorder with only minor improvements from the protocol adaption.

To discuss the last figure of merit, the relative transfer phase $\varphi$, it is illustrative to plot a sample (without protocol adaption) for fixed disorder strength $p$ in a 2D scatter plot, see Supplementary Fig. S2 (c), where points encode pairs $(\varphi_r, \varphi_l)$ with $\varphi_r = \varphi$ the phase of the right-hand edge mode and $\varphi_l$ its left-hand counterpart,

$$\langle 1, \mathbf{0}, 0 | U_\tau(\bar{w}_{\max}) | 1, \mathbf{0}, 0 \rangle = \sqrt{\mathcal{E} - \mathcal{O}}\, e^{i\varphi_l}. \qquad (S20)$$

As expected, the trivial setup shows random, uncorrelated phases even for weak disorder, irrespective of whether the PH symmetry is present or not. For PH breaking disorder, the topological setup cannot sustain locked phases either, as shown in the lower left panel of Supplementary Fig. S2 (c) for very weak disorder $p = 0.005$. However, even for strong PH symmetric disorder [$p = 0.1$ in the upper left panel of Supplementary Fig. S2 (c)] the phases are locked for the SSH based transfer at four discrete points: Most of the population is located at $(\varphi_r = -\pi/2, \varphi_l = 0)$ and $(-\pi/2, \pi)$; the number of samples gathering at $(\pi/2, \pi)$, and $(\pi/2, 0)$ is considerably lower, see the exemplary ratios in Supplementary Fig. S2 (c). Note that both $\varphi_r$ and $\varphi_l$ are ill defined whenever their corresponding overlap vanishes; however, due to numerics and the disorder, this actually never happens.

It is easily checked that a single transfer accumulates a phase of $-\pi/2$ if the length $L$ of the chain is odd and $+\pi/2$ if it is even (see also Subsec. V A below); in the following, we discuss the results for the odd $L = 5$ setup used for Supplementary Fig. S2: As a consequence of the still perfect bulk-edge decoupling in the presence of disorder, we can express its effect by simply shifting the protocol parameter $\bar{w}_{\max} + \delta\bar{w}_{\max}$ of the clean system by a realization-dependent value $\delta\bar{w}_{\max}$. If a single, clean, and perfectly tuned transfer leaves $\varphi_l$ undefined due to the vanishing left-hand population, sampling in the vicinity of this parameter yields $\varphi_l = 0$ and $\pi$ with about the

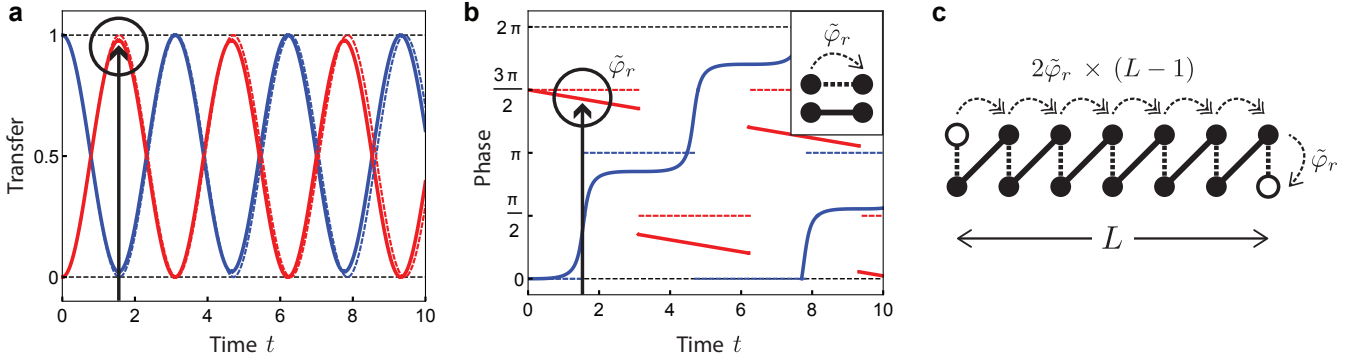

Supplementary Figure S3. *Toy model.* **(a)** Population transfer $\tilde{\mathcal{O}}_l = |\langle l| \tilde{U}_{\text{single}}(t) |l\rangle|^2$ (blue) and $\tilde{\mathcal{O}}_r = |\langle r| \tilde{U}_{\text{single}}(t) |l\rangle|^2$ (red) of a simple Rabi model without detuning $\mu = 0$ (dashed) and with detuning $\mu = 0.3$ (solid) starting in the "left" mode $|l\rangle$ for coupling $A = 1$ as a function of time $t$. The first optimal transfer is slightly reduced and shifted in time for the PH breaking system ($\mu \neq 0$), see circle. **(b)** The corresponding phases $\tilde{\varphi}_l$ (blue) and $\tilde{\varphi}_r$ (red) of the overlaps. Note that for PH symmetric setups $\tilde{\varphi}_r = 3/2\,\pi$ is fixed for the time of first optimal transfer (dashed line in circle). When PH symmetry is broken (here $\mu > 0$), the dynamical phase $e^{-i\frac{\mu t}{2}}$ induces a $t$- and $\mu$-dependent phase shift, $\tilde{\varphi}_r = 3/2\,\pi + \delta\tilde{\varphi}_r(\mu, t)$ (solid line in circle). **(c)** Macroscopic transfers from one edge mode to the other can be thought of as concatenated single transfers, accumulating a total of $(2L-1)$ transfer phases $\tilde{\varphi}_r$. If the latter are kept fixed by PH symmetry, the transfer phase $\varphi_r = L\pi + \pi/2$ is also fixed and only shifts in steps of $\pi$ with $L$.

same probability whereas $\varphi_r = -\pi/2$ is fixed and stable due to the plateau of $\mathcal{O} \approx 1$. This explains the two dominant phase combinations in Supplementary Fig. S2 (c). The rarely sampled combination $(\pi/2, \pi)$ is due to samples with strong disorder where a complete Rabi cycle returns the population to the left edge mode such that $\varphi_r$ jumps by $\pi$. In extremely rare cases, an additional half cycle transfers the population back to the right edge mode which leads to the last phase combination $(\pi/2, 0)$.

For chains of even length, all statements remain true but for the sign flip $\varphi_r \to -\varphi_r$.

In conclusion, detrimental effects of disorder on the transfer phase $\varphi$ are negligible for the topological setup—if the PH symmetry is preserved—due to the rareness of double transfers for weak disorder and the irrelevance of the left-hand phase for a reasonably well tuned complete transfer to the right-hand edge mode.

### B. Symmetry protection

To understand why the particle-hole symmetry is responsible for the fixed transfer phase $\varphi_r = \pm\pi/2$, it is instructive to consider as a toy model a single, local coupling of two adjacent modes $b_l$ and $b_r$ described by the time-reversal invariant Rabi Hamiltonian

$$\tilde{H}_{\text{single}} = \begin{pmatrix} 0 & A \\ A & \mu \end{pmatrix} = A\left(|l\rangle\langle r| + |r\rangle\langle l|\right) + \mu |r\rangle\langle r| \tag{S21}$$

where $A \in \{\bar{w}, \bar{t}\}$ and the local chemical potential ("detuning") $\mu \neq 0$ breaks the PH symmetry $U_C^{\text{SSH}}$ explicitly. The time evolution of (S21) reads

$$\tilde{U}_{\text{single}}(t) = \mathcal{T}\exp\left[-it\tilde{H}_{\text{single}}\right] = \begin{pmatrix} U_{ll} & U_{lr} \\ U_{rl} & U_{rr} \end{pmatrix} \tag{S22}$$

with matrix elements

$$U_{ll} = \langle l| \tilde{U}_{\text{single}}(t) |l\rangle = e^{-i\frac{\mu t}{2}}\left[\cos\left(\frac{\Omega}{2}t\right) + \frac{i\mu}{\Omega}\sin\left(\frac{\Omega}{2}t\right)\right] \equiv \sqrt{\tilde{\mathcal{O}}_l}\, e^{i\tilde{\varphi}_l} \tag{S23a}$$

$$U_{rl} = \langle r| \tilde{U}_{\text{single}}(t) |l\rangle = e^{-i\frac{\mu t}{2}}\left[-\frac{2iA}{\Omega}\sin\left(\frac{\Omega}{2}t\right)\right] \equiv \sqrt{\tilde{\mathcal{O}}_r}\, e^{i\tilde{\varphi}_r} \tag{S23b}$$

and Rabi frequency $\Omega = \sqrt{4A^2 + \mu^2}$ ($U_{lr} = U_{rl}$ and in $U_{rr}$ the relative sign in the sum changes). The overlaps $\tilde{\mathcal{O}}_\alpha$ and phases $\tilde{\varphi}_\alpha$ for $\alpha = l$ (blue) and $\alpha = r$ (red) are plotted in Fig. S3 (a) and (b) for a PH symmetric (dashed) and

breaking (solid) system. In a PH symmetric setup, the phases are discrete, $\tilde{\varphi}_l = \pm\pi$ and $\tilde{\varphi}_r = \pm\pi/2$, and robust against (weak) disorder in the coupling strength $A$ [see also Fig. S2 (c)]. One finds perfect Rabi oscillations,

$$U_C^{\text{SSH}} \tilde{H}_{\text{single}} (U_C^{\text{SSH}})^\dagger = -H_{\text{single}} \quad \Rightarrow \quad \tilde{U}_{\text{single}}(t) = \begin{pmatrix} \cos(At) & -i\sin(At) \\ -i\sin(At) & \cos(At) \end{pmatrix}. \tag{S24}$$

In contrast, PH breaking ($\mu \neq 0$) renders $\tilde{\varphi}_{r,l}$ time- and $\mu$-dependent due to the dynamical phase $e^{-i\frac{\mu t}{2}}$ and, in addition, varies the relative phase $\varphi_r - \varphi_l$ between both edge modes with $\mu$ and $t$, see Eq. (S23a).

It is now illustrative to think of the *macroscopic* transfer from the left to the right edge mode of a length-$L$ chain as a sequence of elementary transfers between adjacent bosonic modes $b_I$, see Fig. S3 (c). The accumulated phase for *PH symmetric* systems is then (modulo $2\pi$)

$$\varphi_r = 2\tilde{\varphi}_r(L-1) + \tilde{\varphi}_r = (2L-1)\tilde{\varphi}_r = \frac{3}{2}(2L-1)\pi = L\pi + \frac{\pi}{2} \tag{S25}$$

which matches the result in Eq. (S31) below. For *PH breaking* disorder, the phases $\tilde{\varphi}_r \to \tilde{\varphi}_I$ for local hopping becomes site dependent because the chemical potential $\mu \to \mu_I$ varies from site to site. Then the total transfer phase $\varphi_r$ is determined by the sum of $2L-1$ random variables (modulo $2\pi$) and the fixed phase $\varphi_r = \pm\pi/2$ is lost, as shown in Fig. S2 (c).

## V. DIAGONALIZATION OF $H_{\text{bSSH}}$

### A. Exact edge modes in the thermodynamic limit

In order to derive the exact edge modes for $L \to \infty$, it is convenient to recast the Hamiltonian $\hat{H}_{\text{bSSH}}$ in the single-particle subspace spanned by $|I\rangle \equiv b_I^\dagger |0\rangle$. Let $w_i \equiv \bar{w}$ and set the remaining couplings $t_i \equiv \bar{t}$ to one. If we shift the edge mode energies to zero (by setting the local mode frequencies $\omega_I \equiv 0$), the single particle Hamiltonian reads

$$H_{\text{bSSH}} = \bar{w} \sum_{i=1}^{L} |i\rangle \langle \bar{i}| + \sum_{i=1}^{L-1} |\bar{i}\rangle \langle i+1| + \text{h.c.}. \tag{S26}$$

Looking sharply at this Hamiltonian (and possibly some numerical results) suggests the following form of the left and right edge modes:

$$|\xi_l\rangle = \mathcal{N} \sum_{i=1}^{L} (-\bar{w})^{i-1} |i\rangle \tag{S27a}$$

$$|\xi_r\rangle = \mathcal{N} \sum_{i=1}^{L} (-\bar{w})^{i-1} |\overline{L-i+1}\rangle \tag{S27b}$$

Here, $\mathcal{N} = \sqrt{(1-\bar{w}^2)/(1-\bar{w}^{2L})}$ is the normalizing factor. These states are motivated by the observation that (1) the mode weight decays exponentially with the distance form the corresponding edge, (2) the local modes contribute with an alternating sign, and (3) only every other local mode carries relevant weight.

All we have to do is to show that $|\xi_{l,r}\rangle$ become degenerate zero energy eigenstates of $H_{\text{bSSH}}$ for $L \to \infty$. This follows by straightforward calculations,

$$H_{\text{bSSH}} |\xi_l\rangle = \bar{w} \mathcal{N} \sum_{i=1}^{L} (-\bar{w})^{i-1} |\bar{i}\rangle + \mathcal{N} \sum_{i=1}^{L-1} (-\bar{w})^{i} |\bar{i}\rangle \tag{S28a}$$

$$= -\mathcal{N} \sum_{i=1}^{L-1} (-\bar{w})^{i} |\bar{i}\rangle + \mathcal{N} \sum_{i=1}^{L-1} (-\bar{w})^{i} |\bar{i}\rangle + \bar{w}\mathcal{N}(-\bar{w})^{L-1} |\overline{L}\rangle \tag{S28b}$$

$$= -(-\bar{w})^{L} \sqrt{\frac{1-\bar{w}^2}{1-\bar{w}^{2L}}} |\overline{L}\rangle \xrightarrow{L \to \infty} 0, \tag{S28c}$$

in the topological phase for $0 \leq \bar{w} < 1$. The same holds for the right edge mode $|\xi_r\rangle$. Obviously $\langle \xi_l | \xi_r \rangle = 0$ for all $L$, but as long as $L < \infty$, $H_{\text{bSSH}}$ couples both states with an overlap exponentially small in $L$,

$$\langle \xi_r | H_{\text{bSSH}} | \xi_l \rangle = -(-\bar{w})^L \frac{1-\bar{w}^2}{1-\bar{w}^{2L}} \tag{S29a}$$

$$\langle \xi_{r,l} | H_{\text{bSSH}} | \xi_{r,l} \rangle = 0, \tag{S29b}$$

so that in the $\{|\xi_l\rangle, |\xi_r\rangle\}$-subspace the Hamiltonian takes the form

$$\tilde{H}_{\text{bSSH}} = -(-\bar{w})^L \frac{1-\bar{w}^2}{1-\bar{w}^{2L}} \begin{pmatrix} 0 & 1 \\ 1 & 0 \end{pmatrix} \tag{S30}$$

so that the degeneracy is lifted by $\Delta E_{\text{edge}} \sim \bar{w}^L (1-\bar{w}^2)/(1-\bar{w}^{2L})$ in this naïve approximation. Below we derive a much more rigorous result for the edge mode splitting in finite systems, but let us first mention that equation (S30) explains the observed even-odd effect of the relative transfer phases: For even chain length $L$, the couplings in $\tilde{H}_{\text{bSSH}}$ are negative, which gives rise to phase of $+\pi/2$ after a complete Rabi cycle,

$$\tilde{U} = \mathcal{T} \exp\left[-i \int_0^\tau \mathrm{d}t\, \mathcal{F}(t) \tilde{H}_{\text{bSSH}}\right] = (-1)^L \begin{pmatrix} 0 & i \\ i & 0 \end{pmatrix} \tag{S31}$$

for $\bar{w}^L(1-\bar{w}^2)/(1-\bar{w}^{2L}) \int_0^\tau \mathrm{d}t\, \mathcal{F}(t) = \pi/2$. Compare this result with Eq. (S25).

### B. Exact diagonalization

> To streamline mathematical expressions, we set $\delta = 0$ ($\Leftrightarrow \omega_I \equiv 0$), $\bar{t} = 1$ and rename $\bar{w}$ to $t$ in Subsections V B, V C and V D (there is no time $t$ involved until Sec. VI).

Here we derive a closed expression for the characteristic polynomial of $H_{\text{bSSH}}$ and use it to find an asymptotically exact expression for the edge mode splitting $\Delta E_{\text{edge}}$ in finite chains of length $L$. The characteristic polynomial $P[H_{\text{bSSH}}](\lambda) = \det(H_{\text{bSSH}} + \lambda \mathbb{1})$ can be calculated recursively via Laplace's formula (recall that we have open boundary conditions and cannot simply diagonalize by Fourier transformation).

To this end, we introduce the class of "even" $2L \times 2L$ matrices

$$H_L^+ \equiv \begin{pmatrix} \lambda & t & & & \\ t & \lambda & 1 & & \\ & & \ddots & & \\ & & 1 & \lambda & t \\ & & & t & \lambda \end{pmatrix} \tag{S32}$$

describing an SSH chain of length $L$, and its "odd" $(2L-1) \times (2L-1)$ descendant

$$H_L^- \equiv \begin{pmatrix} \lambda & 1 & & & \\ 1 & \lambda & t & & \\ & & \ddots & & \\ & & 1 & \lambda & t \\ & & & t & \lambda \end{pmatrix} \tag{S33}$$

which describes our bosonic SSH network of length $L$ where the left mode $b_1$ has been deleted (note that in our bosonic setup this is even physically realizable, though not necessary for the following discussion; in contrast to the fermionic parent theory).

Laplace's formula yields the recursion

$$\det H_L^+ = \lambda \det H_L^- - t^2 \det H_{L-1}^+ \tag{S34a}$$

$$\det H_L^- = \lambda \det H_{L-1}^+ - \det H_{L-1}^-. \tag{S34b}$$

If we insert the second equation in the first, we find

$$\det H_L^+ = (\lambda^2 - t^2) \det H_{L-1}^+ - \lambda \det H_{L-1}^- \tag{S35a}$$

$$\det H_L^- = \lambda \det H_{L-1}^+ - \det H_{L-1}^- \tag{S35b}$$

which is the recursive definition of two coupled polynomial sequences that can be compactly written in vectorial form as

$$\begin{pmatrix} \det H_L^+ \\ \det H_L^- \end{pmatrix} = \begin{pmatrix} \lambda^2 - t^2 & -\lambda \\ \lambda & -1 \end{pmatrix} \begin{pmatrix} \det H_{L-1}^+ \\ \det H_{L-1}^- \end{pmatrix}. \tag{S36}$$

To find an explicit expression for $\det H_L^+$ and $\det H_L^-$, we diagonalize the matrix with the transformation

$$\begin{pmatrix} \chi_L^+ \\ \chi_L^- \end{pmatrix} \equiv \begin{pmatrix} -\frac{\lambda}{\eta_t(\lambda)} & \frac{1}{2}\left[1 + \frac{1+\lambda^2-t^2}{\eta_t(\lambda)}\right] \\ +\frac{\lambda}{\eta_t(\lambda)} & \frac{1}{2}\left[1 - \frac{1+\lambda^2-t^2}{\eta_t(\lambda)}\right] \end{pmatrix} \begin{pmatrix} \det H_L^+ \\ \det H_L^- \end{pmatrix}, \tag{S37}$$

defining new sequences $\chi_L^+$ and $\chi_L^-$ where $\eta_t(\lambda) \equiv \sqrt{[1-(\lambda^2-t^2)]^2 - 4t^2}$. This yields the decoupled recursion

$$\begin{pmatrix} \chi_L^+ \\ \chi_L^- \end{pmatrix} = \begin{pmatrix} P^+ & 0 \\ 0 & P^- \end{pmatrix} \begin{pmatrix} \chi_{L-1}^+ \\ \chi_{L-1}^- \end{pmatrix}, \tag{S38}$$

with eigenvalues $P^\pm = -\frac{1}{2}\left[1 - (\lambda^2 - t^2) \pm \eta_t(\lambda)\right]$, the solutions of which read

$$\chi_L^\pm = (P^\pm)^{L-1} \chi_1^\pm. \tag{S39}$$

To determine the initial values $\chi_1^\pm$, we calculate

$$\det H_1^+ = \det \begin{pmatrix} \lambda & t \\ t & \lambda \end{pmatrix} = \lambda^2 - t^2 \tag{S40a}$$

$$\det H_1^- = \det (\lambda) = \lambda \tag{S40b}$$

and apply the transformation equation (S37). This yields

$$\det \chi_1^\pm = \mp \frac{\lambda}{\eta_t(\lambda)}(\lambda^2 - t^2) + \frac{\lambda}{2}\left[1 \pm \frac{1+\lambda^2-t^2}{\eta_t(\lambda)}\right]. \tag{S41}$$

Finally, the inverse transformation

$$\begin{pmatrix} \det H_L^+ \\ \det H_L^- \end{pmatrix} = \begin{pmatrix} \frac{1+\lambda^2-t^2-\eta_t(\lambda)}{2\lambda} & \frac{1+\lambda^2-t^2+\eta_t(\lambda)}{2\lambda} \\ 1 & 1 \end{pmatrix} \begin{pmatrix} \chi_L^+ \\ \chi_L^- \end{pmatrix} \tag{S42}$$

yields the closed expression

$$P[H_{\text{bSSH}}] = \det H_L^+ = \frac{1+\lambda^2-t^2-\eta_t(\lambda)}{2\lambda} \cdot \chi_L^+ + \frac{1+\lambda^2-t^2+\eta_t(\lambda)}{2\lambda} \cdot \chi_L^- \tag{S43}$$

which can be massaged into the form

$$P[H_{\text{bSSH}}] = \frac{1+\lambda^2-t^2-\eta_t(\lambda)}{\eta_t(\lambda)}\left[1-(\lambda^2-t^2)+\eta_t(\lambda)\right]^L - \frac{1+\lambda^2-t^2+\eta_t(\lambda)}{\eta_t(\lambda)}\left[1-(\lambda^2-t^2)-\eta_t(\lambda)\right]^L. \tag{S44}$$

Note that this indeed is a polynomial in $\lambda$, despite the square root in $\eta_t(\lambda)$ (see below).

The spectrum of the SSH/Majorana chain with open boundary conditions and chain length $L$ is then determined by its roots:

$$\boxed{\frac{1+(\lambda^2-t^2)-\eta_t(\lambda)}{1+(\lambda^2-t^2)+\eta_t(\lambda)} = \left[\frac{1-(\lambda^2-t^2)-\eta_t(\lambda)}{1-(\lambda^2-t^2)+\eta_t(\lambda)}\right]^L} \tag{S45}$$

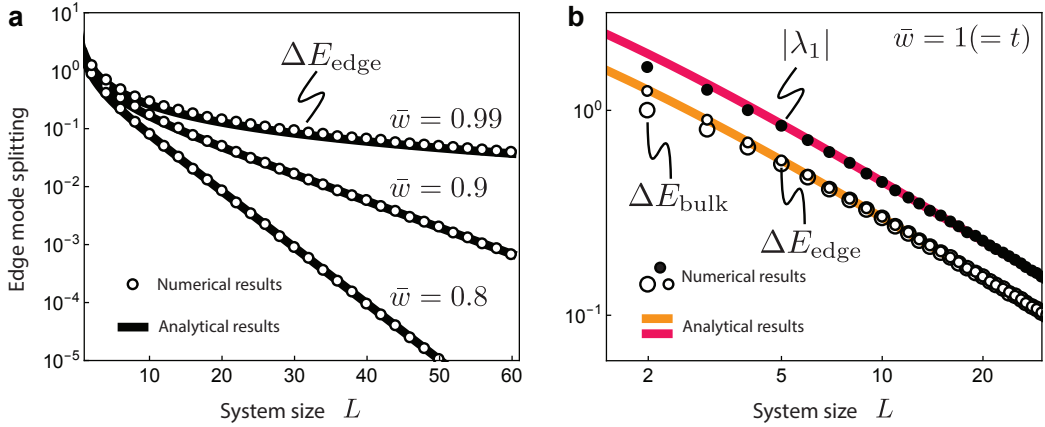

Supplementary Figure S4. *Asymptotic vs. numerical results.* (a) Comparison of numerical results (circles) for the edge mode splitting $\Delta E_{\text{edge}}$ and the analytically derived asymptotically exact expression (black bold lines) equation (S48) in the topological phase for $\bar{w} = t = 0.8, 0.9, 0.99$. Close to the phase transition (large edge mode splitting), small deviations between (exact) numerical and (approximate) analytical results are visible. Note that the observed deviations from the expected exponential decay for small systems ($L \sim 5$) are captured well by our analytical expression. (b) Numerical values for $\Delta E_{\text{edge}}$ (small circles), $\Delta E_{\text{bulk}}$ (large circles) and the asymptotic expressions derived in the text (yellow line) at the phase transition $\bar{w} = 1 = \bar{t}$. Note that the spectrum becomes linear for $L \to \infty$ close to the band crossing and therefore $\Delta E_{\text{edge}} = \Delta E_{\text{bulk}}$ for the asymptotic expressions. In addition, we plot numerical values for $|\lambda_1|$ (black bullets) and compare it with the asymptotic expression. For $L \gtrsim 10$ finite size effects are negligible. Compare the (critical) algebraic decay in (b) with the (gapped) exponential decay in (a).

### C. Asymptotic edge mode splitting

Here we derive an asymptotic expression for the edge mode splitting $\Delta E_{\text{edge}}$ as a function of the coupling strength $t$ in the topological phase, i.e., $0 \leq t < 1$. An application of the binomial theorem allows us to rewrite equation (S44) as the sum

$$P[H_{\text{bSSH}}] = \sum_{n=0}^{\lfloor L/2 \rfloor} \left( \frac{1 + (\lambda^2 - t^2)}{1 - (\lambda^2 - t^2)} \cdot \frac{L - 2n}{2n + 1} - 1 \right) \times \binom{L}{2n} \left[ 1 - (\lambda^2 - t^2) \right]^{L-2n} \times \left\{ \left[ 1 - (\lambda^2 - t^2) \right]^2 - 4t^2 \right\}^n \quad \text{(S46)}$$

where the polynomial nature in $\lambda$ is apparent. Finding the roots of this polynomial, or, equivalently, solving equation (S45), is a non-trivial task as it can be recast as a transcendental equation (see below).

As we are interested in the edge mode energies which are, at least for $L$ large and/or $t$ not to close to unity, nearly zero energy eigenstates, we might assume that these nearly vanishing roots are determined by quadratic terms of $P[H_{\text{bSSH}}]$ up to minor corrections due to a polynomial of degree four. Collecting summands in equation (S46) of degree zero and two, and evaluating the sums yields

$$P[H_{\text{bSSH}}] = -2^{L-1} \left( t^2 + t^{2L} \right) + 2^{L-1} \frac{t^2 + Lt^{2L} - 1 - L}{t^2 - 1} \cdot \lambda^2 + \mathcal{O}(\lambda^4).$$

This can alternatively be found by taylor expanding equation (S44) up to second order at $\lambda = 0$. Solving $P[H_{\text{bSSH}}] = 0$ in this approximation yields the solutions

$$\lambda_\pm = \pm \frac{t^L(t^2 - 1)}{\sqrt{1 - (1 + L)t^{2L} + Lt^{2+2L}}} \quad \text{(S47)}$$

so that the edge mode splitting can be approximated by

$$\Delta E_{\text{edge}} = \frac{2t^L(1 - t^2)}{\sqrt{1 - (1 + L)t^{2L} + Lt^{2+2L}}} \sim 2(1 - t^2) e^{-L/\xi} \quad \text{(S48)}$$

with $\xi = -1/\log t$. As expected, for $L \to \infty$ and/or $t \to 0$, the edge mode splitting vanishes exponentially in $L$: $\Delta E_{\text{edge}} \sim t^L$. Equation (S48) also tells us that the gap closes again for $t \to 1$ where the topological phase transition

occurs (note that in this case the exponential decay is replaced by an algebraic one). Interestingly, equation (S48) also predicts deviations of the exponential decay for small systems that are not too close to the topological transition. These can indeed be verified by numerically diagonalizing $H_{\text{bSSH}}$, as shown in Supplementary Fig. S4 (a).

### D. Analytic spectrum

Here we derive transcendental equations from equation (S45) that implicitly determine the exact spectrum of $H_{\text{(b)SSH}}$ and $H_{\text{(b)MC}}$ (for open boundary conditions) and allow us to find asymptotically exact expressions for the scaling of $\Delta E_{\text{edge}}$ and $\Delta E_{\text{bulk}}$ close to and at the phase transition. It will become graphically transparent how the edge modes emerge or vanish for $t$ crossing the critical value 1.

The discriminating quantity appearing in equation (S45) is $\eta_t(\lambda) = \sqrt{[1-(\lambda^2-t^2)]^2 - 4t^2}$. There are two qualitatively different regimes for $\lambda$: The one for which $\eta_t$ is purely imaginary and the one for which it is real. $\eta_t$ is purely imaginary if ($t \geq 0$)

$$[1-(\lambda^2-t^2)]^2 - 4t^2 \leq 0 \qquad \Leftrightarrow \qquad |1+t^2-\lambda^2| \leq 2t. \tag{S49}$$

In the first case, we have $1+t^2-\lambda^2 \geq 0 \Leftrightarrow |\lambda| \leq \sqrt{1+t^2}$ and

$$1+t^2-\lambda^2 \leq 2t \qquad \Leftrightarrow \qquad |\lambda| \geq \sqrt{(1-t)^2} = |1-t|, \tag{S50}$$

and in the second case, we have $1+t^2-\lambda^2 < 0 \Leftrightarrow |\lambda| > \sqrt{1+t^2}$ and

$$-1-t^2+\lambda^2 \leq 2t \qquad \Leftrightarrow \qquad |\lambda| \leq \sqrt{(1+t)^2} = |1+t|. \tag{S51}$$

Combining both results yields

$$|1-t| \leq |\lambda| \leq |1+t| \tag{S52}$$

with $\eta_t \equiv iy$ and $y \equiv \sqrt{4t^2 - [1-(\lambda^2-t^2)]^2} \in \mathbb{R}$. In the complementary region,

$$|\lambda| < |1-t| \qquad \text{or} \qquad |\lambda| > |1+t| \tag{S53}$$

we have $\eta_t \in \mathbb{R}$.

Let us have a look at equation (S45) for the two cases separately:

- For $|1-t| \leq |\lambda| \leq |1+t|$, equation (S45) takes the form

$$\frac{x_1 - iy}{x_1 + iy} = \left[\frac{x_2 - iy}{x_2 + iy}\right]^L \qquad \Leftrightarrow \qquad \frac{z_1^*}{z_1} = \left[\frac{z_2^*}{z_2}\right]^L \tag{S54}$$

where we introduced $x_1 \equiv 1 - t^2 + \lambda^2$ and $x_2 \equiv 1 + t^2 - \lambda^2$. With $z_k \equiv x_k + iy \equiv |z_k|e^{i\varphi_k}$, this reads

$$e^{-2i\varphi_1} = e^{-2Li\varphi_2} \qquad \Leftrightarrow \qquad \varphi_1 = L\varphi_2 + \pi\mathbb{Z}. \tag{S55}$$

If we define the angular arctangent function $\text{atan}(y/x)$ as

$$\arg(x+iy) = \text{atan}(y/x) \equiv \begin{cases} \arctan\left(\frac{y}{x}\right) & \text{for } x > 0 \\ \arctan\left(\frac{y}{x}\right) + \pi & \text{for } x < 0 \end{cases}, \tag{S56}$$

equation (S45) becomes

$$A(\lambda) \equiv \text{atan}\left[\frac{\sqrt{4t^2 - [1-(\lambda^2-t^2)]^2}}{1-t^2+\lambda^2}\right] \stackrel{!}{=} L\,\text{atan}\left[\frac{\sqrt{4t^2 - [1-(\lambda^2-t^2)]^2}}{1+t^2-\lambda^2}\right] + \pi\mathbb{Z} \equiv \mathbb{B}(\lambda) \tag{S57}$$

which is valid for $|1-t| \leq |\lambda| \leq |1+t|$ and implicitly determines (almost, see below) the complete spectrum of $H_{\text{bSSH}}$.

13

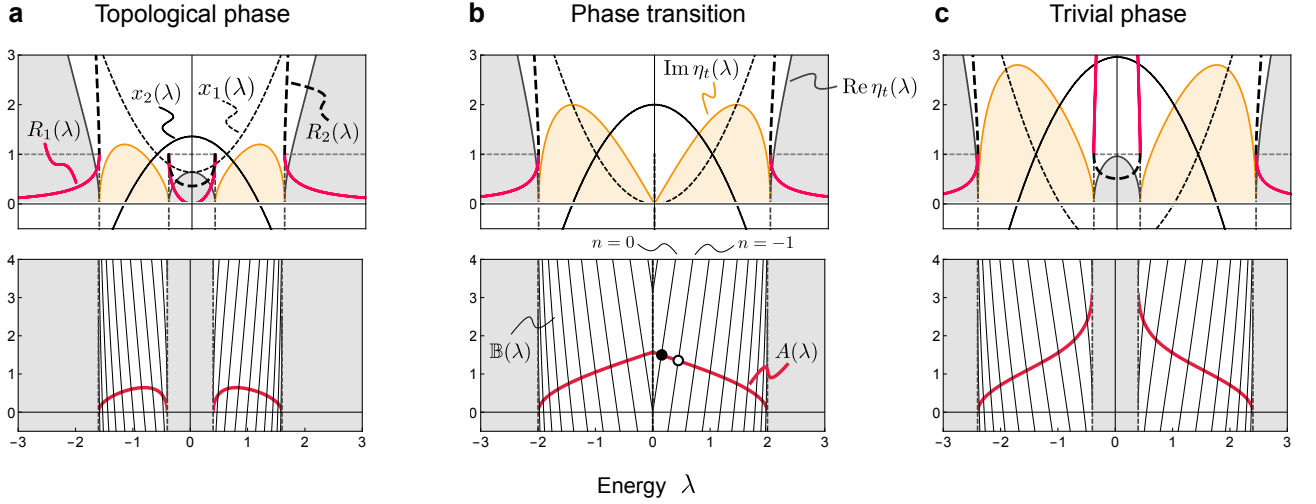

Supplementary Figure S5. *Analytical spectrum — Bulk modes.* Illustration of the analytical quantities defined in the text (Subsec. V D) for couplings **(a)** $t = 0.6$ (topological phase), **(b)** $t = 1.0$ (phase transition), and **(c)** $t = 1.4$ (trivial phase). In the upper row, the quantities $\operatorname{Im}\eta_t(\lambda)$, $\operatorname{Re}\eta_t(\lambda)$, $x_1(\lambda)$, $x_2(\lambda)$, $R_1(\lambda)$, and $R_2(\lambda)$ are plotted. Note that $R_2 > 1$ and $R_1 < 1$ above and below the bands ($|\lambda| > |1+t|$) whereas **(a)** $R_1 < 1$ and $R_2 < 1$ in the band gap ($|\lambda| < |1-t|$) in the topological phase for $t < 1$ but **(c)** $R_1 > 1$ and $R_2 < 1$ in the trivial phase for $t > 1$. In the lower row the quantities $A(\lambda)$ and $\mathbb{B}(\lambda)$ are shown. Each intersection of $A(\lambda)$ with one of the branches of $\mathbb{B}(\lambda)$ corresponds to an eigenenergy of a bulk mode.

- For $|\lambda| < |1-t|$ or $|\lambda| > |1+t|$, both sides of equation (S45) are real which motivates the definition

$$R_k \equiv \frac{x_k - \eta_t}{x_k + \eta_t} \tag{S58}$$

so that equation (S45) takes the simple form

$$R_1 = R_2^L. \tag{S59}$$

The solvability of this equation (for some $L$) is determined by the modulus of $R_k$:

$$|R_k| \lessgtr 1 \quad \Leftrightarrow \quad |x_k - \eta_t| \lessgtr |x_k + \eta_t| \quad \Leftarrow \quad x_k \gtrless 0 \tag{S60}$$

Here we used that $\eta_t \geq 0$ for all allowed $\lambda$. This boils down to the conditions

$$|R_1| \lessgtr 1 \quad \Leftrightarrow \quad 1 - t^2 + \lambda^2 \gtrless 0 \quad \Leftrightarrow \quad \lambda^2 \gtrless t^2 - 1 \tag{S61a}$$

$$|R_2| \lessgtr 1 \quad \Leftrightarrow \quad 1 + t^2 - \lambda^2 \gtrless 0 \quad \Leftrightarrow \quad |\lambda| \lessgtr \sqrt{1+t^2}. \tag{S61b}$$

These ranges must be combined with the allowed intervals for $\lambda$.

First, let $|\lambda| > |1+t|$. Clearly $|1+t| \geq \sqrt{1+t^2}$, so we have $|\lambda| > \sqrt{1+t^2}$ and therefore $|R_2| > 1$. Furthermore $\lambda^2 > (1+t)^2 \geq t^2 - 1$ and we find $|R_1| < 1$. Combined this reads for $|\lambda| > |1+t|$:

$$|R_1| < 1 < |R_2| \leq |R_2|^L \quad \Rightarrow \quad \forall_L : R_1 \neq R_2^L \tag{S62}$$

We conclude that there are no additional solutions for $|\lambda| > |1+t|$, independent of $t$.

The other allowed interval $|\lambda| < |1-t|$ is more interesting. Clearly $\sqrt{1+t^2} \geq |1-t|$, so we have $|\lambda| < |1-t| \leq \sqrt{1+t^2}$ which leads to $|R_2| < 1$. Now comes the crucial step: In the trivial phase we have $t > 1$ which allows us to estimate $(1-t)^2 = t^2 - 1 + 2(1-t) \leq t^2 - 1$ and thereby $\lambda^2 < t^2 - 1$ which yields $|R_1| > 1$. Following the same argument as above, it follows that there are no additional solutions in the range $|\lambda| < |1-t|$. Then all solutions are determined by equation (S57) and we identify the intervals $|1-t| \leq |\lambda| \leq |1+t|$ with the PH-symmetric energy bands gapped by $2|1-t|$. As we just proved, there exist no states in this gap.

In the topological phase for $t < 1$, it follows trivially $\lambda^2 \geq 0 > t^2 - 1$ and therefore $|R_1| < 1$. Note that the statement $|R_2| < 1$ remains valid since it does not depend on $t$. This opens the possibility for non-trivial

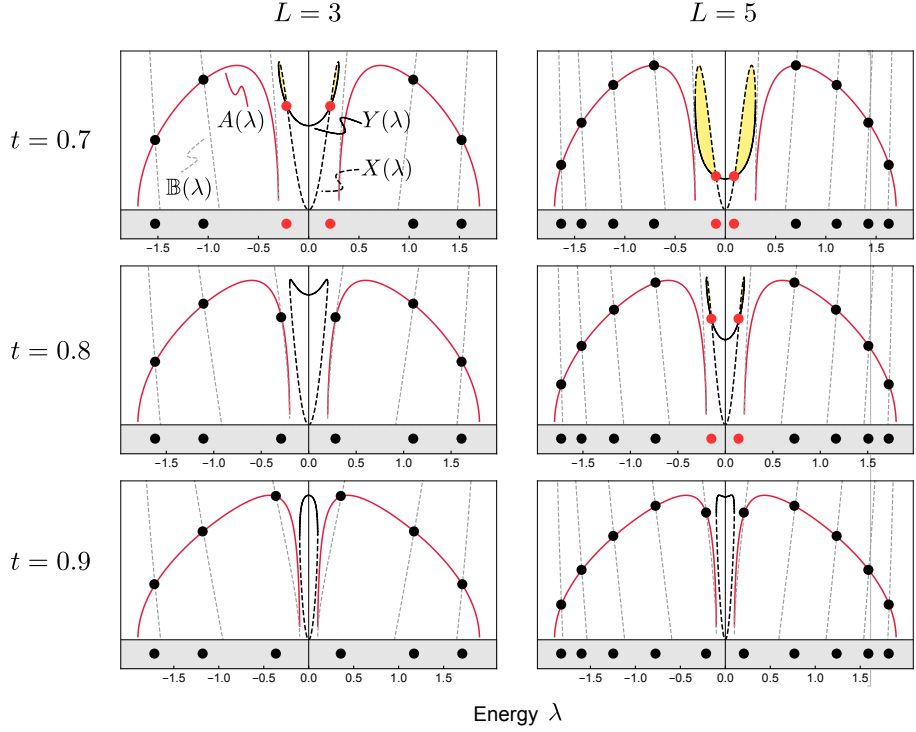

Supplementary Figure S6. *Analytical spectrum — Edge modes.* We plot $A(\lambda)$, $\mathbb{B}(\lambda)$, $X(\lambda)$, and $Y(\lambda)$ for three topological couplings $t = 0.7, 0.8, 0.9$ and two chain lengths $L = 3, 5$. Intersections of $A(\lambda)$ and $\mathbb{B}(\lambda)$ (black bullets) represent bulk eigenmodes. Intersections of $X(\lambda)$ and $Y(\lambda)$ (red bullets) correspond to edge modes. All intersections are projected onto the energy axis to illustrate the spectrum. Note that at $t = 0.8$ there are only bulk modes for the $L = 3$ setup while the $L = 5$ setup already "relabelled" two of the bulk modes as edge modes.

solutions of equation (S59). It is easy to see that at the lower band edge, $|\lambda| = |1 - t|$, one has $\eta_t(\lambda) = 0$, hence $R_1 = 1 = R_2$. Furthermore $R_1(\lambda = 0) \propto 1 - t^2 - (1 - t^2) = 0$ but $R_2(\lambda = 0) \propto 1 + t^2 - (1 - t^2) \neq 0$. Together with $R_1, R_2 \geq 0$, this guarantees at least one pair $\pm\lambda_0$ of additional solutions for $L$ large enough since $\lim_{L \to \infty} R_2^L(\lambda) = 0$ for $|\lambda| < |1 - t|$. These solutions, of course, are lost by equation (S57) to make up for the fixed dimension of the Hilbert space. Monotonicity arguments show that there is indeed just a single pair of additional solutions $\pm\lambda_0$ for $t < 0$ — these are the edge states. Their energy is determined by the equation

$$X(\lambda) \equiv \left[1 + \lambda^2 - t^2 - \eta_t(\lambda)\right]\left[1 - \lambda^2 + t^2 + \eta_t(\lambda)\right]^L \stackrel{!}{=} \left[1 + \lambda^2 - t^2 + \eta_t(\lambda)\right]\left[1 - \lambda^2 + t^2 - \eta_t(\lambda)\right]^L \equiv Y(\lambda) \quad \text{(S63)}$$

which is solvable for $|\lambda| < |1 - t|$ with $0 \leq t < 1$ by the above arguments for $L$ large enough. Interestingly, the critical coupling $t^*$ for which the edge mode solutions appear depends on the chain length $L$ and one finds $t^*(L) < t_{\text{crit}} = 1$ with $\lim_{L \to \infty} t^*(L) = t_{\text{crit}}$, see Supplementary Fig. S6.

We illustrate the relevant quantities of this discussion in Supplementary Fig. S5 for three parameters $t$ below, at, and above the critical value. In Supplementary Fig. S6 we show the size-dependent emergence of the edge mode solutions in the topological phase for three different couplings and two system sizes.

### E. Scaling at the critical point

> To simplify calculations, we set $\delta = 0$ ($\Leftrightarrow \omega_I \equiv 0$) and $\bar{t} = 1$. However, we revert the notation from the technical previous parts, $t \to \bar{w}$, in Subsections V E, V F and V G.

Let $\Delta \bar{w} = \bar{t} - \bar{w}$. Here we consider a chain at criticality, $\Delta \bar{w} = 0$, and set $\bar{t} = 1$. Then, equation (S57) reads

$$\operatorname{atan}\left[\frac{\sqrt{4-[2-\lambda^2]^2}}{\lambda^2}\right] \stackrel{!}{=} L \operatorname{atan}\left[\frac{\sqrt{4-[2-\lambda^2]^2}}{2-\lambda^2}\right] + n\pi \qquad \text{for} \qquad n \in \mathbb{Z} \tag{S64}$$

which is valid for all $0 < |\lambda| < 2$ since $|\bar{w} - 1| = 0$ and $|1 + \bar{w}| = 2$. Inspection shows [see Supplementary Fig. S5 (b)] that the solutions $\pm \lambda_0$ of minimal absolute value (the ones which eventually become edge states for $\Delta \bar{w} > 0$) can be found for $n = 0$ while the next pair of eigenvalues $\pm \lambda_1$ (evolving into the lower band edge for $\Delta \bar{w} > 0$) is specified by $n = -1$; there are only solutions for $-L < n \le 0$.

To evaluate $\Delta E_{\text{edge}} = 2|\lambda_0|$ and $\Delta E_{\text{bulk}} = |\lambda_1 - \lambda_0|$, we have to solve equation (S64) for $n = 0, -1$. As we already know that the gap closes, we can expect $\lambda_0, \lambda_1 \to 0$ for $L \to \infty$. Thus we may expand equation (S64) into linear order of $\lambda$ to find asymptotically exact expressions for both eigenenergies:

$$\frac{\pi}{2} - \frac{\lambda}{2} + \mathcal{O}(\lambda^3) = L\lambda + \mathcal{O}(\lambda^3) + n\pi \qquad \text{for} \qquad n \in \mathbb{Z} \tag{S65}$$

We have

$$\lambda_0 \sim \frac{\pi}{2L+1} \sim \frac{\pi}{2L} \tag{S66a}$$

$$\lambda_1 \sim \frac{3\pi}{2L+1} \sim \frac{3\pi}{2L} \tag{S66b}$$

for $L \to \infty$, and therefore for the time scales

$$\boxed{\Delta E_{\text{edge}}^{-1} \sim \frac{2L+1}{2\pi} \sim \frac{L}{\pi} \qquad \text{and} \qquad \Delta E_{\text{bulk}}^{-1} \sim \frac{2L+1}{2\pi} \sim \frac{L}{\pi}.} \tag{S67}$$

We compare these expressions with numerical results in Supplementary Fig. S4 (b). There it becomes apparent that "$L \to \infty$" can roughly be read as $L \gtrsim 10$.

### F. Scaling away from the critical point

Here we consider the case $|\Delta \bar{w}| > 0$. Recall that the lower band edge is $|\Delta \bar{w}|$ which motivates the new energy variable $\delta = \lambda - |\Delta \bar{w}|$ to measure the distance of bulk modes ($\lambda > |\Delta \bar{w}|$) from the lower band edge.

If we define the LHS argument

$$\alpha_L \equiv \frac{\sqrt{\delta(\delta + 2|\Delta \bar{w}|)[4(1 - \Delta \bar{w}) - \delta(\delta + 2|\Delta \bar{w}|)]}}{2\Delta \bar{w} + \delta(\delta + 2|\Delta \bar{w}|)} \tag{S68}$$

and the RHS argument

$$\alpha_R \equiv \frac{\sqrt{\delta(\delta + 2|\Delta \bar{w}|)[4(1 - \Delta \bar{w}) - \delta(\delta + 2|\Delta \bar{w}|)]}}{2(1 - \Delta \bar{w}) - \delta(\delta + 2|\Delta \bar{w}|)}, \tag{S69}$$

equation (S57) takes the form

$$\operatorname{atan} \alpha_L = L \operatorname{atan} \alpha_R + \pi \mathbb{Z}. \tag{S70}$$

There is a singularity for $\Delta \bar{w} \to \pm 0$ and $\delta \to 0$ in the sense that the left-hand side is discontinuous

$$\lim_{\delta \to 0} \operatorname{atan} \alpha_L = \begin{cases} 0 & \text{for} \quad \Delta \bar{w} > 0 \text{ (topological)} \\ \frac{\pi}{2} & \text{for} \quad \Delta \bar{w} = 0 \text{ (critical)} \\ \pi & \text{for} \quad \Delta \bar{w} < 0 \text{ (trivial)} \end{cases} \tag{S71}$$

induced by the true singularity of the argument

$$\lim_{\delta \to 0} \frac{\sqrt{\delta(\delta + 2|\Delta \bar{w}|)[4(1 - \Delta \bar{w}) - \delta(\delta + 2|\Delta \bar{w}|)]}}{2\Delta \bar{w} + \delta(\delta + 2|\Delta \bar{w}|)} = \begin{cases} +0 & \text{for} \quad \Delta \bar{w} > 0 \text{ (topological)} \\ \infty & \text{for} \quad \Delta \bar{w} = 0 \text{ (critical)} \\ -0 & \text{for} \quad \Delta \bar{w} < 0 \text{ (trivial)} \end{cases}. \tag{S72}$$

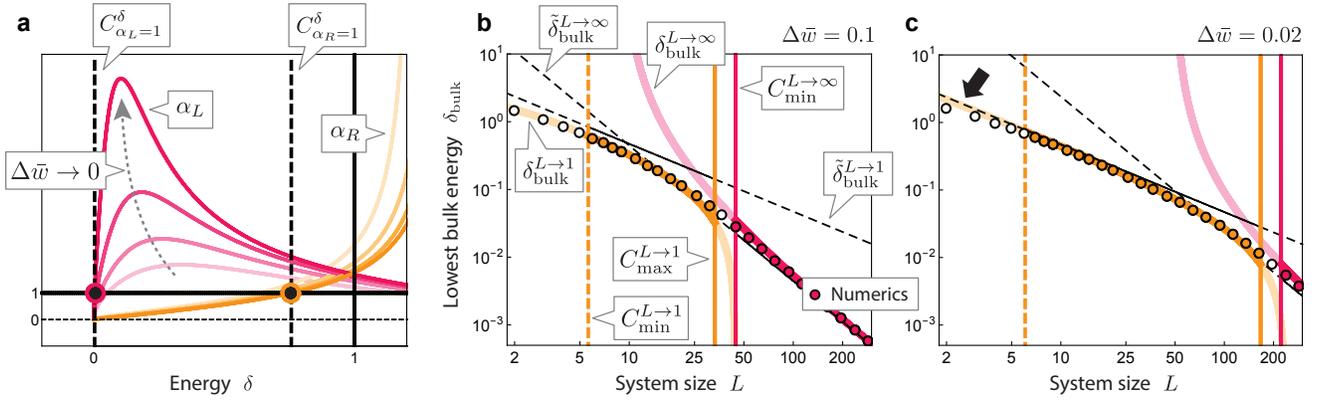

Supplementary Figure S7. *Scaling away from the critical point.* **(a)** We plot the left-hand (right-hand) arguments $\alpha_L$ ($\alpha_R$) defined in equation (S68) and equation (S69) as functions of the energy $\delta = \lambda - |\Delta\bar{w}|$ for couplings $\Delta\bar{w} \to 0$ in the topological phase close to the phase transition. Note that $\alpha_L$ diverges close to the lower band edge $|\Delta\bar{w}|$ ($\delta = 0$) for $\Delta\bar{w} \to 0$ whereas $\alpha_R$ remains finite. **(b)** Lowest bulk energy $\delta_{\text{bulk}} = \lambda_1 - |\Delta\bar{w}|$ in the topological phase for $\Delta\bar{w} = 0.1$ vs. chain length $L$. We compare numerical results (circles) with the analytical ones (solid lines) derived in the text. The latter are valid in their respective $L$-intervals (vertical lines). In the left (yellow) interval, the $1/L$-decay dominates whereas in the right (red) interval the faster $1/L^2$-decay takes over. **(c)** The same for $\Delta\bar{w} = 0.02$, closer to the critical point. Note how the $1/L$-decay dominates for a larger $L$-range. Marked by the arrow, finite-size deviations from the approximate expression $\delta_{\text{bulk}}^{L \to 1}$ become visible; as expected for $L \lesssim C_{\min}^{L \to 1}$.

In contrast, the right-hand side is non-singular,

$$\lim_{\delta \to 0} \operatorname{atan} \alpha_R = 0 \quad \text{for all} \quad \Delta\bar{w}. \tag{S73}$$

Here we used the definition in equation (S56). See Supplementary Fig. S5 [$A(\lambda)$, red curves in the lower panels] for an illustration of these statements. This behaviour is responsible for both the emergence of edge modes for $\Delta\bar{w} > 0$ and the replacement of the asymptotic $1/L$ decay of the bulk modes at $\Delta\bar{w} = 0$ by an $1/L^2$ decay towards the lower band edge $|\Delta\bar{w}|$ at $\Delta\bar{w} \neq 0$.

The discontinuity of the LHS in equation (S57) forbids a consistent expansion for $\delta \to 0$ and $\Delta\bar{w} \to 0$ at the same time. Therefore we consider two cases to simplify equation (S70) separately:

1. $\alpha_L \ll 1$, $\alpha_R \ll 1$, which allows us to solve $\alpha_L (+\pi) = \alpha_R + \pi\mathbb{Z}$ instead of equation (S70).

2. $\alpha_L \gg 1$, $\alpha_R \ll 1$, which allows us to solve $\frac{\pi}{2} = \alpha_R + \pi\mathbb{Z}$ instead of equation (S70).

In addition, we always assume that $\delta \ll 1$ and $|\Delta\bar{w}| \ll 1$, i.e., we consider bulk modes close to the lower band edge and close to criticality. For the simplified equations, we used that the arctangent function can be linearized if we take into account definition (S56) as

$$\operatorname{atan}(y/x) \approx y/x \, (+\pi) \qquad \text{for} \qquad x > 0 \, (x < 0). \tag{S74}$$

### 1. First regime: $\alpha_L \ll 1$

We start with the first case, $\alpha_L \ll 1$, $\alpha_R \ll 1$, which allows to simplify equation (S70) as

$$\alpha_L (+\pi) = \alpha_R + \pi\mathbb{Z} \tag{S75}$$

where the optional $+\pi$ follows in the trivial phase for $\Delta\bar{w} < 0$. We are interested in the bulk modes with lowest energy, i.e., closest to the lower band edge. This yields the two equations

$$\frac{\sqrt{\delta(\delta + 2|\Delta\bar{w}|)[4(1-\Delta\bar{w}) - \delta(\delta + 2|\Delta\bar{w}|)]}}{2\Delta\bar{w} + \delta(\delta + 2|\Delta\bar{w}|)} + \begin{Bmatrix} 0 \\ \pi \end{Bmatrix}$$
$$= L \frac{\sqrt{\delta(\delta + 2|\Delta\bar{w}|)[4(1-\Delta\bar{w}) - \delta(\delta + 2|\Delta\bar{w}|)]}}{2(1-\Delta\bar{w}) - \delta(\delta + 2|\Delta\bar{w}|)} + \begin{Bmatrix} -\pi \\ 0 \end{Bmatrix} \qquad \text{for} \qquad \begin{Bmatrix} \Delta\bar{w} > 0 \\ \Delta\bar{w} < 0 \end{Bmatrix} \tag{S76}$$

which are valid for $\delta \ll \min\{C^\delta_{\alpha_L=1}(\Delta\bar{w}), C^\delta_{\alpha_R=1}(\Delta\bar{w})\}$ with yet to be determined functions $C^\delta_{\alpha_L=1}$ and $C^\delta_{\alpha_R=1}$ that determine the range of validity for the linearization of the LHS and RHS arctangent functions. Note that the scaling of the bulk modes is the same for $\Delta\bar{w} \gtrless 0$. For the choice of the correct value of $\pi\mathbb{Z}$ in equation (S75), see Supplementary Fig. S5.

We have to solve the equation

$$\frac{\pi}{\sqrt{\delta(\delta+2|\Delta\bar{w}|)[4(1-\Delta\bar{w})-\delta(\delta+2|\Delta\bar{w}|)]}} = \frac{L}{2(1-\Delta\bar{w})-\delta(\delta+2|\Delta\bar{w}|)} - \frac{1}{2\Delta\bar{w}+\delta(\delta+2|\Delta\bar{w}|)} \quad (S77)$$

which reduces to

$$\pi\sqrt{1-\Delta\bar{w}}\,\Delta\bar{w} = \sqrt{\delta} \cdot \sqrt{2|\Delta\bar{w}|}(L\Delta\bar{w}+\Delta\bar{w}-1) \quad (S78)$$

in lowest order of $\delta$. Note that this requires the additional condition $\delta \ll 1$ so that our total range of validity reads now $\delta \ll \min\{1, C^\delta_{\alpha_L=1}(\Delta\bar{w}), C^\delta_{\alpha_R=1}(\Delta\bar{w})\}$.

Solving for $\delta$ yields

$$\delta^{L\to\infty}_{\text{bulk}} = \frac{\pi^2}{2}\frac{(1-\Delta\bar{w})|\Delta\bar{w}|}{(L\Delta\bar{w}+\Delta\bar{w}-1)^2} \stackrel{L\to\infty}{\sim} \frac{\pi^2(1-\Delta\bar{w})}{2|\Delta\bar{w}|}\frac{1}{L^2} \equiv \tilde{\delta}^{L\to\infty}_{\text{bulk}} \quad (S79)$$

valid for $0 < |\Delta\bar{w}| < 1$.

To sum it up, in the gapped phases ($|\Delta\bar{w}| > 0$), the *bulk* modes closest to the band edge have the asymptotic energy

$$E^{L\to\infty}_{\text{bulk}} = \Delta\bar{w} + \delta^{L\to\infty}_{\text{bulk}} = |\Delta\bar{w}|\left[1 + \frac{\pi^2/2}{\frac{\Delta\bar{w}^2}{1-\Delta\bar{w}}L^2 - 2\Delta\bar{w}L + (1-\Delta\bar{w})}\right] \quad (S80a)$$

$$\sim |\Delta\bar{w}| + \frac{\pi^2(1-\Delta\bar{w})}{2|\Delta\bar{w}|}\frac{1}{L^2} \quad (L\to\infty). \quad (S80b)$$

Note that due to the exponential decay of the edge mode energy $\Delta E_{\text{edge}}$, one has $E^{L\to\infty}_{\text{bulk}} \approx \Delta E_{\text{bulk}}$ in the topological phase.

We are left with the determination of $C^\delta_{\alpha_{R/L}=1}(\Delta\bar{w})$, i.e., the range of $\delta$ for which these relations are valid. We define $C^\delta_{\alpha_{R/L}=1}(\Delta\bar{w})$ as the smallest positive solutions for $\delta$ where $\alpha_R$ and $\alpha_L$ are equal to one:

- The condition $\alpha_L \stackrel{!}{=} 1$ reduces to

$$\delta(\delta+2|\Delta\bar{w}|)[4(1-\Delta\bar{w})-\delta(\delta+2|\Delta\bar{w}|)] = [2\Delta\bar{w}+\delta(\delta+2|\Delta\bar{w}|)]^2 \quad (S81)$$

which yields the smallest positive solution

$$C^\delta_{\alpha_L=1}(\Delta\bar{w}) = \left[(\Delta\bar{w}-1)^2 - \sqrt{2\Delta\bar{w}^2-4\Delta\bar{w}+1}\right]^{1/2} - |\Delta\bar{w}| \quad (S82a)$$

$$\sim (\sqrt{2}-1)\Delta\bar{w} \quad (S82b)$$

$$\approx 0.4\,\Delta\bar{w} \quad (S82c)$$

where the linear terms follow for $0 < \Delta\bar{w} \ll 1$, i.e., close to criticality. If we take into account the behaviour of $\alpha_L$ for $\delta \to 0$, see Supplementary Fig. S7 (a), we can conclude that for $0 \leq \delta < C^\delta_{\alpha_L=1}(\Delta\bar{w})$ it is $\alpha_L \ll 1$ and the linearization of the LHS arctangent is valid. Contrary, for $C^\delta_{\alpha_L=1}(\Delta\bar{w}) < \delta \ll 1$ we conclude that $\alpha_L \gg 1$ for $\Delta\bar{w} \ll 1$.

- The condition $\alpha_R \stackrel{!}{=} 1$ reduces to

$$\delta(\delta+2|\Delta\bar{w}|)[4(1-\Delta\bar{w})-\delta(\delta+2|\Delta\bar{w}|)] = [2(\Delta\bar{w}-2)+\delta(\delta+2|\Delta\bar{w}|)]^2 \quad (S83)$$

which yields the smallest positive solution

$$C^\delta_{\alpha_R=1}(\Delta\bar{w}) = \left[\Delta\bar{w}^2 + (2-\sqrt{2})(1-\Delta\bar{w})\right]^{1/2} - |\Delta\bar{w}| \quad (S84a)$$

$$\sim \sqrt{2-\sqrt{2}} - \left(1+\frac{1}{2}\sqrt{2-\sqrt{2}}\right)\Delta\bar{w} \quad (S84b)$$

$$\approx 0.8 - 1.4\,\Delta\bar{w} \quad (S84c)$$

where the linear terms follow for $0 < \Delta\bar{w} \ll 1$, i.e., close to criticality. If we take into account the smooth behaviour of $\alpha_R$ for $\delta \to 0$, see Supplementary Fig. S7 (a), we can conclude that for $0 \leq \delta \ll C_{\alpha_R=1}^{\delta}(\Delta\bar{w})$ it is $\alpha_R \ll 1$ and the linearization of the RHS arctangent is valid.

To be self-consistent, we have to plug in our solution (S79) into the upper bounds on $\delta$, equation (S82c), equation (S84c) and the additional constraint $\delta \ll 1$,

$$\frac{\pi^2}{2} \frac{(1-\Delta\bar{w})|\Delta\bar{w}|}{(L\Delta\bar{w} + \Delta\bar{w} - 1)^2} \ll \min\{0.4\,\Delta\bar{w}, 0.8 - 1.4\,\Delta\bar{w}, 1\} = 0.4\,\Delta\bar{w} \tag{S85}$$

for $\Delta\bar{w} \to 0$. We see that the decisive bound is given by the constraint $\alpha_L \ll 1$ for small $\Delta\bar{w}$.

Solving

$$\frac{\pi^2}{2} \frac{(1-\Delta\bar{w})\Delta\bar{w}}{(L\Delta\bar{w} + \Delta\bar{w} - 1)^2} = (\sqrt{2}-1)\,\Delta\bar{w} \tag{S86}$$

leads to the condition on the system size

$$L \gg \frac{\pi\sqrt{\frac{1}{2}+\frac{1}{\sqrt{2}}}+1}{\Delta\bar{w}} - \frac{1}{8}\sqrt{\frac{1}{2}+\frac{1}{\sqrt{2}}}\pi\Delta\bar{w} - \frac{1}{4}\sqrt{2\sqrt{2}+2}\pi - 1 \sim \frac{\pi\sqrt{\frac{1}{2}+\frac{1}{\sqrt{2}}}+1}{\Delta\bar{w}} \sim \frac{4.5}{\Delta\bar{w}} \equiv C_{\min}^{L\to\infty}(\Delta\bar{w}) \tag{S87}$$

valid for $\Delta\bar{w} \to 0$. See Supplementary Fig. S7 (b) and (c) for an illustration of this range. We immediately see that the quadratic decay of bulk modes towards the band edge is a unique feature of the gapped phases (trivial and topological) that sets in for larger system sizes the closer the system is to criticality. At $\Delta\bar{w} = 0$, there is no quadratic decay left and we end up with the $1/L$ decay that we already derived above.

The scaling derived here is asymptotically correct ($L \to \infty$) for small deviations from criticality, $\Delta\bar{w} \ll 1$. As we are anyway interested in chains driven close to $\Delta\bar{w} \approx 0$, the latter is not restrictive. However, it would be interesting to know the scaling that dominates for small chains, $L \sim 1$.

### 2. Second regime: $\alpha_L \gg 1$

To this end, we assume that $C_{\alpha_L=1}^{\delta}(\Delta\bar{w}) < \delta \ll \min\{1, C_{\alpha_R=1}^{\delta}(\Delta\bar{w})\}$ and again $\Delta\bar{w} \ll 1$. This implies that $\alpha_L \gg 1$ and $\alpha_R \ll 1$. This allows us to approximate equation (S70) with $\frac{\pi}{2} = \alpha_R - \pi$ for the lowest bulk modes. Additionally, we use $\delta \ll 1$ to expand the RHS in second order of $\delta$, and finally, we expand in linear order of $\Delta\bar{w}$.

This yields the quadratic equation

$$9\Delta\bar{w}L\,\delta^2 + 4L(2+\Delta\bar{w})\,\delta + 4(2L\Delta\bar{w} - 3\pi) = 0 \tag{S88}$$

with relevant (positive) solution

$$\begin{aligned}\delta_{\text{bulk}}^{L\to 1} &= \frac{1}{9\Delta\bar{w}}\left[-4 - 2\Delta\bar{w} + 2\sqrt{4 - 17\Delta\bar{w}^2 + \Delta\bar{w}\left(4 + \frac{27\pi}{L}\right)}\right] \\ &\sim \frac{3\pi}{2L} - \Delta\bar{w}\left(1 + \frac{3\pi}{4L} + \frac{81\pi^2}{32L^2}\right) \qquad (\Delta\bar{w} \ll 1) \\ &\sim \frac{3\pi}{2L} \equiv \tilde{\delta}_{\text{bulk}}^{L\to 1} \qquad (\Delta\bar{w} \to 0)\,.\end{aligned} \tag{S89}$$

To be consistent, we have to expand this solution once more into linear order of $\Delta\bar{w}$ to find

$$\boxed{E_{\text{bulk}}^{L\to 1} = \Delta\bar{w} + \delta_{\text{bulk}}^{L\to 1} = \frac{3\pi}{2L} - \Delta\bar{w}\left(\frac{3\pi}{4L} + \frac{81\pi^2}{32L^2}\right)\,.} \tag{S90}$$

Compare this with the result (S66b) in the limit $\Delta\bar{w} \to 0$. Note that this limit works because we defused the singularity by setting the LHS to $\frac{\pi}{2}$. In Supplementary Fig. S7 (b) and (c) we show this approximation and its relation to the asymptotic expression derived above.

Self-consistency demands that

$$0.4\,\Delta\bar{w} < \frac{3\pi}{2L} - \Delta\bar{w}\left(1 + \frac{3\pi}{4L} + \frac{81\pi^2}{32L^2}\right) \ll \min\{0.8 - 1.4\,\Delta\bar{w}, 1\} = 0.8 - 1.4\,\Delta\bar{w} \tag{S91}$$

where we used equation (S84c) and equation (S82c). The left-hand inequality can be solved via the solution of

$$\frac{3\pi}{2L} - \Delta\bar{w}\left(1 + \frac{3\pi}{4L} + \frac{81\pi^2}{32L^2}\right) = (\sqrt{2} - 1)\,\Delta\bar{w} \tag{S92}$$

which reads

$$L < -\frac{27}{16}\pi\Delta\bar{w} + \frac{3\pi}{2\sqrt{2}\Delta\bar{w}} - \frac{3\pi}{4\sqrt{2}} \sim \frac{3\pi}{2\sqrt{2}\Delta\bar{w}} \sim \frac{3.3}{\Delta\bar{w}} \equiv C_{\max}^{L\to 1}(\Delta\bar{w}). \tag{S93}$$

The right-hand inequality can be solved via the solution of

$$\frac{3\pi}{2L} - \Delta\bar{w}\left(1 + \frac{3\pi}{4L} + \frac{81\pi^2}{32L^2}\right) = \sqrt{2 - \sqrt{2}} - \left(1 + \frac{1}{2}\sqrt{2 - \sqrt{2}}\right)\Delta\bar{w} \tag{S94}$$

which reads

$$L \gg \frac{3\pi}{4\sqrt{2-\sqrt{2}}}\left[\sqrt{\frac{9\sqrt{2-\sqrt{2}}\Delta\bar{w} + \Delta\bar{w} - 2}{\Delta\bar{w} - 2}} + 1\right] \sim \frac{3\pi}{2\sqrt{2-\sqrt{2}}} - \frac{27\pi\Delta\bar{w}}{16} \sim 6.2 - 5.3\,\Delta\bar{w} \equiv C_{\min}^{L\to 1}(\Delta\bar{w}). \tag{S95}$$

In Supplementary Fig. S7 (b) and (c) we illustrate the interval bounded from below by equation (S95) and from above by equation (S93). Note that the upper bound diverges for $\Delta\bar{w} \to 0$ as the $1/L$ decay towards $|\Delta\bar{w}|$ takes over from the $1/L^2$ decay. In the limit $\Delta\bar{w} \to 0$, we find the already known result

$$\lim_{\Delta\bar{w}\to 0} E_{\text{bulk}}^{L\to 1} = \tilde{\delta}_{\text{bulk}}^{L\to 1} = \frac{3\pi}{2L} \tag{S96}$$

with a validity range of $6 \lesssim L$. Indeed, for small chains of length $L \lesssim 6$ one finds finite-size deviations from the exact $1/L$ decay predicted for a continuous system, as can be seen in Supplementary Fig. S7 (c), which is not even captured by our more sophisticated expression in equation (S90).

### G. Universal scaling

Here we derive the universal scaling of the eigenenergies in the thermodynamic limit exactly. Recall that for the purpose of state transfer, we have to tune the system closer to the critical point ($\Delta\bar{w} \to 0$) for $L \to \infty$ to allow for optimal scaling of the transfer time ($\tau \sim L$) which requires $\Delta E_{\text{edge}} \sim 1/L$ for the edge mode splitting. In the following, we make these statements rigorous. To this end, we introduce the rescaled variables $\lambda' \equiv L\lambda$ for energies and $\Delta\bar{w}' \equiv L\Delta\bar{w}$ for couplings; we are interested in the $\Delta\bar{w}'$-dependence of $\Delta E'_{\text{bulk}} = |\lambda'_1 - \lambda'_0|$ and $\Delta E'_{\text{edge}} = 2\lambda'_0$ in the thermodynamic limit $L \to \infty$.

We start by rewriting equation (S45) in terms of the new variables,

$$\frac{x'_1 - \eta'_t}{x'_1 + \eta'_t} = \left[\frac{x'_2 - \eta'_t}{x'_2 + \eta'_t}\right]^L \tag{S97}$$

where

$$x'_{1/2} = 1 \pm \frac{1}{L^2}\left[\lambda'^2 - (L - \Delta\bar{w}')^2\right] \tag{S98}$$

and

$$\eta'_t = \frac{1}{L^2}\sqrt{[\Delta\bar{w}'^2 - \lambda'^2][(\Delta\bar{w}' - 2L)^2 - \lambda'^2]}. \tag{S99}$$

Note that we did not introduce the relative energies $\delta' = \lambda' - |\Delta \bar{w}|'$ since we are also interested in the edge mode which lives in the band gap. We can now take the limit $L \to \infty$ of both sides in equation (S97) to find the transcendental equation

$$\boxed{\frac{\Delta \bar{w}' - \sqrt{\Delta \bar{w}'^2 - \lambda'^2}}{\Delta \bar{w}' + \sqrt{\Delta \bar{w}'^2 - \lambda'^2}} = e^{-2\sqrt{\Delta \bar{w}'^2 - \lambda'^2}}.} \tag{S100}$$

Here we used that $\lim_{L \to \infty} L\, x_1' = 2\Delta \bar{w}'$ and

$$\lim_{L \to \infty} L\, \eta_t' = 2\sqrt{\Delta \bar{w}'^2 - \lambda'^2} \tag{S101}$$

which leads to

$$\lim_{L \to \infty} \frac{x_1' - \eta_t'}{x_1' + \eta_t'} = \frac{\Delta \bar{w}' - \sqrt{\Delta \bar{w}'^2 - \lambda'^2}}{\Delta \bar{w}' + \sqrt{\Delta \bar{w}'^2 - \lambda'^2}}. \tag{S102}$$

For the RHS we used the well-known relation $\lim_{L \to \infty} (1 + x/L)^L = \exp(x)$ to derive

$$\lim_{L \to \infty} \left[\frac{x_2' \pm \eta_t'}{2}\right]^L = e^{\pm \sqrt{\Delta \bar{w}'^2 - \lambda'^2} - \Delta \bar{w}'} \tag{S103}$$

and therefore

$$\lim_{L \to \infty} \left[\frac{x_2' - \eta_t'}{x_2' + \eta_t'}\right]^L = e^{-2\sqrt{\Delta \bar{w}'^2 - \lambda'^2}}. \tag{S104}$$

As already mentioned previously, there are two types of solutions for equation (S100): For $|\lambda|' < |\Delta \bar{w}|'$, the above equation defines solutions within the gap (which is bounded by $|\Delta \bar{w}|'$) in the thermodynamic limit and is well-defined in the field of real numbers; this implies the exponential decay of the edge mode splitting. In contrast, for $|\lambda|' > |\Delta \bar{w}|'$ (in-band), the equation becomes complex-valued

$$\frac{\Delta \bar{w}' - i\sqrt{\lambda'^2 - \Delta \bar{w}'^2}}{\Delta \bar{w}' + i\sqrt{\lambda'^2 - \Delta \bar{w}'^2}} = e^{-2i\sqrt{\lambda'^2 - \Delta \bar{w}'^2}} \tag{S105}$$

which can be recast as the transcendental equation over the field of real numbers

$$\boxed{\operatorname{atan}\left[\frac{\sqrt{\lambda'^2 - \Delta \bar{w}'^2}}{\Delta \bar{w}'}\right] = \sqrt{\lambda'^2 - \Delta \bar{w}'^2} + \pi \mathbb{Z}} \tag{S106}$$

encoding an algebraic decay of bulk-modes towards the band edge $|\Delta \bar{w}|$.

Inspection shows that the transmutation of "complex" bulk solutions to "real" edge solutions occurs at $\Delta \bar{w}' = 1$, see Supplementary Fig. S8 (a). Formally, the energy $\lambda_0'$ of the edge mode is determined by

$$\frac{\Delta \bar{w}' - \sqrt{\Delta \bar{w}'^2 - \lambda_{0,e}'^2}}{\Delta \bar{w}' + \sqrt{\Delta \bar{w}'^2 - \lambda_{0,e}'^2}} = e^{-2\sqrt{\Delta \bar{w}'^2 - \lambda_{0,e}'^2}} \qquad \text{for} \qquad 1 < \Delta \bar{w}' \tag{S107}$$

and

$$\operatorname{atan}\left[\frac{\sqrt{\lambda_{0,b}'^2 - \Delta \bar{w}'^2}}{\Delta \bar{w}'}\right] = \sqrt{\lambda_{0,b}'^2 - \Delta \bar{w}'^2} \qquad \text{for} \qquad 0 \leq \Delta \bar{w}' \leq 1. \tag{S108}$$

In contrast, the energy of the lowest bulk mode is determined by

$$\operatorname{atan}\left[\frac{\sqrt{\lambda_1'^2 - \Delta \bar{w}'^2}}{\Delta \bar{w}'}\right] = \sqrt{\lambda_1'^2 - \Delta \bar{w}'^2} - \pi \qquad \text{for all} \qquad 0 \leq \Delta \bar{w}'. \tag{S109}$$

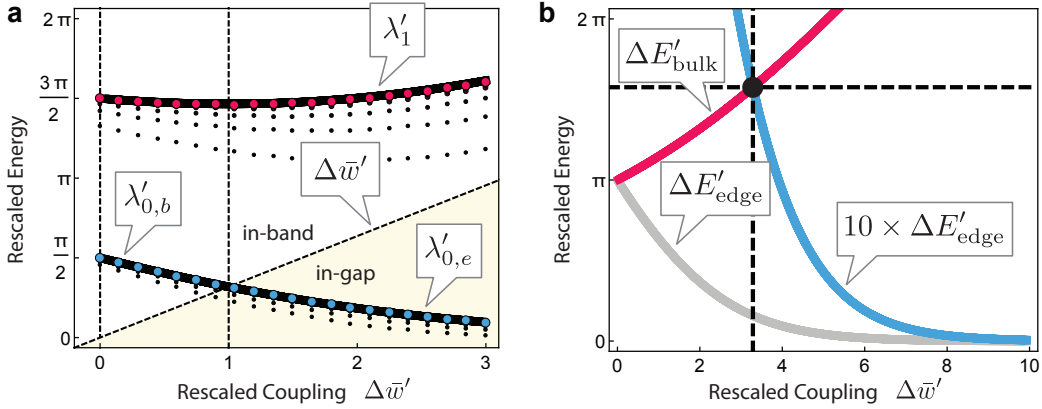

Supplementary Figure S8. *Universal scaling.* (a) Numerical (bullets) and analytical (lines) data for the rescaled lowest (edge mode) energy $\lambda'_0 = L\lambda_0$ and the lowest band (bulk mode) energy $\lambda'_1 = L\lambda_1$ as a function of the rescaled coupling $\Delta \bar{w}' = L\Delta \bar{w}$. Bold coloured bullets mark numerical data for a quasi-infinite system of length $L = 200$ whereas finite size effects are evident for smaller systems (small black bullets) starting at $L = 5$. Note that the solutions for $\lambda'_0$ split into the two types $\lambda'_{0,b}$ for $\lambda'_0 > \Delta\bar{w}'$ (in-band) and $\lambda'_{0,e}$ for $\lambda'_0 < \Delta\bar{w}'$ (in-gap). (b) Analytical results for the rescaled bulk-edge separation $\Delta E'_{\text{bulk}} = |\lambda'_1 - \lambda'_0|$ and the rescaled edge-mode splitting $\Delta E'_{\text{edge}} = 2\lambda'_0$. Additionally, we show the scaled edge-mode energy $10 \times \Delta E'_{\text{edge}}$; the intersection of the latter with $\Delta E'_{\text{bulk}}$ determines parameters of fixed bulk-edge energy ratio. We find $\Delta \bar{w}' \approx 3.3$ and $\Delta E'_{\text{bulk}} \approx 5.0$ (black bullet). Note that $\Delta E'_{\text{bulk}} = \pi = \Delta E'_{\text{edge}}$ for $\Delta \bar{w}' = 0$, i.e., at the critical point where the spectrum becomes linear.

These results are illustrated in Supplementary Fig. S8 (a) and compared to finite size numerical results.

The bottom line of this analysis is that we can fix the ratio $R$ of the two relevant energy scales for a transfer, namely $\Delta E_{\text{edge}}$ and $\Delta E_{\text{bulk}}$ if we approach the topological phase transition from within the topological phase as $\Delta \bar{w} = \Delta \bar{w}'/L$ for $L \to \infty$:

$$\Delta E_{\text{bulk}} = R \, \Delta E_{\text{edge}} \quad \Leftrightarrow \quad \lambda'_1 - \lambda'_0 = \Delta E'_{\text{bulk}} = R \, \Delta E'_{\text{edge}} = 2R \, \lambda'_0 \quad \Leftrightarrow \quad \lambda'_1 = (2R+1)\,\lambda'_0. \tag{S110}$$

In conjunction with equation (S100), this constraint implicitly determines $\Delta \bar{w}'$. E.g., for $R = 10$ we find

$$\Delta \bar{w}' \approx 3.3 \tag{S111a}$$
$$\lambda'_1 = 21\,\lambda'_0 \approx 5.2 \tag{S111b}$$
$$\Delta E'_{\text{bulk}} = 20\lambda'_0 \approx 5.0 \tag{S111c}$$
$$\Delta E'_{\text{edge}} = 2\lambda'_0 \approx 0.5 \tag{S111d}$$

which is illustrated in Supplementary Fig. S8 (b). Therefore we have

$$\Delta \bar{w} \sim \frac{3.3}{L} \quad \text{and} \quad \tau \gtrsim 2.0 \cdot L = \Delta E_{\text{edge}}^{-1} \tag{S112}$$

for the given energy ratio $R = 10$. Note that the condition $\tau \gtrsim \Delta E_{\text{edge}}^{-1}$ is merely necessary to facilitate at least one Rabi cycle between the edges; in particular, it allows for optimal scaling $\tau \sim L$. However, adiabatically decoupling bulk modes from the edge subspace is determined by $\Delta E_{\text{bulk}}$ which vanishes also with $L^{-1}$. This motivates the analysis of the next section VI.

## VI. ADIABATICITY

> To streamline mathematical expressions, we replace the calligraphic symbols $\mathcal{P}$ and $\mathcal{F}$ used for pulses in the main text by lower-case letters $p$ and $f$.

We use rigorous bounds on non-adiabatic losses [S10] in conjunction with the previously derived scaling of the edge-mode splitting and bulk gap to tackle the adiabatic bulk-edge decoupling quantitatively.

In the following, we write the SSH chain Hamiltonian in the form

$$H(s) = H_0 + \bar{w}(s)\,H_1 \tag{S113}$$

where $H_0$ ($H_1$) describes the topological (trivial) dimerization of the chain (all couplings set to 1), $0 \leq \bar{w}(s) \leq 1$ encodes the coupling, and $s = a + (b-a)\,t/\tau$ is a dimensionless time with $s \in [a,b]$, whereas $t \in [0,\tau]$. Let $g = \Delta E_{\text{bulk}}/2$ be half the gap separating bulk from edge modes. With our previous results [see also Supplementary Fig. S9 (a)], we can estimate

$$2g(\bar{w}) = \Delta E_{\text{bulk}} \geq \Delta \bar{w} = 1 - \bar{w} \tag{S114}$$

which becomes asymptotically an equality for $L \to \infty$.

In [S10], the following rigorous upper bound on the non-adiabatic losses was derived:

$$1 - \mathcal{E} = \langle \Psi_0 | U_\tau^\dagger Q_0 U_\tau | \Psi_0 \rangle \leq C^2 \tag{S115}$$

with

$$C = \frac{2}{\tau}\left[\frac{\|\dot{H}(a)\|}{g^2(a)} + \frac{\|\dot{H}(b)\|}{g^2(b)}\right] + \frac{2}{\tau}\int_a^b \mathrm{d}s \left(\frac{\|\ddot{H}(s)\|}{g^2(s)} + 7\sqrt{2}\,\frac{\|\dot{H}(s)\|^2}{g^3(s)}\right) \tag{S116}$$

where $Q_0 = \mathbb{1} - |1,\mathbf{0},0\rangle\langle 1,\mathbf{0},0| - |0,\mathbf{0},1\rangle\langle 0,\mathbf{0},1|$ is the projector onto the bulk sector at the beginning ($s = a$) and at the end ($s = b$) and $\|\bullet\|$ is the operator norm. We have $\|H_0\| = 1 = \|H_1\|$ (both describe decoupled dimers with maximum absolute eigenvalue 1) and will consider pulses $\bar{w}(s) \in C^k(\mathbb{R},[0,1])$ with $k \geq 1$ which are compactly supported on an interval $I = [a,b]$ so that

$$C = \frac{2}{\tau}\int_a^b \mathrm{d}s \left(\frac{|\ddot{\bar{w}}(s)|}{g^2(s)} + 7\sqrt{2}\,\frac{|\dot{\bar{w}}(s)|^2}{g^3(s)}\right). \tag{S117}$$

Equation (S114) allows us to estimate

$$C \leq \frac{2}{\tau}\int_a^b \mathrm{d}s \left[2^2\,\frac{|\ddot{\bar{w}}(s)|}{(1-\bar{w}(s))^2} + 7\sqrt{2}\,2^3\,\frac{|\dot{\bar{w}}(s)|^2}{(1-\bar{w}(s))^3}\right]. \tag{S118}$$

To allow for a complete state transfer, we showed earlier that the critical coupling has to be approached as $\bar{w}_{\text{max}} \sim 1 - \Delta\bar{w}'_{\text{min}}/L$, where $\bar{w}_{\text{max}} = \max_{s \in I} \bar{w}(s)$, in combination with an (at least) linearly growing protocol timescale $\tau \gtrsim L$.

Therefore we assume the form

$$\bar{w}(s) = (1 - \Delta\bar{w}'_{\text{min}}/L) \cdot p(s) \tag{S119}$$

for $L > \Delta\bar{w}'_{\text{min}}$ with $p(a) = 0 = p(b)$, $p(c) = 1$, and $p$ monotonically increasing (decreasing) on $[a,c]$ ($[c,b]$), where $c = (a+b)/2$. For the following analysis, it is more convenient to introduce $\bar{p}(s) = 1 - p(s)$ which vanishes during the pulse: $\bar{p}(c) = 0$. Then

$$\boxed{C \leq \frac{C_1}{\tau}\underbrace{\int_a^b \mathrm{d}s\,\frac{|\bar{p}''(s)|}{(\varepsilon_L + \bar{p}(s))^2}}_{\equiv \chi_1} + \frac{C_2}{\tau}\underbrace{\int_a^b \mathrm{d}s\,\frac{|\bar{p}'(s)|^2}{(\varepsilon_L + \bar{p}(s))^3}}_{\equiv \chi_2} \equiv \frac{C_L[\bar{p}]}{\tau}} \tag{S120}$$

with

$$C_1 = \frac{2^3}{1 - \Delta\bar{w}'_{\text{min}}/L} \sim \text{const}, \quad C_2 = \frac{7\sqrt{2}\,2^4}{1 - \Delta\bar{w}'_{\text{min}}/L} \sim \text{const} \quad \text{and} \quad \varepsilon_L = \frac{\Delta\bar{w}'_{\text{min}}}{L - \Delta\bar{w}'_{\text{min}}} \sim \frac{1}{L}. \tag{S121}$$

For $L \to \infty$, $C_L[\bar{p}]$ will diverge due to the vanishing of $\bar{p}(s)$ at $s = c$ and $\lim_{L \to \infty} \varepsilon_L = 0$. In order to keep the bulk losses $1 - \mathcal{E}$ constant for $L \to \infty$, we have to scale the protocol timescale as $\tau \sim L^{1+\alpha_{\bar{p}}}$ with $\alpha_{\bar{p}} \geq 0$ such that the asymptotic order of $C_L[\bar{p}]$ is matched and $\lim_{L \to \infty} C_L[\bar{p}]/L^{1+\alpha_{\bar{p}}} = \text{const}$.

We stress that common knowledge tells us that $\tau \gtrsim 1/g_{\text{min}}^2 \sim L^2$ is needed for adiabaticity, i.e., $\alpha_{\bar{p}} = 1$, since $g_{\text{min}} \approx \varepsilon_L \sim 1/L$, whereas the Lieb-Robinson bound [S11] (and $\Delta E_{\text{edge}} \sim 1/L$) in principle allows for linear scaling: $\alpha_{\bar{p}} = 0$.

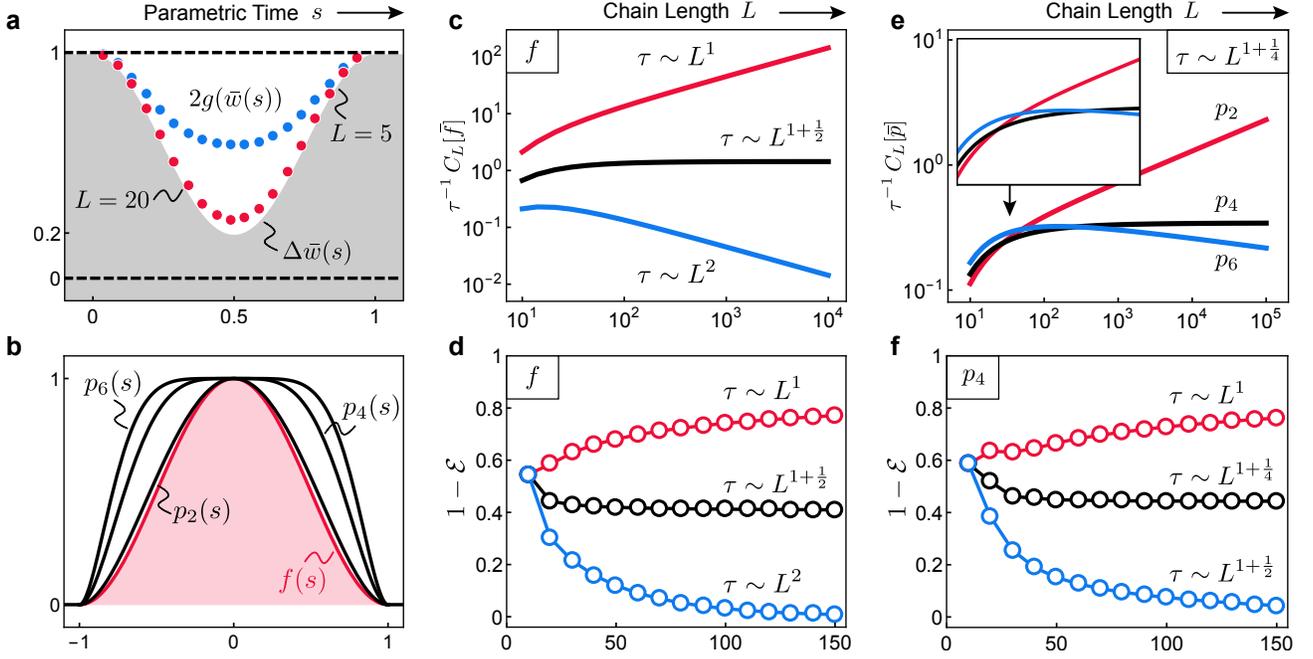

Supplementary Figure S9. *Adiabatic bulk-edge decoupling.* **(a)** Finite-size bulk gap $\Delta E_\text{bulk} = 2g(\bar{w}(s))$ for $\bar{w}(s) = 0.8 \cdot \sin^2(\pi s)$ as a function of parametric time $s \in [0,1]$ for two systems of size $L=5$ (blue bullets) and $L=20$ (red bullets). For comparison, the exact bulk gap $\Delta \bar{w}(s) = 1 - \bar{w}(s)$ in the thermodynamic limit ($L \to \infty$) is drawn as shaded region. Clearly $2g(\bar{w}(s)) \geq \Delta \bar{w}(s)$ for systems of finite size; this lower bound on $g$ is used in the text for the estimation of the bulk losses. **(b)** The three polynomial pulses (black lines) $p_n(s)$ for $n = 2, 4, 6$ as derived in the text, see equation (S142). For comparison, we show the shifted and rescaled pulse $f(s) = \sin^2\left[\frac{\pi}{2}(x-1)\right]$ (red) which was used in the main text. All four pulses are compactly supported on $I = [-1, 1]$ and continuously differentiable on $\mathbb{R}$. **(c)** Numerical evaluations of the upper bounds $C_L[\bar{f}]/\tau$ for $f(s) = \sin^2(\pi s)$ with $\tau = \tau_0 \cdot L^{1+\alpha}$ for $\alpha = 0, \frac{1}{2}, 1$ ($\tau_0 = 10^3$). Obviously $\tau \sim L^{1+\frac{1}{2}}$ leads to a finite upper bound $C$ for $L \to \infty$, i.e., $\alpha_{\bar{f}} = \frac{1}{2}$. **(d)** Numerical simulations of the bulk losses $1 - \mathcal{E}$ with $f(s) = \sin^2(\pi s)$, $\bar{w}_\text{max} = 1 - 3.3/L$ and $\tau = \tau_0 \cdot L^{1+\alpha}$ for $\alpha = 0, \frac{1}{2}, 1$ ($\tau_0$ is normalized such that $\tau = 10$ for $L = 10$). Note that for $\tau \sim L$ the bulk losses increase with $L$ whereas for $\tau \sim L^{1+\frac{1}{2}}$ they converge towards a constant value which can be made arbitrarily small by increasing the prefactor $\tau_0$. **(e)** Numerical evaluations of the upper bounds $C_L[\bar{p}]/\tau$ for polynomials $p_n(s)$ with $n = 2, 4, 6$ as shown in **(b)**. $\tau = \tau_0 \cdot L^{1+\frac{1}{4}}$ ($\tau_0 = 500$) is fixed for all three such that $C_L[\bar{p}_4]/\tau$ converges to a finite value. The inset illustrates the dependence of the prefactors $C_n[q]$ on the pulse shape (that is, $n$): With $n$ the prefactors grow so that the upper bounds intersect when plotted over $L$. Thus it may be beneficial to choose pulse shapes with poor scaling in $L$ for small systems. **(f)** Numerical simulations of $1 - \mathcal{E}$ with the polynomial pulse $p_4(s)$, $\bar{w}_\text{max} = 1 - 3.3/L$ and $\tau = \tau_0 \cdot L^{1+\alpha}$ for $\alpha = 0, \frac{1}{4}, \frac{1}{2}$. For $\tau \sim L$ the bulk losses increase with $L$ whereas for $\tau \sim L^{1+\frac{1}{4}}$ one finds constant losses; for $\tau \sim L^{1+\frac{1}{2}}$ the losses vanish with $L \to \infty$. Compare these results with **(d)**.

In the following, we demonstrate that our pulse $\bar{p}(s) = \cos^2(\pi s) = 1 - \sin^2(\pi s)$ yields a scaling of $\alpha_{\bar{p}} = 1/2$ and thereby already surpasses the naïve estimates for adiabaticity. Subsequently, we present a sequence of pulses $\bar{p}_n(s)$ for which provably $\alpha_{\bar{p}_n} = 1/n$, so that the optimal scaling $\alpha_{\bar{p}} = 0$ can be approached systematically. The question whether there is an optimal pulse $p(s)$ with $\alpha_{\bar{p}} = 0$ is left open. We compare all estimates (which are, after all, only *sufficient* conditions for adiabaticity) with numerical simulations of the bulk losses and find that the actual scaling saturates the upper bounds (up to $L$-independent prefactors).

1. We start with $\bar{p} = \bar{f}(s) = \cos^2(\pi s) = 1 - \sin^2(\pi s)$ for $s \in [0, 1]$, as used for demonstrative purposes in the main text. We have to evaluate

$$C_L[\bar{f}] = \int_0^1 \mathrm{d}s \left[ C_1 \frac{2\pi^2 |\cos(2\pi s)|}{(\varepsilon_L + \cos^2(\pi s))^2} + C_2 \frac{\pi^2 \sin^2(2\pi s)}{(\varepsilon_L + \cos^2(\pi s))^3} \right] \tag{S122}$$

where we set $\Delta \bar{w}'_\text{min} = 3.3$. We evaluate the integrals numerically and plot $C_L[\bar{f}]/\tau$ as a function of $L$ for $\tau = \tau_0 \cdot L^{1+\alpha}$ with $\alpha = 0, \frac{1}{2}, 1$ in Supplementary Fig. S9 (c). To compare the scaling of the rigorous upper bounds on the bulk losses with the real system, we simulate the time evolution for $f(s) = \sin^2(\pi s)$ with

$\tau = \tau_0 \cdot L^{1+\alpha}$ ($\alpha = 0, \frac{1}{2}, 1$) and $\bar{w}_{\max} = 1 - 3.3/L$ and calculate $1 - \mathcal{E}$ at $t = \tau$. Note that we do *not* tune $\bar{w}_{\max}$ or $\tau$ in any way to optimize the *transfer* $\mathcal{O}$ as we are only interested in the adiabatic decoupling of bulk and edge at this point. The numerical losses are shown in Supplementary Fig. S9 (d): Up to finite size effects and $L$-independent prefactors, the rigorous upper bounds capture the scaling of the actual system quite nicely. In particular, the result $\alpha_{\bar{f}} = \frac{1}{2}$ is verified. The latter surpasses the conservative estimate $\tau \gtrsim 1/g_{\min}^2 \sim L^2$ but does not reach optimal (linear) scaling $\alpha_{\bar{f}} = 0$.

2. Let us now follow a more systematic approach and consider pulses of the form

$$\bar{p}_n(s) = s^n \cdot q(s) \qquad \text{for} \qquad s \in I = [-1, 1] \quad \text{and} \quad n \geq 2 \text{ even} \tag{S123}$$

with $s^{-n} \geq q(s) > 0$ on $I$ and $q(\pm 1) = 1$.

We impose the continuity conditions

$$\bar{p}'_n(\pm 1) = \pm n\, q(\pm 1) + q'(\pm 1) = 0 \qquad \Rightarrow \qquad q'(\pm 1) = \mp n \tag{S124}$$

so that the boundary terms in equation (S116) vanish and $\bar{p}_n(s)$ becomes a $C^1$-function if set to 1 outside $I$ (which corresponds to the stable situation of statically decoupled edge modes). One could smoothen the function further by requiring

$$\bar{p}''_n(\pm 1) = n(n-1)\, q(\pm 1) \pm 2n\, q'(\pm 1) + q''(\pm 1) = 0 \qquad \Rightarrow \qquad q''(\pm 1) = n(n+1) \sim n^2 \tag{S125}$$

so that $\bar{p}_n \in C^2$ if extended with 1 to $\mathbb{R}$. This, however, will not change the gist of the statements that follow. The upshot of these considerations is that getting rid of higher derivatives at the critical time $s = 0$ (when the gap is of order $1/L$) has to be paid for by growing derivatives of $q(s)$ at the boundaries of $I$ in order to smoothen the transition into the stationary, decoupled state before and after the pulse.

Note that for the $m$th derivative one has

$$\bar{p}_n^{(m)} = \sum_{k=0}^{m} \binom{m}{k} \frac{n!}{(n-k)!}\, s^{n-k} q^{(m-k)}(s) \tag{S126}$$

so that $\bar{p}_n^{(m)}(0) = 0$ for $m < n$. The motivation is that flattening the pulse close to the critical region at $s = 0$ (where $g \approx g_{\min} \sim 1/L$) may be beneficial for the scaling of $\tau$ with $L$.

In particular,

$$\bar{p}(s) = s^n q(s) \tag{S127a}$$
$$\bar{p}'(s) = n\, s^{n-1} q(s) + s^n q'(s) \tag{S127b}$$
$$\bar{p}''(s) = n(n-1)\, s^{n-2} q(s) + 2n\, s^{n-1} q'(s) + s^n q''(s). \tag{S127c}$$

In equation (S120), this yields

$$\chi_1 \leq n(n-1) \underbrace{\int_I ds\, \frac{|s^{n-2} q(s)|}{(\varepsilon_L + s^n q(s))^2}}_{\chi_{1,1}} + 2n \underbrace{\int_I ds\, \frac{|s^{n-1} q'(s)|}{(\varepsilon_L + s^n q(s))^2}}_{\chi_{1,2}} + \underbrace{\int_I ds\, \frac{|s^n q''(s)|}{(\varepsilon_L + s^n q(s))^2}}_{\chi_{1,3}} \tag{S128}$$

and

$$\chi_2 \leq n^2 \underbrace{\int_I ds\, \frac{|s^{2n-2} q^2(s)|}{(\varepsilon_L + s^n q(s))^3}}_{\chi_{2,1}} + 2n \underbrace{\int_I ds\, \frac{|s^{2n-1} q(s) q'(s)|}{(\varepsilon_L + s^n q(s))^3}}_{\chi_{2,2}} + \underbrace{\int_I ds\, \frac{|s^{2n} (q'(s))^2|}{(\varepsilon_L + s^n q(s))^3}}_{\chi_{3,3}} \tag{S129}$$

where the inequalities are due to the triangle inequality.

If we introduce the minimum $0 < q \equiv \min_{s \in I} |q(s)|$ and the maximum $Q \equiv \max_{s \in I} |q(s)|$ (similarly for derivatives: $Q'$ and $Q''$), the integrals can be estimated straightforwardly:

- The first term of $\chi_1$ reads

$$\chi_{1,1} = \int_{-1}^{1} ds \frac{|s^{n-2}q(s)|}{(\varepsilon_L + s^n q(s))^2} \leq \int_{-1}^{1} ds \frac{s^{n-2}Q}{(\varepsilon_L + s^n q)^2} = \frac{2}{\varepsilon_L^2} \int_{0}^{1} ds \frac{s^{n-2}Q}{(1 + s^n q/\varepsilon_L)^2} \tag{S130a}$$

$$= 2Q\, \varepsilon_L^{-2+1/n+1-2/n}\, q^{-1/n-1+2/n} \int_{0}^{(q/\varepsilon_L)^{\frac{1}{n}}} du \frac{u^{n-2}}{(1+u^n)^2} \tag{S130b}$$

$$\leq 2Q\, \varepsilon_L^{-1-1/n}\, q^{1/n-1} \int_{0}^{\infty} du \frac{u^{n-2}}{(1+u^n)^2} = 2Q\, \varepsilon_L^{-1-1/n}\, q^{1/n-1} \frac{\pi}{n^2 \sin \frac{\pi}{n}} \tag{S130c}$$

$$= \frac{1}{\varepsilon_L} \left[\frac{q}{\varepsilon_L}\right]^{\frac{1}{n}} \frac{2\pi}{n^2 \sin \frac{\pi}{n}} \frac{Q}{q} \tag{S130d}$$

where we used that $s^n \geq 0$ for even $n$. In the second row we substituted $u = (q/\varepsilon_L)^{\frac{1}{n}} s$. Note that the last estimate becomes an equality for $L \to \infty$ since $\varepsilon_L \to 0$.

- The second term reads

$$\chi_{1,2} = \int_{-1}^{1} ds \frac{|s^{n-1}q'(s)|}{(\varepsilon_L + s^n q(s))^2} \leq \int_{-1}^{1} ds \frac{|s|^{n-1}Q'}{(\varepsilon_L + s^n q)^2} = \frac{2}{\varepsilon_L^2} \int_{0}^{1} ds \frac{s^{n-1}Q'}{(1+s^n q/\varepsilon_L)^2} \tag{S131a}$$

$$= 2Q'\, \varepsilon_L^{-2+1/n+1-1/n}\, q^{-1/n-1+1/n} \int_{0}^{(q/\varepsilon_L)^{\frac{1}{n}}} du \frac{u^{n-1}}{(1+u^n)^2} \tag{S131b}$$

$$\leq 2Q'\, \varepsilon_L^{-1}\, q^{-1} \int_{0}^{\infty} du \frac{u^{n-1}}{(1+u^n)^2} = 2Q'\, \varepsilon_L^{-1}\, q^{-1} \frac{1}{n} \tag{S131c}$$

$$= \frac{1}{\varepsilon_L} \frac{2}{n} \frac{Q'}{q}. \tag{S131d}$$

- The third term reads

$$\chi_{1,3} = \int_{-1}^{1} ds \frac{|s^n q''(s)|}{(\varepsilon_L + s^n q(s))^2} \leq \int_{-1}^{1} ds \frac{s^n Q''}{(\varepsilon_L + s^n q)^2} = \frac{2}{\varepsilon_L^2} \int_{0}^{1} ds \frac{s^n Q''}{(1+s^n q/\varepsilon_L)^2} \tag{S132a}$$

$$= 2Q''\, \varepsilon_L^{-2+1/n+1}\, q^{-1/n-1} \int_{0}^{(q/\varepsilon_L)^{\frac{1}{n}}} du \frac{u^n}{(1+u^n)^2} \tag{S132b}$$

$$\leq 2Q''\, \varepsilon_L^{-1+1/n}\, q^{-1-1/n} \int_{0}^{\infty} du \frac{u^n}{(1+u^n)^2} = 2Q''\, \varepsilon_L^{-1+1/n}\, q^{-1-1/n} \frac{\pi}{n^2 \sin \frac{\pi}{n}} \tag{S132c}$$

$$= \frac{1}{\varepsilon_L} \left[\frac{\varepsilon_L}{q}\right]^{\frac{1}{n}} \frac{2\pi}{n^2 \sin \frac{\pi}{n}} \frac{Q''}{q}. \tag{S132d}$$

- The first term of $\chi_2$ reads

$$\chi_{2,1} = \int_{-1}^{1} ds \frac{|s^{2n-2}q^2(s)|}{(\varepsilon_L + s^n q(s))^3} \leq \int_{-1}^{1} ds \frac{s^{2n-2}Q^2}{(\varepsilon_L + s^n q)^3} = \frac{2}{\varepsilon_L^3} \int_{0}^{1} ds \frac{s^{2n-2}Q^2}{(1+s^n q/\varepsilon_L)^3} \tag{S133a}$$

$$= 2Q^2\, \varepsilon_L^{-3+1/n+2-2/n}\, q^{-1/n-2+2/n} \int_{0}^{(q/\varepsilon_L)^{\frac{1}{n}}} du \frac{u^{2n-2}}{(1+u^n)^3} \tag{S133b}$$

$$\leq 2Q^2\, \varepsilon_L^{-1-1/n}\, q^{-2+1/n} \int_{0}^{\infty} du \frac{u^{2n-2}}{(1+u^n)^3} = 2Q^2\, \varepsilon_L^{-1-1/n}\, q^{-2+1/n} \frac{\pi(n-1)}{2n^3 \sin \frac{\pi}{n}} \tag{S133c}$$

$$= \frac{1}{\varepsilon_L} \left[\frac{q}{\varepsilon_L}\right]^{\frac{1}{n}} \frac{\pi(n-1)}{n^3 \sin \frac{\pi}{n}} \left[\frac{Q}{q}\right]^2. \tag{S133d}$$

- The second term reads

$$\chi_{2,2} = \int_{-1}^{1} \mathrm{d}s \, \frac{|s^{2n-1} q(s) q'(s)|}{(\varepsilon_L + s^n q(s))^3} \leq \int_{-1}^{1} \mathrm{d}s \, \frac{|s|^{2n-1} QQ'}{(\varepsilon_L + s^n q)^3} = \frac{2}{\varepsilon_L^3} \int_0^1 \mathrm{d}s \, \frac{s^{2n-1} QQ'}{(1 + s^n q/\varepsilon_L)^3} \tag{S134a}$$

$$= 2QQ' \, \varepsilon_L^{-3+1/n+2-1/n} \, q^{-1/n-2+1/n} \int_0^{(q/\varepsilon_L)^{\frac{1}{n}}} \mathrm{d}u \, \frac{u^{2n-1}}{(1+u^n)^3} \tag{S134b}$$

$$\leq 2QQ' \, \varepsilon_L^{-1} q^{-2} \int_0^\infty \mathrm{d}u \, \frac{u^{2n-1}}{(1+u^n)^3} = 2QQ' \, \varepsilon_L^{-1} q^{-2} \, \frac{1}{2n} \tag{S134c}$$

$$= \frac{1}{\varepsilon_L} \frac{1}{n} \frac{QQ'}{q^2}. \tag{S134d}$$

- The third term reads

$$\chi_{2,3} = \int_{-1}^{1} \mathrm{d}s \, \frac{|s^{2n} (q'(s))^2|}{(\varepsilon_L + s^n q(s))^3} \leq \int_{-1}^{1} \mathrm{d}s \, \frac{s^{2n} Q'^2}{(\varepsilon_L + s^n q)^3} = \frac{2}{\varepsilon_L^3} \int_0^1 \mathrm{d}s \, \frac{s^{2n} Q'^2}{(1 + s^n q/\varepsilon_L)^3} \tag{S135a}$$

$$= 2Q'^2 \, \varepsilon_L^{-3+1/n+2} \, q^{-1/n-2} \int_0^{(q/\varepsilon_L)^{\frac{1}{n}}} \mathrm{d}u \, \frac{u^{2n}}{(1+u^n)^3} \tag{S135b}$$

$$\leq 2Q'^2 \, \varepsilon_L^{-1+1/n} \, q^{-2-1/n} \int_0^\infty \mathrm{d}u \, \frac{u^{2n}}{(1+u^n)^3} = 2Q'^2 \, \varepsilon_L^{-1+1/n} \, q^{-2-1/n} \, \frac{\pi(n+1)}{2n^3 \sin \frac{\pi}{n}} \tag{S135c}$$

$$= \frac{1}{\varepsilon_L} \left[ \frac{\varepsilon_L}{q} \right]^{\frac{1}{n}} \frac{\pi(n+1)}{n^3 \sin \frac{\pi}{n}} \left[ \frac{Q'}{q} \right]^2. \tag{S135d}$$

Combining these results in equation (S128) and (S129) with equation (S120) yields the final upper bound for bulk losses

$$\begin{aligned}
C_L[\bar{p}_n] \, \varepsilon_L \;\leq\; & \left[ \frac{q}{\varepsilon_L} \right]^{\frac{1}{n}} \frac{2\pi \, C_1 \, n(n-1)}{n^2 \sin \frac{\pi}{n}} \frac{Q}{q} + \frac{2 \, C_1 \, 2n}{n} \frac{Q'}{q} + \left[ \frac{\varepsilon_L}{q} \right]^{\frac{1}{n}} \frac{2\pi \, C_1}{n^2 \sin \frac{\pi}{n}} \frac{Q''}{q} \\
& + \left[ \frac{q}{\varepsilon_L} \right]^{\frac{1}{n}} \frac{\pi(n-1) \, C_2 \, n^2}{n^3 \sin \frac{\pi}{n}} \left[ \frac{Q}{q} \right]^2 + \frac{C_2 \, 2n}{n} \frac{QQ'}{q^2} + \left[ \frac{\varepsilon_L}{q} \right]^{\frac{1}{n}} \frac{\pi(n+1) \, C_2}{n^3 \sin \frac{\pi}{n}} \left[ \frac{Q'}{q} \right]^2
\end{aligned} \tag{S136a}$$

$$\stackrel{L \to \infty}{\leq} \left[ \frac{q}{\varepsilon_L} \right]^{\frac{1}{n}} \frac{3\pi \, (n-1)}{n \sin \frac{\pi}{n}} \left[ 2C_1 \frac{Q}{q} + C_2 \left( \frac{Q}{q} \right)^2 \right] \tag{S136b}$$

where the last line describes the dominant term for $L \to \infty$. In conclusion, we have

$$\sqrt{1 - \mathcal{E}} \stackrel{L \to \infty}{\leq} \frac{1}{\tau \, \varepsilon_L^{1+\frac{1}{n}}} \frac{3\pi \, (n-1) \, q^{\frac{1}{n}}}{n \sin \frac{\pi}{n}} \left[ 2C_1 \frac{Q}{q} + C_2 \left( \frac{Q}{q} \right)^2 \right] \equiv \frac{C_n[q]}{\tau \, \varepsilon_L^{1+\frac{1}{n}}}. \tag{S137}$$

With $\varepsilon_L \sim \frac{1}{L}$ it follows that $\tau \sim L^{1+\frac{1}{n}}$ is sufficient to keep the bulk losses constant for $L \to \infty$ if a pulse of the form $\bar{p}_n$ is used instead of $\bar{f}(s) = \cos^2(\pi s)$.

There are a few comments in order:

1. If we taylor $\bar{f}(s) = \cos^2(\pi s)$ around its minimum at $s = 1/2$,

$$\cos^2(\pi s) = \left( s - \frac{1}{2} \right)^2 \cdot \underbrace{\left[ \pi^2 - \frac{\pi^4}{3} \left( s - \frac{1}{2} \right)^2 + \ldots \right]}_{>0 \text{ for } s \in [0,1]}, \tag{S138}$$

we immediately conclude that $\alpha_{\bar{f}} = \frac{1}{2}$ since $\bar{f}(s)$ is of the form $\bar{p}_2(s)$ for appropriately chosen $q(s)$ (and shifted/rescaled $s$).

2. It is important to stress that the coefficient

$$C_n[q] = \frac{3\pi (n-1) q^{\frac{1}{n}}}{n \sin \frac{\pi}{n}} \left[ 2C_1 \frac{Q}{q} + C_2 \left(\frac{Q}{q}\right)^2 \right] \tag{S139}$$

is independent of $L$ but *does* depend on the pulse shape via $n$ and $Q/q$: First, for $n \to \infty$, we have

$$\frac{\pi (n-1)}{n \sin \frac{\pi}{n}} \to n. \tag{S140}$$

The better scaling comes at the price of larger upper bounds, i.e., larger time scales $\tau$ to begin with. Secondly, $Q$ tends to diverge with $n \to \infty$ as well. We already showed that continuous differentiability at the beginning and end of the pulse implies (at least) $|q'(\pm 1)| = n$ and therefore $Q' \geq n$, and $|q''(\pm 1)| \sim n^2 \Rightarrow Q'' \geq n^2$ if $\bar{p}_n \in C^2$ is required. Note that this blows up the coefficients of the sub-leading terms in equation (S136a).

The interplay of $C_n[q]$ and the $L^{1+\frac{1}{n}}$-scaling can lead to the situation depicted in Supplementary Fig. S9 (e) where it is beneficial for small $L$ to choose $n$ smaller despite the inferior $L$-scaling, simply because the prefactors can be prohibitively large when $L$ is not yet large enough.

3. Constructing possible $q(s)$ for given $n$ so that $p_n(s) = 1 - \bar{p}_n(s)$ is compactly supported on $[-1, 1]$ and $k$-times continuously differentiable on $\mathbb{R}$ is easily accomplished with the polynomial ansatz

$$1 - p_n(s) = \sum_{j=n}^{D} \rho_j s^j \tag{S141}$$

for large enough $D$. Solving for $\{\rho_j\}$ yields possible solutions ($k = 1$)

$$p_2(s) = 1 - x^2(2 - 1x^2) \tag{S142a}$$
$$p_4(s) = 1 - x^4(3 - 2x^2) \tag{S142b}$$
$$p_6(s) = 1 - x^6(4 - 3x^2) \tag{S142c}$$
$$\vdots$$

which are plotted in Supplementary Fig. S9 (b) and compared with $f(s) = \sin^2(\pi s)$. In Supplementary Fig. S9 (f) we show numerical results for $1 - \mathcal{E}$ for the polynomial pulse $p_4(s)$.

---

[S1] Kitaev, A. Y. Unpaired majorana fermions in quantum wires. *Physics-Uspekhi* **44**, 131–136 (2001).
[S2] Wigner, E. P. On the statistical distribution of the widths and spacings of nuclear resonance levels. *Mathematical Proceedings of the Cambridge Philosophical Society* **47**, 790 (1951).
[S3] Wigner, E. P. On the distribution of the roots of certain symmetric matrices. *The Annals of Mathematics* **67**, 325 (1958).
[S4] Dyson, F. J. The threefold way. algebraic structure of symmetry groups and ensembles in quantum mechanics. *Journal of Mathematical Physics* **3**, 1199–1215 (1962).
[S5] Altland, A. & Zirnbauer, M. R. Nonstandard symmetry classes in mesoscopic normal-superconducting hybrid structures. *Physical Review B* **55**, 1142–1161 (1997).
[S6] Schnyder, A. P., Ryu, S., Furusaki, A. & Ludwig, A. W. W. Classification of topological insulators and superconductors in three spatial dimensions. *Physical Review B* **78**, 195125 (2008).
[S7] Kitaev, A. Periodic table for topological insulators and superconductors. In *AIP Conference Proceedings*, vol. 1134, 22–30 (AIP, 2009).
[S8] Schnyder, A. P., Ryu, S., Furusaki, A. & Ludwig, A. W. W. Classification of topological insulators and superconductors. In *AIP Conference Proceedings*, vol. 1134, 10–21 (AIP, 2009).
[S9] Ryu, S., Schnyder, A. P., Furusaki, A. & Ludwig, A. W. W. Topological insulators and superconductors: tenfold way and dimensional hierarchy. *New Journal of Physics* **12**, 065010 (2010).
[S10] Jansen, S., Ruskai, M.-B. & Seiler, R. Bounds for the adiabatic approximation with applications to quantum computation. *Journal of Mathematical Physics* **48**, 102111 (2007).
[S11] Lieb, E. H. & Robinson, D. W. The finite group velocity of quantum spin systems. *Communications in Mathematical Physics* **28**, 251–257 (1972).



\end{document}